\documentclass[fleqn,usenatbib]{mnras}

\usepackage{newtxtext,newtxmath}

\usepackage[T1]{fontenc}

\DeclareRobustCommand{\VAN}[3]{#2}
\let\VANthebibliography\thebibliography
\def\thebibliography{\DeclareRobustCommand{\VAN}[3]{##3}\VANthebibliography}

\usepackage{graphicx}	
\usepackage{amsmath}	

\usepackage{multirow}
\usepackage{xcolor}
\usepackage{caption}
\usepackage{subcaption}
\usepackage{threeparttable}
\usepackage{rotating}
\usepackage{rotfloat}
\usepackage{float}
\usepackage{moresize}
\usepackage{wrapfig}
\usepackage{multicol}
\usepackage[normalem]{ulem}
\usepackage{epstopdf}
\usepackage{longtable}
\usepackage{natbib}
\usepackage[export]{adjustbox}
\usepackage{tabularx}
\usepackage{tablefootnote}
\usepackage{url}
\usepackage[utf8]{inputenc}

\newcommand{\kms}{km\,s$^{-1}$}
\newcommand{\ms}{m~s$^{-1}$}

\newcommand{\rearth}{$R_{\oplus}$}

\newcommand{\logrhk}{$\rm log\,R^{\prime}_\mathrm{HK}$}

\title[TOI-1472, TOI-1648]{Precise mass and radius determination for two new and one known Neptune-sized planets around G Dwarf hosts}

\author[I. Carleo et al.]{
Ilaria Carleo,$^{1,2}$\thanks{E-mail: ilaria.carleo@inaf.it}
Grzegorz Nowak,$^{3}$
Felipe Murgas,$^{2,4}$
Enric Pall{\'e},$^{2,4}$
Gaia Lacedelli,$^{2,4}$
Thomas Masseron,$^{2,4}$
\newauthor Emily W. Wong,$^{5}$
Amadeo Castro-Gonz{\'a}lez,$^{5}$
Dawid Jankowski,$^{3}$
Patrick Eggenberger,$^{5}$
James S. Jenkins,$^{6,7}$
\newauthor Krzysztof Go\'zdziewski,$^{3}$
Vincent Bourrier,$^{5}$
Douglas R. Alves,$^{8,6}$
José I. Vines,$^{9}$
Keivan G.\ Stassun,$^{10}$
\newauthor Matteo Brogi,$^{11,1}$
Sergio Messina,$^{12}$
Catherine A. Clark,$^{13}$
Karen A.\ Collins,$^{14}$
Hans J.~Deeg,$^{2,4}$
Elise Furlan,$^{13}$
\newauthor Davide Gandolfi,$^{11}$
Samuel Geraldía González,$^{2,4}$
Coel Hellier,$^{15}$
Artie P. Hatzes,$^{16}$
Steve~B.~Howell,$^{17}$
\newauthor Judith Korth,$^{5}$
Emil Knudstrup,$^{18,24}$
Jorge Lillo-Box,$^{19}$
John H. Livingston,$^{20,21,22}$
Jaume Orell-Miquel,$^{23}$
\newauthor Carina Persson,$^{24}$
Seth Redfield,$^{25}$
Boris Safonov,$^{24}$
David Baker,$^{25}$
Rafael Delfin Barrena Delgado,$^{2,4}$
\newauthor Allyson Bieryla,$^{14}$
Andrew Boyle,$^{13,28}$
Pau Bosch-Cabot,$^{29}$
Núria Casasayas Barris,$^{2,4}$
Stavros Chairetas,$^{2}$
\newauthor David~R.~Ciardi,$^{13}$
Akihiko Fukui,$^{30,2}$
Pere Guerra,$^{29}$
Kiyoe Kawauchi,$^{31}$
Florence Libotte,$^{2,32,33}$
\newauthor Michael B.~Lund,$^{13}$
Rafael Luque,$^{34}$
Eduardo Lorenzo Martín Guerrero de Escalante,$^{2}$
Bob Massey,$^{35}$
\newauthor Edward J. Michaels,$^{36}$
Giuseppe Morello,$^{34,37}$
Norio Narita,$^{30,20,2}$
Hannu Parvianien,$^{2,4}$
Richard P. Schwarz,$^{14}$
\newauthor Avi Shporer,$^{38}$
Monika Stangret,$^{2,4}$
Cristilyn N. Watkins$^{39}$
\\
$^{1}$INAF -- Osservatorio Astrofisico di Torino, Via Osservatorio 20, I-10025, Pino Torinese, Italy\label{inafoato}\\
$^{2}$Instituto de Astrof\'{i}sica de Canarias (IAC), 38205 La Laguna, Tenerife, Spain \label{iac}\\ 
$^{3}$Institute of Astronomy, Faculty of Physics, Astronomy and Informatics, Nicolaus Copernicus University, Grudziądzka 5, 87-100 Toruń, Poland \label{umk}\\ 
$^{4}$Departamento de Astrof\'isica, Universidad de La Laguna (ULL), E-38206 La Laguna, Tenerife, Spain \label{ull}\\ 
$^{5}$Astronomical Observatory, University of Geneva, Chemin Pegasi 51b, CH-1290 Versoix, Switzerland \label{unige}\\  
$^{6}$Centro de Astrofísica y Tecnologías Afines (CATA), Casilla 36-D, 7591245, Santiago, Chile \label{cata}\\ 
$^{7}$Núcleo de Astronomía, Facultad de Ingeniería y Ciencias, Universidad Diego Portales, Av. Ejército 441 Santiago, Chile \label{udp}\\  
$^{8}$Departamento de Astronomía, Universidad de Chile, Casilla 36-D, 7591245, Santiago, Chile \label{uchile}\\
$^{9}$Instituto de Astronom\'ia, Universidad Cat\'olica del Norte, Angamos 0610, 1270709, Antofagasta, Chile\\
$^{10}$Department of Physics \& Astronomy, Vanderbilt University, Nashville, TN, USA \label{vanderbilt}\\ 
$^{11}$Dipartimento di Fisica, Universit{\'a} degli Studi di Torino, via Pietro Giuria 1, I-10125, Torino, Italy \label{unito}\\ 
$^{12}$INAF - Osservatorio Astrofisico di Catania, Via S. Sofia 78, I-95123 Catania, Italy \label{inaf-catania} \\
$^{13}$NASA Exoplanet Science Institute, IPAC, California Institute of Technology, Pasadena, CA 91125 USA \label{IPAC}\\ 
$^{14}$Center for Astrophysics \textbar \ Harvard \& Smithsonian, 60 Garden Street, Cambridge, MA 02138, USA \label{harvard}\\ 
$^{15}$Astrophysics Group, Keele University, Staffs ST5 5BG, U.K. \label{keele}\\ 
$^{16}$Th\"uringer Landessternwarte Sternwarte 5 D-07778, Tautenburg, Germany \label{tautenburg}\\ 
$^{17}$NASA Ames Research Center, Moffett Field, CA 94035, USA \label{ames}\\ 
$^{18}$Stellar Astrophysics Centre, Department of Physics and Astronomy, Aarhus University, Ny Munkegade 120, 8000 Aarhus C, Denmark \\ 
$^{19}$Centro de Astrobiolog\'ia (CAB, CSIC-INTA), Depto. de Astrof\'isica, ESAC campus, 28692, Villanueva de la Ca\~nada (Madrid), Spain \label{cab}\\ 
$^{20}$Astrobiology Center, 2-21-1 Osawa, Mitaka, Tokyo 181-8588, Japan \label{Tokyo}\\ 
$^{21}$National Astronomical Observatory of Japan, 2-21-1 Osawa, Mitaka, Tokyo 181-8588,  Japan.\label{NRAO}\\ 
$^{22}$Department of Astronomy, The Graduate University for Advanced Studies (SOKENDAI), 2-21-1 Osawa, Mitaka, Tokyo, Japan \label{Sokendai}\\  
$^{23}$The University of Texas at Austin, 2515 Speedway, Stop C1402, Austin, Texas 78712-1206, USA \label{utexas}\\ 
$^{24}$Chalmers University of Technology, Department of Space, Earth and Environment, Onsala Space Observatory, SE-439 92 Onsala, Sweden \label{OSO}\\ 
$^{25}$Astronomy Department and Van Vleck Observatory, Wesleyan University, Middletown, CT 06459, USA \label{Wesleyan}\\ 
$^{26}$Sternberg Astronomical Institute, Lomonosov Moscow State University, 119992 Universitetskii prospekt 13, Moscow, Russia \label{Moscow}\\ 
$^{27}$Physics Department, Austin College, Sherman, TX 75090, USA \label{AustinCollege}\\ 
$^{28}$Department of Physics and Astronomy, University of North Carolina at Chapel Hill, Chapel Hill, NC 27599, USA \label{unc}\\ 
$^{29}$Observatori Astronòmic Albanyà, Camí de Bassegoda S/N, Albanyà 17733, Girona, Spain \label{Girona}\\ 
$^{30}$Komaba Institute for Science, The University of Tokyo, 3-8-1 Komaba, Meguro, Tokyo 153-8902, Japan \label{UniTokyo}\\ 
$^{31}$Department of Physical Sciences, Ritsumeikan University, Kusatsu, Shiga 525-8577, Japan \label{Ritsumeikan}\\  
$^{32}$Sabadell Astronomical Society, 08206 Sabadell, Barcelona, Spain \label{sabadell}\\  
$^{33}$Europlanet Society, Department of Planetary Atmospheres of the Royal Belgian Institute for Space Aeronomy, B-1180 Brussels, Belgium \label{RoyalBelgian}\\ 
$^{34}$Instituto de Astrofísica de Andalucía (IAA-CSIC), Glorieta de la Astronomía s/n, 18008 Granada, Spain  \label{Andalucia}\\ 
$^{35}$Villa '39 Observatory, Landers, CA 92285, USA \label{Villa}\\ 
$^{36}$Waffelow Creek Observatory, 10780 FM 1878, Nacogdoches, TX 75961, USA \label{Waffelow}\\ 
$^{37}$INAF - Osservatorio Astronomico di Palermo, Piazza del Parlamento, 1, 90134 Palermo, Italy. \label{inaf-Palermo}\\  
$^{38}$Department of Physics and Kavli Institute for Astrophysics and Space Research, Massachusetts Institute of Technology, Cambridge, MA 02139, USA \label{Kavli}\\ 
$^{39}$Bozeman, MT 59718, USA \label{Bozeman} 
}
\date{Accepted XXX. Received YYY; in original form ZZZ}

\pubyear{\the\year{}}

\begin{document}
\label{firstpage}
\pagerange{\pageref{firstpage}--\pageref{lastpage}}
\maketitle

\begin{abstract}
As part of the KESPRINT collaboration, we present the discovery and characterization of three exoplanets in the sub-Neptune to super-Neptune regime, spanning key regions of the exo-Neptunian landscape. TOI-1472\,c and TOI-1648\,b are newly discovered sub-Neptunes, while TOI-1472\,b is a previously known super-Neptune for which we provide an improved mass measurement. These planets have orbital periods of 6--15 days and radii of 2.5--4.1~R$_\oplus$, probing regions where planet formation and atmospheric evolution remain poorly understood. We combine TESS transit photometry with ground-based radial velocities to determine precise masses, radii, and orbital properties. TOI-1472\,b has a mass of $18.0^{+0.84}_{-0.85}$~M$_\oplus$ and a radius of $4.06 \pm 0.10$~R$_\oplus$, TOI-1472\,c has a mass of $21.1^{+0.96}_{-0.99}$~M$_\oplus$ and a radius of $3.33 \pm 0.08$~R$_\oplus$, and TOI-1648\,b has a mass of $7.4^{+1.1}_{-1.3}$~M$_\oplus$ and a radius of $2.54^{+0.14}_{-0.12}$~R$_\oplus$. The planets exhibit a range of eccentricities (0.041--0.178), indicating diverse evolutionary histories. TOI-1648\,b, with a high Transmission Spectroscopy Metric (TSM~$\sim$59), is a promising target for atmospheric characterization. Together, these three planets provide precise constraints on the structure, composition, and dynamical evolution of small to intermediate-sized exoplanets, enriching our understanding of the exo-Neptunian landscape.

\end{abstract}




\section{Introduction} \label{sec:intro}
Small exoplanets (R$_{\rm P}$ $\textless$ 4 R$_{\rm \oplus}$) present many interesting population properties, which still remain widely unexplained to the best of our knowledge. The \textit{Kepler} mission confirmed the earlier finding reported by \citet{2011arXiv1109.2497M}, who first hinted at the high occurrence rate of sub-Neptunes in close-in orbits (within $\sim$1~AU), and revealed a bimodal radius distribution with a gap around 1.7 R$_{\oplus}$, known as the ``radius valley'' (\citealt{FulPet18,Bergeretal2018}). One of the interpretations is attributed to atmospheric evolution \citep{Daietal2019,Beanetal2021}; planets below the gap (super-Earths) were fully stripped of their primordial hydrogen-dominated atmospheres\footnote{Although we note that recent detections revealed the existence of a handful of low-density super-Earths inconsistent with rocky compositions, likely requiring the presence of a significant amount of volatiles in their atmospheres or interiors \citep[e.g.][]{2021MNRAS.501.4148L,2022A&A...664A.199L,2023A&A...675A..52C,2023A&A...678A..80P}.}, while planets above the gap are sub-Neptunes, which can be divided into gas-rich super-Earths and water worlds. Gas-rich super-Earths (i.e., Earth-like rocky-iron cores) are thought to form in the inner regions of the protoplanetary disk, which retained their primordial atmospheres without being substantially enriched in volatiles (\citealt{IkomaHori2012,LeeChiang2016}). Water worlds should form beyond the snowline by the accretion of water and rocks -- i.e., they should be a scaled-up version of icy moons in the solar system --, followed by inward migration \citep{ZengLietal2019,Venturinietal2020}. \cite{Luque22}, using a refined sample of small planets orbiting M-dwarf hosts, proposed that the radius valley is determined more by the interior composition of exoplanets than by the loss of atmospheric mass. A recent investigation employing atmospheric mass-loss models \citep{Rogersetal2023} has successfully replicated the results reported by \cite{Luque22} by incorporating spontaneous mass loss (called ``boil-off'') following disk dispersal (\citealt{IkomaHori2012,OwenWu2016}). Overall, sub-Neptunes are situated at the intersection of a modeling degeneracy in interior structure; comprising rock, ices, and/or gas, they frequently present numerous combinations of these materials capable of aligning with the mass and radius data of individual planets.

At larger radii (4 $\textless$ R$_{\rm P}$ $\textless$ 10 R$_{\rm \oplus}$), other intriguing aspects appear, such as a dearth of Neptune-sized and super-Neptune-sized planets in the closest orbits: the so-called Neptunian desert \citep[e.g.,][]{Benitez2011,SzaboKiss2011,Mazehetal2016}. Beyond the desert lies the savanna, where such planets are more commonly found \citep{2023A&A...669A..63B}. Between these two regions sits the Neptunian ridge, a transitional zone marked by a notable concentration of Neptune-sized planets with orbital periods between $\sim$3.2 and $\sim$5.7 days \citep{CastroGonzalez2024}. This accumulation aligns closely with the well-known hot Jupiter pile-up \citep{2003A&A...407..369U}, hinting at possible links in the dynamical histories of intermediate-sized and gas giant planets. Notably, planets located within the desert and ridge tend to exhibit higher densities than those in the surrounding savanna \citep{CastroGonzalez2024b}. This density contrast could reflect differences in formation or migration pathways, or enhanced atmospheric loss processes operating in the desert and ridge environments.

In this paper, we report the discovery of two new sub-Neptunes, TOI-1472\,c and TOI-1648\,b, and present an improved mass measurement for the previously published super-Neptune TOI-1472\,b \citep{Polanskietal2024}. These systems are of particular interest because they occupy key regions of the small-planet and Neptune-sized planet populations, including the radius valley and the Neptunian desert, where formation and evolution processes remain poorly understood. Precise mass and radius determinations for these planets provide important constraints on their interior compositions and atmospheric properties, allowing us to test models of planetary structure, atmospheric loss, and dynamical evolution. 

In the following sections, we present the Transiting Exoplanet Survey Satellite (TESS; \citealt{2015JATIS...1a4003R}) photometry, ground-based follow-up observations, spectroscopy, and high-resolution imaging (Section \ref{sec:obs}), stellar modeling (Section \ref{sec: Stellar modelling}), frequency analysis (Section \ref{sec:freq}), transit and radial velocity joint fits with planetary system modeling (Section \ref{sec:planet}), and detailed system characterization (Section \ref{sec:characterization}), followed by our conclusions in Section \ref{sec:concl}.

\section{Observations} \label{sec:obs}
\subsection{Photometric data}

\subsubsection{TESS photometry}
\label{sec:TESS_phot}
TOI-1472 (TIC 306955329, stellar properties in Table \ref{t:star_param}) was observed by TESS between October 2019 and November 2024. TOI-1648 (TIC 376353509, Table \ref{t:star_param}) has been observed between November 2019 and June 2024. Observations of the targets were conducted with a 2-minute cadence. Table~\ref{table:tess_obs} summarizes the corresponding TESS sectors, CCDs, and camera configurations. 

The two candidates, TOI-1472.01 and TOI-1648.01, were identified through the Science Processing Operations Center (SPOC) transit search, which employed an adaptive, noise-compensating matched filter technique \citep{Jenkins02,Jenkins10,Jenkins2020kepler}. This process flagged a Threshold Crossing Event (TCE) for each target. Subsequently, a preliminary transit model incorporating limb darkening was fitted to the detected signals \citep{Li:DVmodelFit2019}, followed by a comprehensive battery of diagnostic tests to assess the planetary nature of the transits \citep{Twicken:DVdiagnostics2018}. Independently, the Quick Look Pipeline (QLP) at the Massachusetts Institute of Technology (MIT) also detected the transit signatures in the full-frame image (FFI) data \citep{Huang2020QLP1,Huang2020QLP2}. These vetting results were carefully reviewed by the TESS Science Office (TSO), and the signals were confirmed across multiple observing sectors, consistently passing all data validation diagnostics.

The additional planet in the TOI-1472 system, discovered later through our radial velocity campaign, was not identified by the transit pipelines during this stage and thus was not designated as a candidate.

TESS photometric data for these targets were processed by both the MIT QLP and the SPOC pipeline \citep{jenkins2016}, each applying simple aperture photometry (SAP; \citealt{Twicken10}) to generate light curves. Instrumental systematics were further mitigated using the Presearch Data Conditioning (PDCSAP) algorithm \citep{Smith2012_PDCSAP,Stumpe2012}. For our analysis, we retrieved the SPOC PDCSAP light curves from the Mikulski Archive for Space Telescopes (MAST\footnote{\url{https://mast.stsci.edu/}}) and used these corrected data in our transit modeling (see Section~\ref{sec:planet}).

\begin{figure*}
\centering
\includegraphics[width=0.43\linewidth,trim=10 10 8 0,clip]{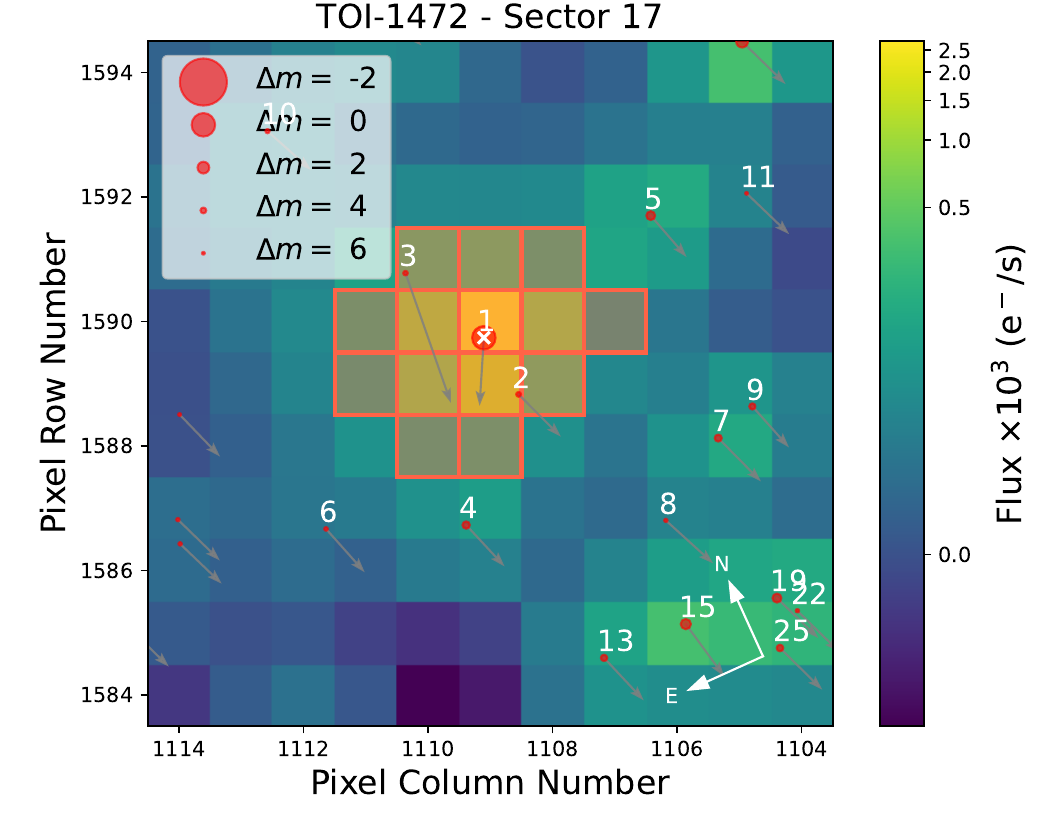}
\includegraphics[width=0.43\linewidth,trim=10 10 8 0,clip]{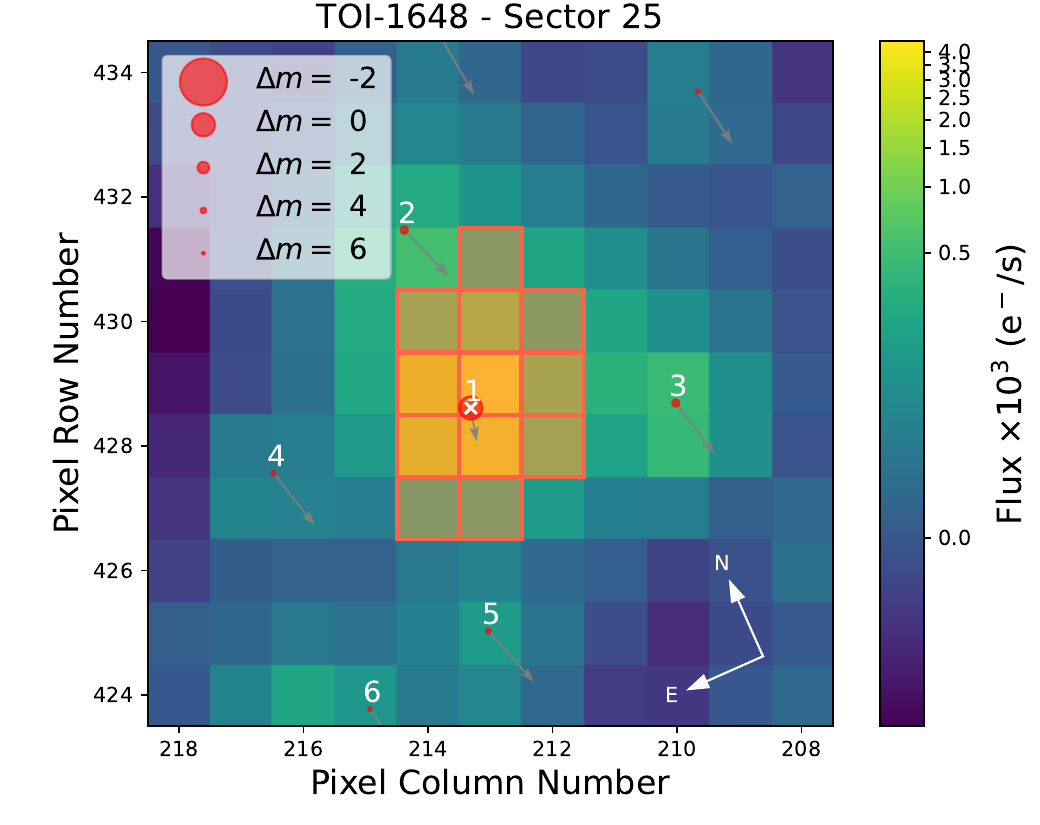}
\includegraphics[width=0.43\linewidth,trim=0 0 0 0,clip]{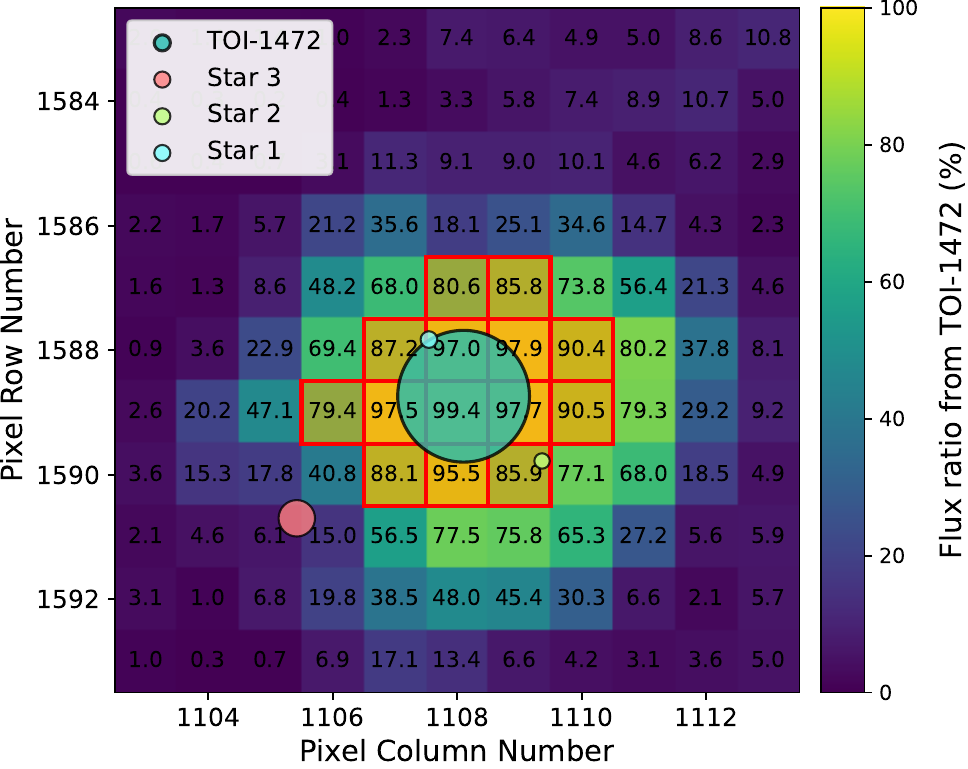}
\includegraphics[width=0.43\linewidth,trim=0 0 0 0,clip]{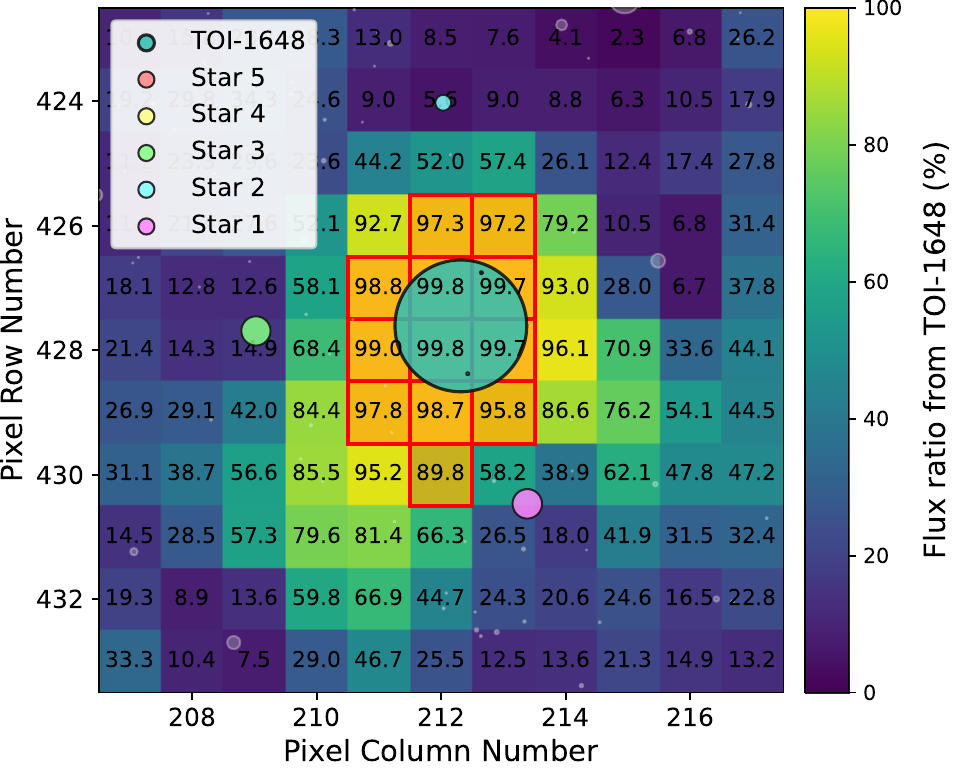}
\caption{\textit{Upper}: TESS TPF (Target Pixel Files) of Sector 17 for TOI-1472, and Sector 25 for TOI-1648. The color scale represents electron counts per pixel. Orange squares mark the pixels selected by the TESS pipeline for aperture photometry. Gaia DR3 sources are overlaid as circles with sizes corresponding to their G-band magnitude difference relative to the target, as indicated in the legend. This visualization was created using the {\tt tpfplotter} code \citep{aller2020}. Gray arrows show the proper motion vectors of all sources in the field. \textit{Bottom}: TESS heatmaps for the two targets, produced with \texttt{TESS-cont} \citep{CastroGonzalez2024b}, illustrating the percentage of the target star’s flux contained within each pixel. } \label{fig:tpfplotter}
\end{figure*}

We investigated potential flux contamination from nearby sources in the TESS data for our targets by first overlaying Gaia Data Release 3 \citep{2023A&A...674A...1G} catalog stars onto the Target Pixel Files (TPFs) using the \texttt{tpfplotter} tool \citep{aller2020} (see Figure~\ref{fig:tpfplotter}). Subsequently, we employed \texttt{TESS-cont} \citep{CastroGonzalez2024b} to quantify the contribution of these neighboring sources to the total flux within the SPOC photometric apertures.

For TOI-1472, the three most contaminant sources are TIC\,306955328, TIC\,306950348, and TIC\,306950347 (star$\#$2, star$\#$3, and star$\#$5 in Figure~\ref{fig:tpfplotter}, respectively), with total flux contributions of 1.34$\%$, 0.97$\%$, and 0.57$\%$, respectively. For TOI-1648, the three most contaminant sources are TIC 376353519 (star$\#$2), TIC\,376353488 (star$\#$5), and TIC\,376353502 (star$\#$3), with contributions of 0.45$\%$, 0.12$\%$, and 0.10$\%$, respectively. Following \citet{2018AJ....156..277L}, \citet{2018AJ....156...78L}, \citet{2021MNRAS.508..195D}, and \citet{2022MNRAS.509.1075C}, and considering the transit depths of the planetary candidates estimated by the SPOC pipeline, we find that  potential deep eclipses of 17$\%$-86$\%$ (i.e. $<$ 100$\%$) in the nearby sources of TOI-1472 and TOI-1648 could generate the observed transit signals. We note that we discard this possibility through follow-up transit (Sect.~\ref{sec:TFOP_lc}) and spectroscopic (Sect.~\ref{sec:spectroscopic_data}) observations and that the PDCSAP data already account for flux dilution, so no additional corrections were necessary.

\begin{table}
   \caption[]{Stellar properties of TOI-1472 and TOI-1648.}
     \label{t:star_param}
     \small
     \centering
       \begin{tabular}{lccl}
         \hline
         \noalign{\smallskip}
         Parameter   &  TOI-1472 & TOI-1648  & Ref \\
         \noalign{\smallskip}
         \hline
         \noalign{\smallskip}
$\alpha$ (ICRS, 2016.0)          &  00:56:27.42  	&  03:01:36.28    &  Gaia DR3\\
$\delta$ (ICRS, 2016.0)          &  +48:38:15.06 	&  +69:13:47.72   &   Gaia DR3\\
$\mu_{\alpha}$ (mas/yr)  &   103.399 	& 16.187  &    Gaia DR3\\
$\mu_{\delta}$ (mas/yr)  &   -13.622  	&  5.429   &  Gaia DR3\\
RV     (km\,s$^{-1}$)     &   -15.047$\pm$0.309  	& -32.177$\pm$0.144    &  Gaia DR3  \\
$\pi$  (mas)             &   8.1964  	&  14.3348   &  Gaia DR3\\
Distance (pc) & 121.3097 & 69.6704 & Gaia DR3\\
\noalign{\medskip}
$V$ (mag)                  &   11.302$\pm$0.014  	& 10.43$\pm$0.006    &   TIC v1.8\\
$B$ (mag)                &  12.206$\pm$0.174  	&   11.448$\pm$0.068  &  TIC v1.8 \\
$G$ (mag)                  &     11.114$\pm$0.001	&  10.1848$\pm$0.0003   &   Gaia DR3 \\
$G_{\rm BP}$-$G_{\rm RP}$ (mag)    &   1.073  	&  1.173   &    Gaia DR3\\
TESS (mag)              &   10.551$\pm$0.006 	&   9.583$\pm$0.006  &  TIC v1.8\\
J$_{\rm 2MASS}$ (mag)    &  9.794$\pm$0.021  	&  8.804$\pm$0.032   &  TIC v1.8 \\
H$_{\rm 2MASS}$ (mag)    &   9.384$\pm$0.023 	&  8.296$\pm$0.044   &   TIC v1.8 \\
K$_{\rm 2MASS}$ (mag)    &   9.277$\pm$0.021 	&  8.183$\pm$0.018   &  TIC v1.8 \\
\noalign{\medskip}
$S_{\rm MW}$             & 0.30$\pm$0.01   	&  0.25$\pm$0.01   &  This work \\
$\log R^{'}_{\rm HK}$    &    -4.79$\pm$0.02  	&  -4.95$\pm$0.01   & This work \\ 
\noalign{\medskip}
         \hline
      \end{tabular}
\end{table}

\begin{table*}
\caption{Summary of TESS observations for the two targets.}
\centering
\begin{tabular}{lcccc}
\hline\hline
 Target  &  Sector  & CCD Number  & Camera       & Cadence [min]  \\[1mm]

\hline
{\it  TOI-1472 }      & 17 &  4  & 2 & 2 \\
      & 57 &  4  & 2 & 2\\
      & 58 &  3  & 2 & 2 \\ 
      & 84 &  4  & 2 & 2 \\ 
      & 84 &  3  & 2 & 2 \\[1mm]

\hline
{\it  TOI-1648 }      & 18 &  2  & 1 & 2 \\
      & 19 &  2  & 2 & 2 \\
      & 25 &  4  & 4 & 2 \\
      & 52 &  4  & 3  & 2 \\
      & 59 &   2 & 2  & 2 \\
      & 79 &  3  & 4 &  2\\      
      & 86 &  2  & 2 & 2 \\[1mm]

\hline

\end{tabular}

\label{table:tess_obs}
\end{table*}

\subsubsection{TFOP follow-up light curves}
\label{sec:TFOP_lc}

To accurately identify the true sources of the TESS signals, correct for any blending effects on transit depths, assess possible wavelength-dependent variations in transit depth, and refine the transit timing, we conducted ground-based photometric follow-up of the fields surrounding TOI-1472 and TOI-1648. This was carried out through the TESS Follow-up Observing Program (TFOP; \citealt{collins:2019})\footnote{\url{https://tess.mit.edu/followup}}. Transit observations were scheduled using the {\tt TESS Transit Finder}, a tailored version of the {\tt Tapir} software package \citep{Jensen:2013}. 

We observed three transit windows of TOI-1472.01 using the Austin College 0.6\,m telescope at Adams Observatory in Sherman, TX, USA (Adams), the 0.35\,m telescope at Waffelow Creek Observatory (WCO) in Nacogdoches, TX, USA, and the Las Cumbres Observatory Global Telescope \citep[LCOGT;][]{Brown:2013} 2\,m Faulkes Telescope North at Haleakala Observatory on Maui, Hawai'i. The 2\,m LCOGT telescope is equipped with the MuSCAT3 multi-band imager \citep{Narita:2020}. We observed three transit windows of TOI-1648.01 from the LCOGT 1\,m network node at McDonald Observatory near Fort Davis, Texas, USA (LCO-McD). 

Images obtained with LCOGT telescopes were processed using the standard LCOGT {\tt BANZAI} pipeline \citep{McCully:2018}, while differential photometry was performed with {\tt AstroImageJ} \citep{Collins:2017}. This software was also utilized to calibrate and extract differential photometry from data collected by other participating observatories. For all follow-up light curves, photometric apertures were carefully chosen as small circular regions that excluded flux from the nearest known Gaia DR3 neighbors of the target stars. The resulting light curve data are publicly accessible on the {\tt EXOFOP-TESS} platform\footnote{\url{https://exofop.ipac.caltech.edu/tess/}}, summarized in Table \ref{table:carleo-SG1-phot-obs}, and jointly modeled as detailed in Section \ref{sec:planet}.

\begin{table*}
\caption{Summary of TFOP Ground-based Lightcurve Follow-up}
\centering
\begin{tabular}{lllccl}
\hline\hline
Telescope & Date  & Filter  &  Phot. Aper.      & Nearest Gaia DR3  & Transit  \\
          & [UTC] &         &  Radius [arcsec]  & neighbor [arcsec] & Coverage \\
\hline

{\it  TOI-1472 b}\\
\hline
Adams 0.6\,m    & 2019-12-18   & $\mathrm{I_c}$ &  8.7  & 17.0 & full \\
WCO 0.35\,m     & 2020-10-12   & Sloan $g'$     &  5.3  & 17.0 & full \\
LCO MuSCAT3 2.0\,m      & 2021-01-09   & g,r,i,z        &  6.8  & 17.0 & full \\[2mm] 

\hline

{\it  TOI-1648 b}\\
\hline
LCO-McD 1.0\,m & 2020-08-13   & Pan-STARRS $z_s$ & 5.4  &  15.7  & full \\
LCO-McD 1.0\,m & 2020-09-04   & Sloan $i'$       & 10.9 &  15.7  & full \\
LCO-McD 1.0\,m & 2020-09-04   & Sloan $i'$       & 10.5 &  15.7  & full \\[2mm]

\hline

\end{tabular}

\label{table:carleo-SG1-phot-obs}
\end{table*}

\subsubsection{Ground-based archival data}

We examined publicly accessible archival data (already processed lightcurves) from the ASAS-SN survey \citep{Shappee:2014,Kochanek2017,Hart:2023}, the WASP transit survey \citep{2006PASP..118.1407P}, and the Zwicky Transient Facility (ZTF; \citealt{Bellm:2019,Masci:2019}) to search for rotational variability in the host stars.

For TOI-1472, ASAS-SN has 490 observations spanning $\sim$2030 days in $g$ band, and 200 observations spanning $\sim$1440 days in $V$ band. No significant peak in the Generalized Lomb–Scargle (GLS) periodogram \citep{Zech09} for the time series was found in the ASAS-SN data sets. The WASP photometry has one season of coverage of TOI-1472, spanning 130 days in 2007, amounting to 4000 photometric data points. Analyzing the light curve using methods from \citet{2011PASP..123..547M} we find no rotational modulation, with a 95\%-confidence upper limit of 2 mmag. For this star there was no ZTF data.

TOI-1648 has ASAS-SN data, no WASP coverage, and no ZTF time series. The ASAS-SN observations consist of 480 $g$ band data points spanning $\sim$1990 days, while the $V$ band observations have 180 observations spanning $\sim$2500 days. As was the case for TOI-1648, we did not find any significant peak in the periodogram of the time series that could belong to the rotation period of the star.

\subsection{Spectroscopic data}
\label{sec:spectroscopic_data}

The two stars were observed as part of the {\tt KESPRINT} collaboration\footnote{\url{www.kesprint.science}}, which aims to measure the masses of small and intermediate-sized planets. Observations were conducted using the HARPS-N visible spectrograph \citep{Cosentinoetal2014} at the Telescopio Nazionale Galileo (TNG) in La Palma, Spain, under the programs CAT19A\_162, CAT21A\_119, CAT22A\_111 (PI: Nowak), ITP19\_1 (PI: Palle), CAT20B\_80 (PI: Casasayas), CAT23A\_52, CAT23B\_74 (PI: Carleo).

\textbf{TOI-1472}.
We collected 52 HARPS-N RVs for TOI-1472 between  13 January 2020 UT and  13 February 2023 UT, with an exposure time of 1800 s, and an average signal-to-noise (SNR) per pixel of 48. We used the offline version of HARPS-N data reduction software (DRS) through the Yabi web application (\citealt{yabi}) installed at IA2 Data Center\footnote{\url{https://www.ia2.inaf.it}} in order to extract the RVs and activity indicators. For TOI-1472 we used a K5 mask template, as it better matches the star's spectral type, and a cross-correlation function (CCF) width of 30\,km\,s$^{-1}$. We obtained an average precision of 1.7\,m\,s$^{-1}$. The list of RVs is presented in Table~\ref{tab:rvdata_toi1472}, together with the chromospheric activity indexes, S-index and \logrhk.

For the analysis, we also incorporated 22 HIRES RV measurements (with a typical RV uncertainty of 1.7 m\,s$^{-1}$) from \cite{Polanskietal2024}, who reported the discovery of planet b.

\textbf{TOI-1648}.
We collected 53 HARPS-N RVs between 10 October 2020 UT and 10 March 2023 UT with exposure time of 1800 s and average SNR of 67. As for TOI-1472, we used a K5 mask and 30\,km\,s$^{-1}$ of CCF width to extract the RVs with Yabi. We obtained an average RV uncertainty of 1.2 m/s. Table \ref{tab:rvdata_toi1648} lists the RVs and activity indexes for TOI-1648. 

Activity-related parameters such as \logrhk, the S-index, bisector span, cross-correlation function (CCF) contrast, and CCF full width at half maximum (FWHM) were extracted from the HARPS-N Data Reduction Software (DRS). For \logrhk we adopted the B-V color calculated from the magnitude values in Table \ref{t:star_param}. Additional diagnostics—including the chromatic index (CRX), differential line width (dLW), H$\alpha$ index, and sodium doublet lines (Na$_1$ and Na$_2$)—were derived using the \texttt{serval} pipeline \citep{Zechmeister2018}.

\subsection{High Resolution Imaging}
To evaluate potential contamination from bound or unbound stellar companions that could bias the derived planetary radii \citep{ciardi2015, FurlanHowell2017, FurlanHowell2020, lillo-box12, lillo-box14b, lillo-box24}, we followed standard validation procedures by obtaining high-resolution imaging of the TOIs using optical speckle, lucky imaging, and/or near-infrared adaptive optics techniques.

	\subsubsection{Optical Speckle Imaging}

TOI-1472 and TOI-1648 were observed on 09 September 2021 UT and 29 October 2020 UT, respectively, using the speckle polarimeter mounted on the 2.5-m telescope at the Caucasian Observatory of the Sternberg Astronomical Institute (SAI), Lomonosov Moscow State University. Observations were performed with an Andor iXon 897 Electron Multiplying CCD detector \citep{Safonov2017}. For TOI-1472, imaging was conducted in the $I_\mathrm{c}$ band, yielding an angular resolution of 0.083$^{\prime\prime}$. For TOI-1648 a filter with a bandpass of 50 nm centered on 625~nm was used with an angular resolution of 0.063$^{\prime\prime}$. No nearby stellar companions were detected in either case. At separations of 0.25$^{\prime\prime}$ and 1.0$^{\prime\prime}$, the achieved contrast limits were $\Delta I_\mathrm{c}=4.1^m$ and $5.3^m$ for TOI-1472, and $\Delta I_\mathrm{c}=4.6^m$ and $6.0^m$ for TOI-1648.

\begin{figure}
\centering
\includegraphics[width=0.95\linewidth]{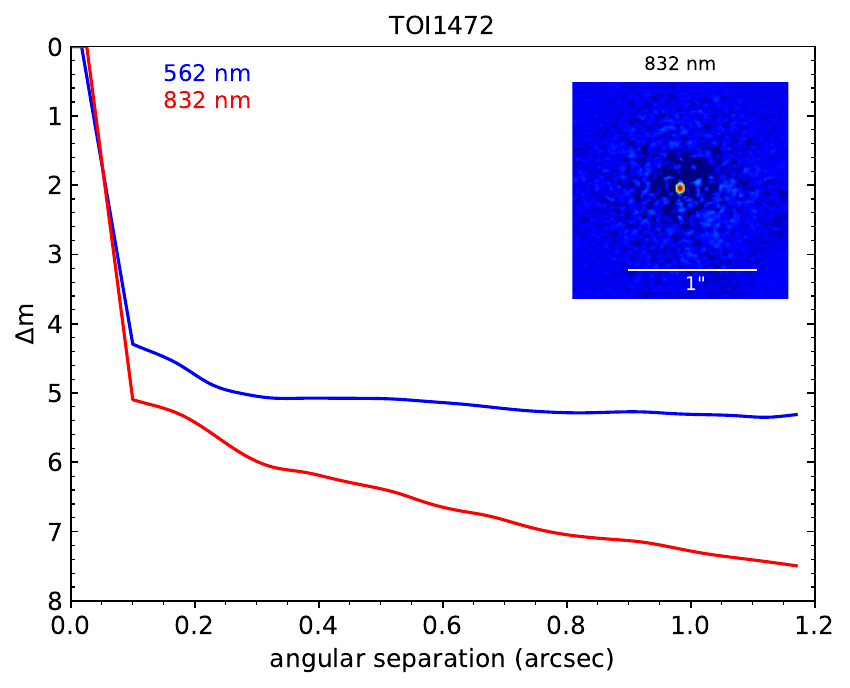}
\includegraphics[width=0.95\linewidth]{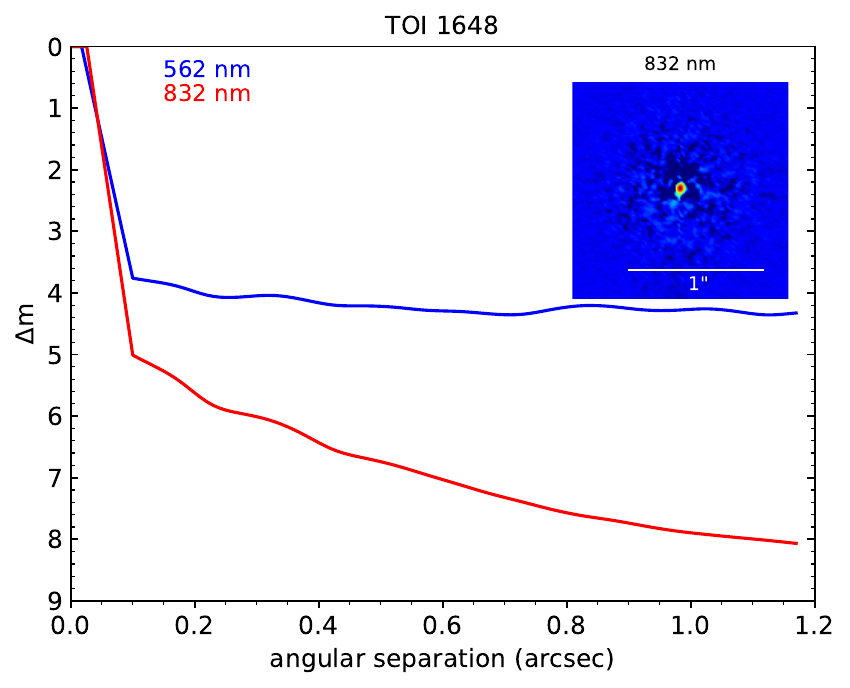}
\caption{Reconstructed ‘Alopeke speckle images and $5\sigma$ contrast limits from simultaneous diffraction-limited exposure sequences using the 562 nm filter on the blue camera and the 832 nm filter on the red camera. No nearby sources are detected. \label{fig:speckleim}}
\end{figure}

Both TOI-1472 and TOI-1648 were also observed on 03 December 2020 UT and 22 October 2021 UT respectively using the ‘Alopeke high-resolution speckle instrument on the Gemini North 8-m telescope \citep{Scottetal2021}. ‘Alopeke provides simultaneous speckle imaging in two bands (562 nm and 832 nm) with output data products including a reconstructed image with robust contrast limits on companion detections. Five sets of 1000 X 0.06 second images were obtained and processed along with the PSF standards HR 0189 (TOI-1472) and HR 0829 (TOI-1648) in our standard reduction pipeline (see \citealt{howell2011}) for each star. Figure \ref{fig:speckleim} shows the final 5-sigma magnitude contrast curves and the 832 nm reconstructed speckle image. We find that TOI-1472 and TOI-1648 are single stars with no companion brighter than 5-8 magnitudes below that of the target star from the 8-m telescope diffraction limit (20 mas) out to 1.2\arcsec. At the distance of TOI-1472 (d\,=\,122 pc) and TOI-1648 (d\,=\,70 pc) these angular limits correspond to spatial limits of 2.4 to 146 au and 1.4 to 84 au, respectively.

\subsubsection{Lucky Imaging}
We observed TOI-1648 using the AstraLux lucky-imaging camera (\citealt{hormuth08}) installed at the 2.2-m telescope in Calar Alto (Almer{\'i}a, Spain). Observations were performed on the night of 26 February 2020 under the AstraLux follow-up of TESS planet candidates program (PI: J. Lillo-Box). The observations and data analysis corresponding to this target are detailed in \cite{lillo-box24}. In summary, we found no stellar companions around this target with a 95\% sensitivity limit at 0.5\arcsec of 4.3 mag fainter than the target. This implies a probability of an undetected source capable of mimicking the transit signal is 0.29\% (\citealt{lillo-box14b}).


\subsubsection{Near-Infrared AO Imaging}
Observations of TOI-1472 and TOI-1648 were made on 28 May 2020 UT and 09 September 2020 UT with the NIRC2 instrument on Keck-II (10m) behind the natural guide star AO system \citep{wizinowich2000}. TOI-1472 was also observed with the ShARCS camera on the Shane 3-m at Lick Observatory (Dressing et al., submitted). The pixel scales of the ShARCS and NIRC2 instruments are $0.033\arcsec$ and $0.009942\arcsec$ per pixel, respectively. Observations at Lick Observatory were taken using a standard 5-point quincunx dither pattern, while the Keck observations employed a 3-point dither pattern designed to avoid the lower left quadrant of the detector. The reduced science frames were subsequently combined into final mosaiced images, achieving angular resolutions of approximately $0.3\arcsec$ for ShARCS and $0.05\arcsec$ for NIRC2.
	
To assess the sensitivity of the final combined adaptive optics (AO) images, we injected artificial sources at regular azimuthal intervals of $20^\circ$ around the primary target, placing them at radial distances corresponding to integer multiples of the target's full width at half maximum (FWHM) \citep{furlan2017}. The flux of each simulated source was adjusted until it was detected with a $5\sigma$ significance level using standard aperture photometry. At each separation, the $5\sigma$ contrast limit was calculated as the average detection threshold of all injected sources, while the associated uncertainty was derived from the root mean square (rms) variation of the detection limits across the azimuthal positions.

\section{Stellar modelling} \label{sec: Stellar modelling}
\subsection{Method 1: \texttt{BACCHUS}+PARAM}\label{subsec:thomas} 
We performed a consistent spectroscopic analysis of the stellar parameters using the updated version of the \texttt{BACCHUS} code \citep{2016ascl.soft05004M, 2022ApJS..262...34H}, which is based on MARCS model atmospheres \citep{2008A&A...486..951G}. The analysis was carried out using co-added HARPS-N spectra. Stellar effective temperatures (T$_{\rm eff}$) were determined by minimizing any correlation between the abundances of \ion{Fe}{I} lines and their excitation potentials. Surface gravities ($\log g$) were inferred by enforcing ionization equilibrium between \ion{Fe}{I} and \ion{Fe}{II} lines. Microturbulent velocities ($\xi_t$) were obtained by ensuring that there was no trend between iron abundances and the equivalent widths of the lines. The metallicity ([Fe/H]) was computed as the mean abundance from \ion{Fe}{I} lines.

To estimate projected stellar rotational velocities ($v \sin i$), we measured the average broadening of iron lines in the HARPS-N spectra, after accounting for instrumental and thermal broadening. Macroturbulent velocity was neglected due to the cool temperatures of the stars. For all targets, this analysis yielded only upper limits on $v \sin i$ ($<$3.5 km s$^{-1}$), consistent with slow rotation periods ($\gtrsim$13 days).

Stellar masses and radii were subsequently derived using the PARAM Bayesian framework \citep{2012MNRAS.427..127B,2017MNRAS.467.1433R}, incorporating spectroscopic temperatures and luminosities derived from updated \textit{Gaia} parallaxes. Since PARAM does not account for systematic differences between various stellar evolution models, we employed both MESA and PARSEC isochrones to estimate the impact of model-dependent uncertainties. We added the discrepancy between the two sets of outputs to the formal PARAM errors to provide a more realistic uncertainty estimate. While this approach reduces model-related systematics, it may still underestimate the total error budget for stellar radius and luminosity, as discussed by \citet{Tayar2022}. The final stellar parameters are summarized in Table~\ref{Table: stellar spectroscopic parameters}.


\subsection{Method 2: \texttt{SPECIES}+ARIADNE} 
We independently derived stellar atmospheric and bulk parameters using a two-step approach that combines the \texttt{SPECIES} code \citep{SotoJenkins2018} with the ARIADNE framework \citep{VinesJenkins2022}. \texttt{SPECIES} was applied to the co-added high-resolution spectra to extract fundamental spectroscopic parameters, which were then used as priors in ARIADNE— a Bayesian Model Averaging (BMA) tool designed to infer stellar properties of nearby stars.

\texttt{SPECIES} derives T$_{\rm eff}$, $\log g$, [Fe/H], and $\xi_t$ by first measuring the equivalent widths of selected Fe I and Fe II absorption lines. These measurements are used in conjunction with interpolated ATLAS9 stellar atmosphere models \citep{castelli2004new} and the radiative transfer code MOOG \citep{sneden1973nitrogen} to iteratively solve for atmospheric properties. The method enforces excitation and ionization equilibrium by minimizing any dependence of iron abundances on excitation potential and reduced equivalent width ($W/\lambda$), assuming local thermodynamic equilibrium (LTE).

In parallel, $v \sin i$ is estimated by comparing observed absorption features with synthetic line profiles, adjusted using empirical calibrations based on stellar temperature. These spectroscopic outputs are then fed into ARIADNE, which incorporates multiple stellar evolution models under a BMA framework to yield posterior distributions for global stellar parameters such as radius and mass, while marginalizing over model-dependent uncertainties.

The obtained values are listed in Table \ref{Table: stellar spectroscopic parameters}

\subsection{Method 3: Spectral Energy Distribution}\label{sec:sed} 

\begin{figure}
\centering
\includegraphics[width=0.95\linewidth,trim=80 70 50 50,clip]{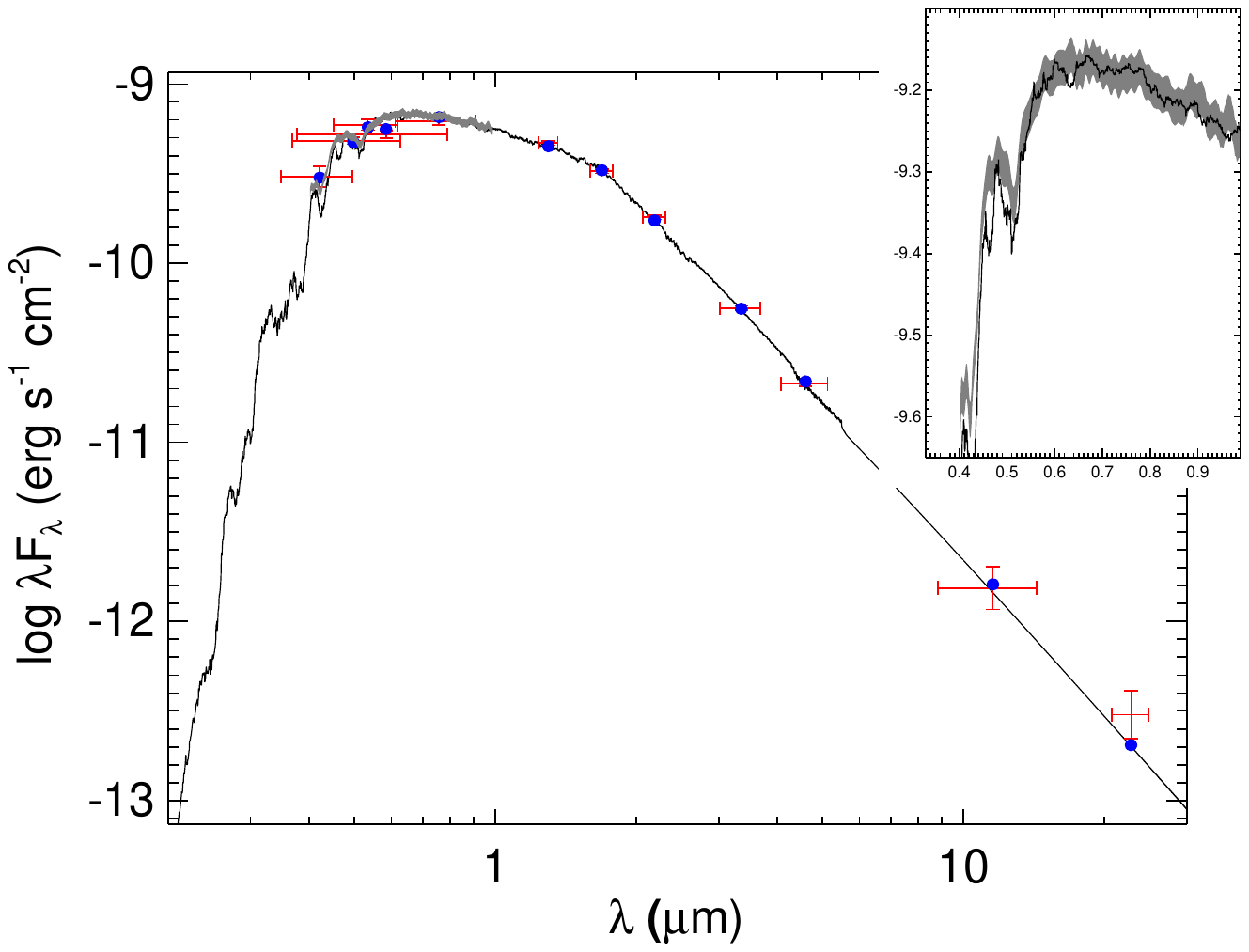}
\includegraphics[width=0.95\linewidth,trim=80 70 50 50,clip]{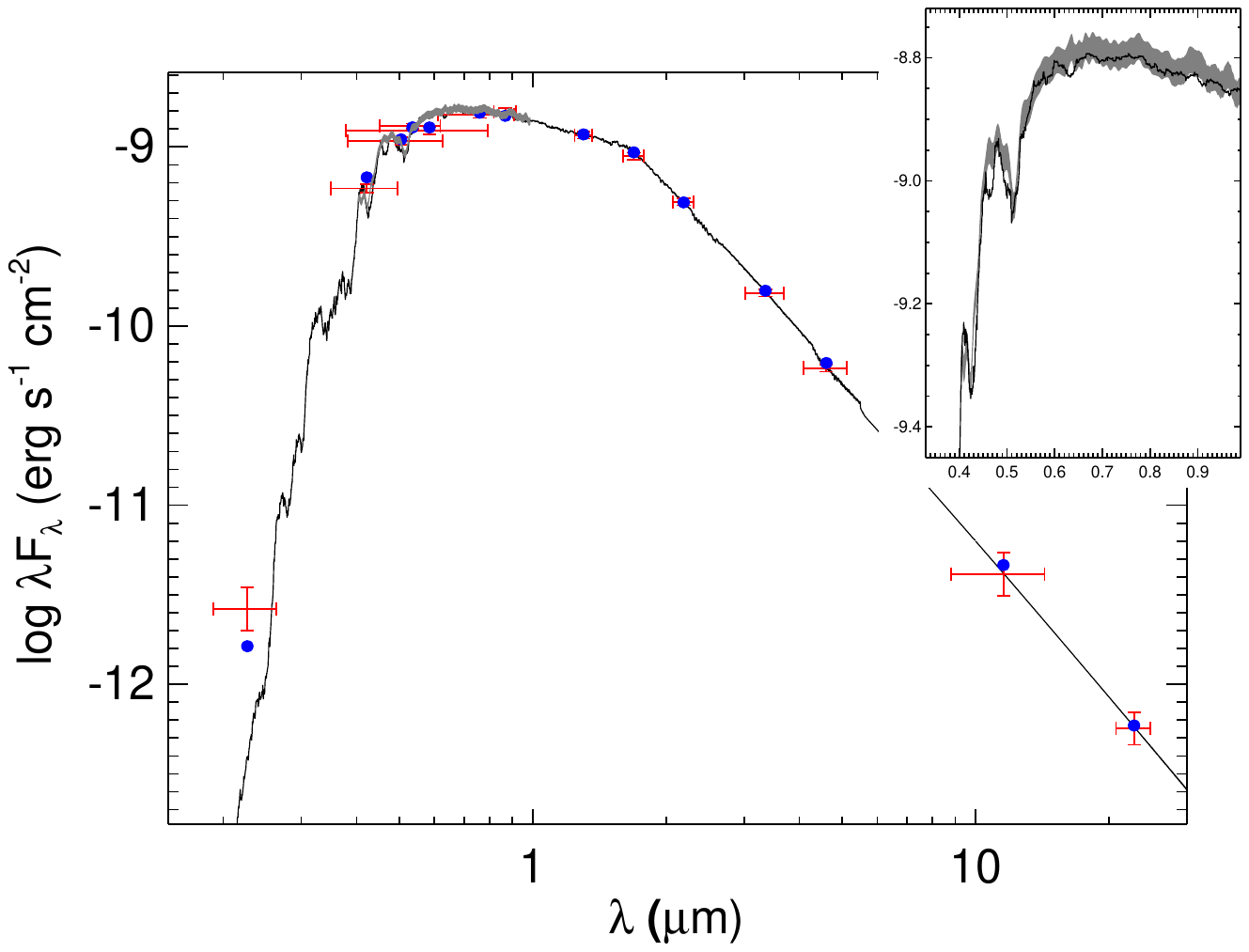}
\caption{Spectral energy distributions of TOI-1472 (top) and TOI-1648 (bottom). The red markers indicate the observed photometric data, with horizontal bars denoting the effective bandwidth of each filter. Blue points represent the synthetic fluxes derived from the best-fitting PHOENIX stellar atmosphere model, shown in black. In the inset, the absolute flux-calibrated \textit{Gaia} spectrum is displayed as a gray shaded region. \label{fig:sed}}
\end{figure}

We performed a broadband spectral energy distribution (SED) analysis for each star, incorporating the precise parallaxes from \textit{Gaia} DR3 to derive empirical estimates of stellar radii \citep{Stassun:2016,Stassun:2017,Stassun:2018}. Photometric data were compiled from several sources: near-infrared $JHK_S$ bands from \textit{2MASS}, mid-infrared W1–W4 from \textit{WISE}, optical $G_{\rm BP}$ and $G_{\rm RP}$ magnitudes from \textit{Gaia}, and ultraviolet fluxes from \textit{GALEX} where available. The absolute flux-calibrated \textit{Gaia} spectra were included in the analysis. This photometric coverage extends across a wide wavelength baseline, typically from 0.4 to 10~$\mu$m and, in some cases, reaching from 0.2 to 20~$\mu$m (see Figure~\ref{fig:sed}).

The observed SEDs were modeled using synthetic spectra from the PHOENIX atmosphere grid \citep{Husser:2013}, adopting the stellar effective temperature, metallicity, and surface gravity from our spectroscopic analysis. Interstellar extinction ($A_V$) was treated as a free parameter in the fit, but constrained not to exceed the line-of-sight reddening predicted by the Galactic dust maps of \citet{Schlegel:1998}.

By integrating the best-fit, dereddened model SED, we computed the bolometric flux ($F_{\rm bol}$) received at Earth. Combining this flux with the \textit{Gaia} parallax yields the bolometric luminosity ($L_{\rm bol}$), and the stellar radius ($R_\star$) was then calculated via the Stefan–Boltzmann law. Stellar masses were inferred using empirical calibrations of \citet{Torres:2010}.

Chromospheric activity indicators ($R'_{\rm HK}$) were used to estimate stellar ages using the empirical age-activity relations of \citet{Mamajek:2008}. The related age uncertainties are the formal uncertainty in the adopted relation itself and do not include potential systematic errors between different relations, which are typically of the order of 1 Gyr. These same calibrations also provide gyrochronology-based estimates of stellar rotation periods.

Figure~\ref{fig:sed} displays the SED fits for each star. A summary of the derived physical parameters is provided in Table~\ref{Table: stellar spectroscopic parameters}, with individual values discussed below.

\textbf{TOI-1472}: The best fit has $A_V$\,=\,$0.03 \pm 0.03$, with a reduced $\chi^2$ of 1.3, $F_{\rm bol}$\,=\,$9.10 \pm 0.21 \times 10^{-10}$ erg~s$^{-1}$~cm$^{-2}$, $L_{\rm bol}$\,=\,$0.4225 \pm 0.0099$~L$_\odot$, and estimated rotation period of $37 \pm 2$~d. 
\textbf{TOI-1648}: The best fit has $A_V$\,=\,$0.01 \pm 0.01$, with a reduced $\chi^2$ of 1.4, $F_{\rm bol}$\,=\,$2.254 \pm 0.019 \times 10^{-9}$ erg~s$^{-1}$~cm$^{-2}$, $L_{\rm bol}$\,=\,$0.3420 \pm 0.0028$~L$_\odot$, and estimated rotation period of $45 \pm 2$~d.

We computed the galactic U, V, W Local Rest System (LRS) velocities for both TOI-1472 and TOI-1648 using the Gaia DR3 astrometry and RV values given in Table \ref{t:star_param}.   We then used these absolute space velocities to compute the galactic dynamic stellar population membership of the star using the formalism originally developed by \cite{JohnsonSoderblom1987}.  Both stars are clear dynamical members of the galactic thin disk population.  This agrees with their [Fe/H] abundances as given in Table \ref{Table: stellar spectroscopic parameters}.

It is noteworthy that the three methods do not yield consistent results for all parameters. This discrepancy arises primarily from the different prescriptions of microturbulence adopted by each code, which subsequently propagate into the determination of other stellar parameters. The microturbulence parameter is empirical in nature, introduced to account for the effects of unresolved velocity fields on the derived abundances, and can therefore, in principle, be adjusted with some degree of freedom. Nevertheless, the values inferred by BACCHUS are more closely aligned with those typically observed in stars of comparable spectral type, which motivates our preference for adopting the BACCHUS results in the determination of the planetary characteristics. It is worth emphasizing that when the BACCHUS parameters are used as input for the ARIADNE analysis, the resulting stellar mass and radius are in excellent agreement with those obtained using PARAM. Furthermore, irrespective of the adopted set of parameters, the inferred stellar mass and radius remain consistent within the quoted uncertainties.

\begin{table*}
\centering
 \caption{Spectroscopic  parameters for TOI-1472 and TOI-1648, as derived in this work. }   
\begin{tabular}{llccccccc }
\hline \hline \noalign{\smallskip} \noalign{\smallskip}
    \multicolumn{7}{c}{TOI-1472} \\ \noalign{\smallskip}  
 \hline
     \noalign{\smallskip} \noalign{\smallskip}
Method  & $T_\mathrm{eff}$  & $\log g_\star$ & [Fe/H] & $\xi_t$  & Mass & Radius &   Age  \\  
& (K)  &(cgs)& (dex) & km/s & (M$_{\odot}$)& (R$_{\odot}$)& (Gyr)    \\
\noalign{\smallskip}
     \hline
\noalign{\smallskip} 
\texttt{BACCHUS}+PARAM$^a$  &  5100$\pm$50  & 4.5$\pm$0.1  & 0.27$\pm$0.09 & 0.9$\pm$0.1  & 0.87$\pm$0.06  &  0.84$\pm$0.02  &   \ldots    \\

SPECIES+ARIADNE   & 5106$\pm$50  & 4.2$\pm$0.1  &   0.19$\pm$0.04   & 0.3$\pm$0.2 & 0.89$\pm$0.03  & 0.84$\pm$0.01 & 3.1$_{-2.2}^{+4.3}$   \\

SED &  \ldots   & \ldots   & \ldots  & \ldots & 0.90$\pm$0.05 & 0.83$\pm$0.02 & 3.1$\pm$0.4   \\ \noalign{\smallskip} 
\hline 
  \noalign{\smallskip} \noalign{\smallskip}
    \multicolumn{7}{c}{TOI-1648} \\ \noalign{\smallskip}
 \hline
     \noalign{\smallskip} \noalign{\smallskip}
\texttt{BACCHUS}+PARAM$^a$  & 4850$\pm$50  &  4.5$\pm$0.1  &   0.20$\pm$0.09 & 0.8$\pm$0.1  & 0.83$\pm$0.06  & 0.81$\pm$0.02  &  \ldots  \\

SPECIES+ARIADNE   &   5125$_{-40}^{+126}$   &  4.7$\pm$0.2   &  0.12$\pm$0.04    & 1.2$\pm$0.1  &  0.88$_{-0.07}^{+0.02}$  & 0.79$\pm$0.01 & 0.5$_{-0.5}^{+7.7}$ $^b$ \\ 

SED &  \ldots   &\ldots   &\ldots  &\ldots  & 0.82$\pm$0.05 & 0.83$\pm$0.02 & 5.7$\pm$0.4 \\ \noalign{\smallskip}

\hline 
\end{tabular}  
\label{Table: stellar spectroscopic parameters}
\parbox{0.9\linewidth}{\vspace{2pt}\footnotesize
$^{a}$Adopted values for the joint fit in Section~\ref{sec:planet}.  
$^{b}$The young limit of this age can be ruled out by the lack of stellar rotation signal in the TESS data (see Section~\ref{sec:freq}).
}
\end{table*}

\section{Frequency analysis} \label{sec:freq}

To explore periodic signals within the TESS photometry, radial velocity (RV) data, and stellar activity indicators, we conducted a frequency analysis using Generalized Lomb-Scargle (GLS) periodograms for the two systems. 

\begin{figure*}
 \centering
 \begin{subfigure}[h]{0.49\textwidth}
     \centering
     \includegraphics[width=\textwidth]{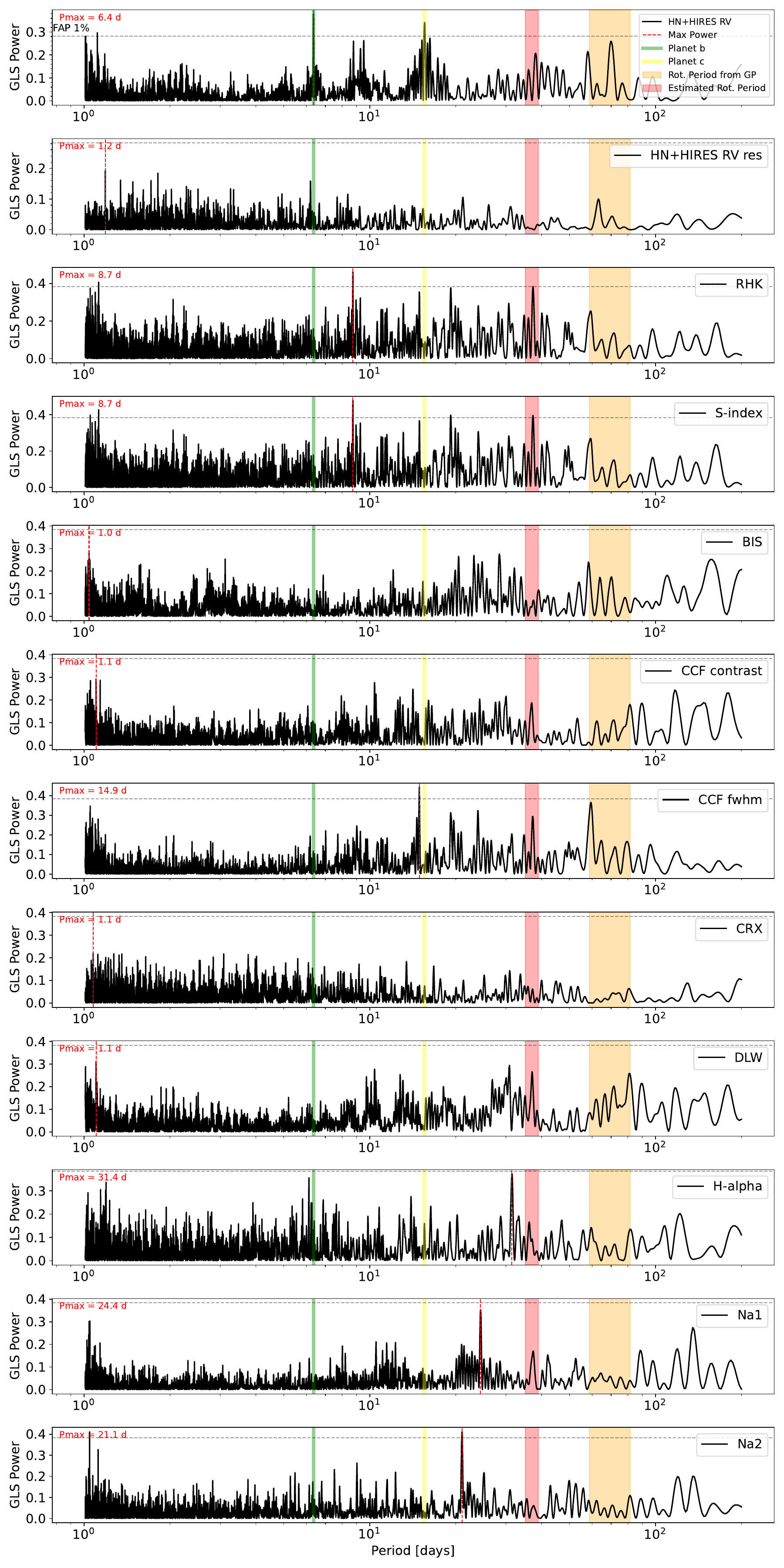}
 \end{subfigure}
 \hfill
 \begin{subfigure}[h]{0.49\textwidth}
     \centering
     \includegraphics[width=\textwidth]{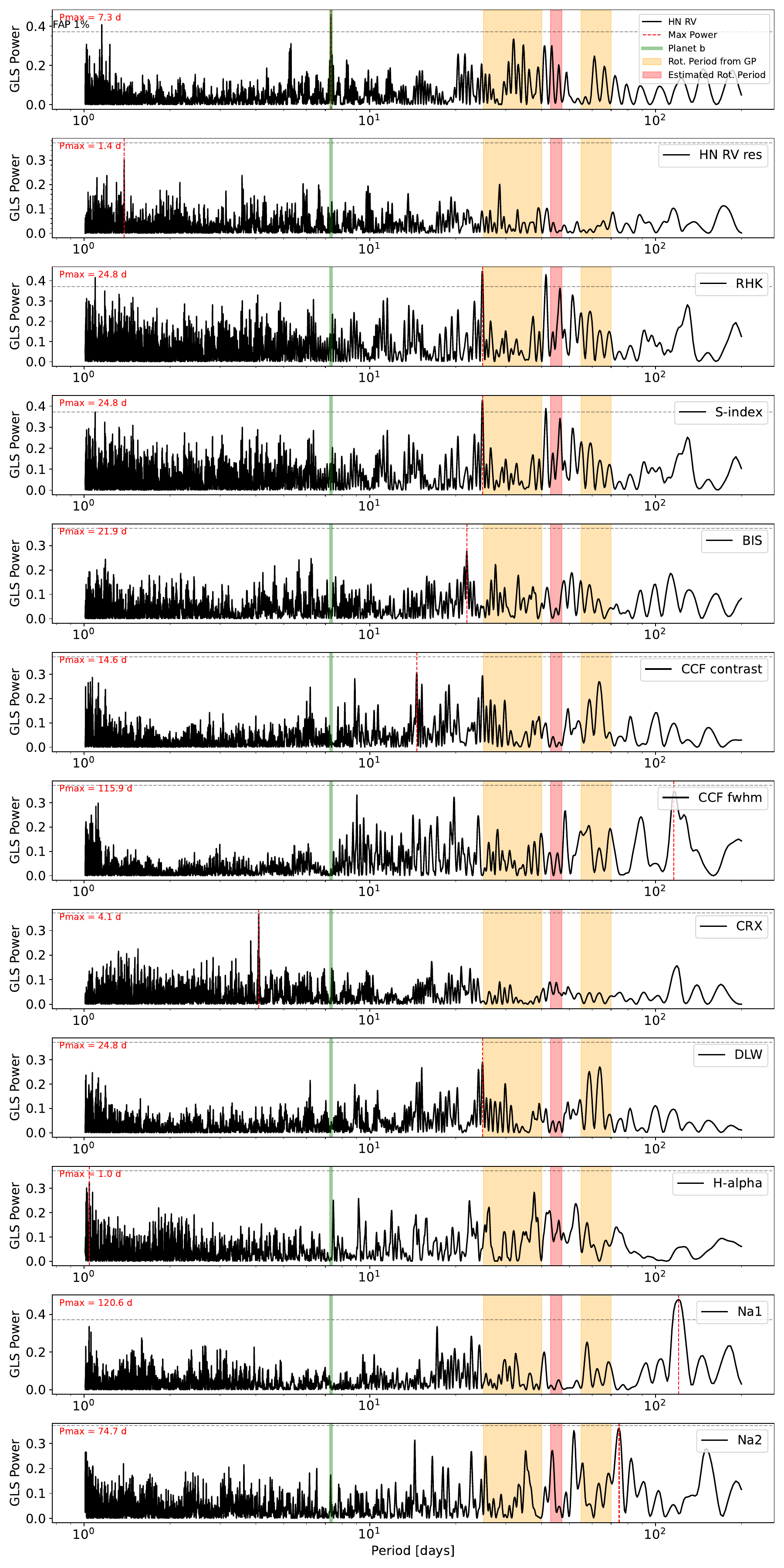}
 \end{subfigure}
    \caption{GLS periodograms for TOI-1472 (left), TOI-1648 (right). Periodograms are shown for the RV data, the residuals after subtracting the fitted planetary signal, and all stellar activity indicators derived from the HARPS-N DRS (\logrhk, S-index, bisector span, CCF contrast, and CCF FWHM) and from the \texttt{serval} pipeline (CRX, dLW, H$\alpha$, and the Na$_1$ and Na$_2$ lines). For TOI-1472, where two datasets (HARPS-N and HIRES) are available, the RVs have been shifted to a common median RV prior to computing the GLS periodogram. The orange shaded area denotes the stellar rotation period inferred from the joint Gaussian Process (GP) analysis using wide period priors (2–100 days), as detailed in Section \ref{sec:planet}. The red shaded area marks the rotation period derived from the SED-based analysis (Section \ref{sec:sed}). The red dashed line indicates the location of the dominant peak in each periodogram, while the green and yellow lines denote the orbital periods of planets b and c, respectively. For TOI-1648, the rotational modulation exhibits a double-peaked structure in the GP posterior.}
    \label{fig:periodogram1472-1648}
\end{figure*}

We searched for signs of stellar rotation in the TESS light curves, but none of the targets exhibited significant periodic modulation. In contrast, the RV data revealed prominent peaks (above the False Alarm Probability = 1\%) aligned with the orbital periods of the detected planets (see Figure \ref{fig:periodogram1472-1648}). The periodograms of the activity indicators showed a range of peaks, with varying significance across different diagnostics. 

Given the lack of a consistent rotation period across activity tracers, we adopted a Gaussian Process (GP) model in the joint RV-photometry fit (Section \ref{sec:planet}) with a broad, non-informative prior on the stellar rotation period, spanning from 2 to 100 days. This approach accommodates the uncertainty and diversity of the activity signals. In the periodogram figures, two shaded bands are used to guide interpretation: the red band marks the rotation period estimated from the spectral energy distribution (Section \ref{sec:sed}), while the orange band reflects the posterior distribution of the GP-inferred rotation period.

\begin{figure*}
 \centering
     \includegraphics[width=0.89\textwidth]{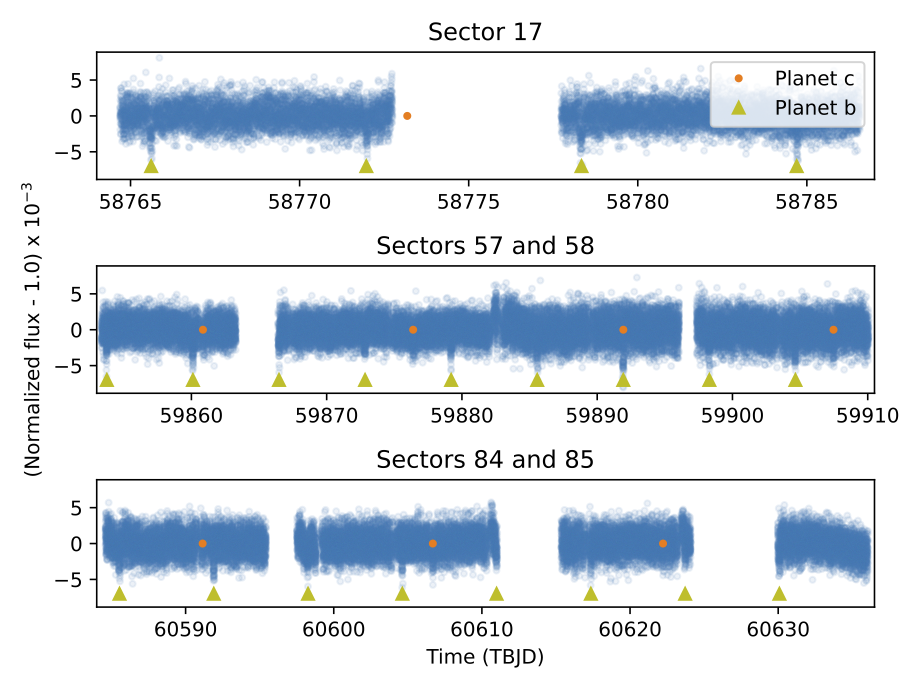}
     \caption{TESS time series for TOI-1472 during sectors 17, 57, 58, 84 and 85. The green triangle indicates the transits of TOI-1472\,b, while the orange dots indicate the planet c's transits. }
     \label{fig:toi1472-2planetsTESS}
\end{figure*}

For TOI-1472, the GP-inferred rotation period is nearly twice the value estimated from the SED. Both values, however, are consistent with prominent peaks observed in the activity periodograms, supporting their physical plausibility. Additionally, for TOI-1472, the periodogram reveals a highly significant peak (above FAP = 1\%) at approximately 15 days, which does not correspond to any features in the periodograms of the stellar activity indicators. A visual inspection of the TESS data further revealed additional transits, providing evidence for a second planet in the system. Figure~\ref{fig:toi1472-2planetsTESS} shows the TOI-1472 TESS time series in Sectors 17, 57, and 58. TESS did not detect planet c earlier because its transits during Sector 17 coincided with data gaps. In Sectors 57 and 58, although two full transits are visible, one transit is only partial, and another is completely blended with the transit of planet b, resulting in a deeper combined signal. These factors contributed to the initial non-detection of planet c by the SPOC pipeline.

In the case of TOI-1648, the GP posterior shows a bimodal distribution with peaks around 32 and 64 days, whereas the SED-based estimate lies between these two values. All these periods are consistent with signals in the activity indicators, suggesting possible harmonics. Motivated by the presence of a significant peak around 120 days in some of the activity indices, the prior range on the stellar rotation period was extended from 2–100 days to 2–200 days. With this broader range, an additional peak in the GP posterior emerges around 160 days (five times the rotation period of 32 days). This long-period signal may correspond to longer-timescale activity modulations or harmonics.

\section{Planet system parameters: joint fit}\label{sec:planet}

To derive the planetary system parameters, we conducted a comprehensive joint fit of the radial velocity (RV) measurements alongside the photometric transit data. This analysis was performed using the {\tt PyORBIT} package\footnote{\url{https://github.com/LucaMalavolta/PyORBIT}} \citep{Malavoltaetal2016,Malavoltaetal2018}. For modeling the transit light curves, we employed the \texttt{batman} code \citep{batman}, which implements the quadratic limb-darkened transit model from \citet{2002ApJ...580L.171M}. Each transit included a local polynomial trend to account for residual systematics, as the TESS light curves had not been pre-detrended. The fitted transit parameters included the time of inferior conjunction ($T_c$), orbital period ($P$), impact parameter ($b$), orbital eccentricity ($e$), and argument of periastron ($\omega$), using the \citet{Eastman2013} parameterization of eccentricity as $\sqrt{e}\cos\omega$ and $\sqrt{e}\sin\omega$. Quadratic limb darkening coefficients were parameterized following \citet{Kipping2013}. Additional free parameters comprised the scaled planetary radius ($R_{P}$/$R_{\star}$), and the stellar mass ($M_{\star}$) and radius ($R_{\star}$). Gaussian priors on the stellar mass and radius were imposed based on the spectroscopic analysis presented in Section \ref{subsec:thomas}. Limb darkening priors were derived from {\tt PyLDTk}\footnote{\url{https://github.com/hpparvi/ldtk}} \citep{Parviainen2015,Husser2013}, taking into account the bandpasses of the different instruments, and inflating the uncertainties to 0.1 to accommodate possible discrepancies between observed and theoretical limb darkening \citep{PatelandEspinoza2022}. The impact parameter was treated as a free parameter, and no dilution factor was applied since the PDCSAP fluxes were already corrected for crowding effects (see Section \ref{sec:obs}).

Instrumental offsets, as well as additional sources of noise such as stellar activity and systematics, were accounted for by including individual RV zero-point offsets and jitter terms in the model. Parameter exploration was carried out using the dynamic nested sampling algorithm implemented in \texttt{dynesty}\footnote{\url{https://github.com/joshspeagle/dynesty}} \citep{Speagle2020,Koposovetal2022}, with 1000 live points to efficiently sample the parameter space. For each system, we compared fits both with and without Gaussian Process (GP) regression. When included, the GP modeling utilized the \texttt{george} package \citep{george}, applying a quasi-periodic kernel as described by \citet{2015ApJ...808..127G}. In this kernel, the hyperparameters are: $h$, the amplitude of correlated noise; $\theta$, representing the stellar rotation period; $\omega$, the inverse length scale related to the evolution of active regions; and $\lambda$, the decay timescale of correlations. Model comparison was based on Bayesian evidence (log$\mathcal{Z}$) computed through nested sampling. In all cases, the GP-enhanced model was strongly preferred, with a difference in log-evidence ($\Delta \log \mathcal{Z}$) exceeding 3, indicating decisive support for including stellar activity modeling.

\begin{figure}
 \centering
 \hfill
 \begin{subfigure}[h]{0.49\textwidth}
     \centering
     \includegraphics[width=\textwidth]{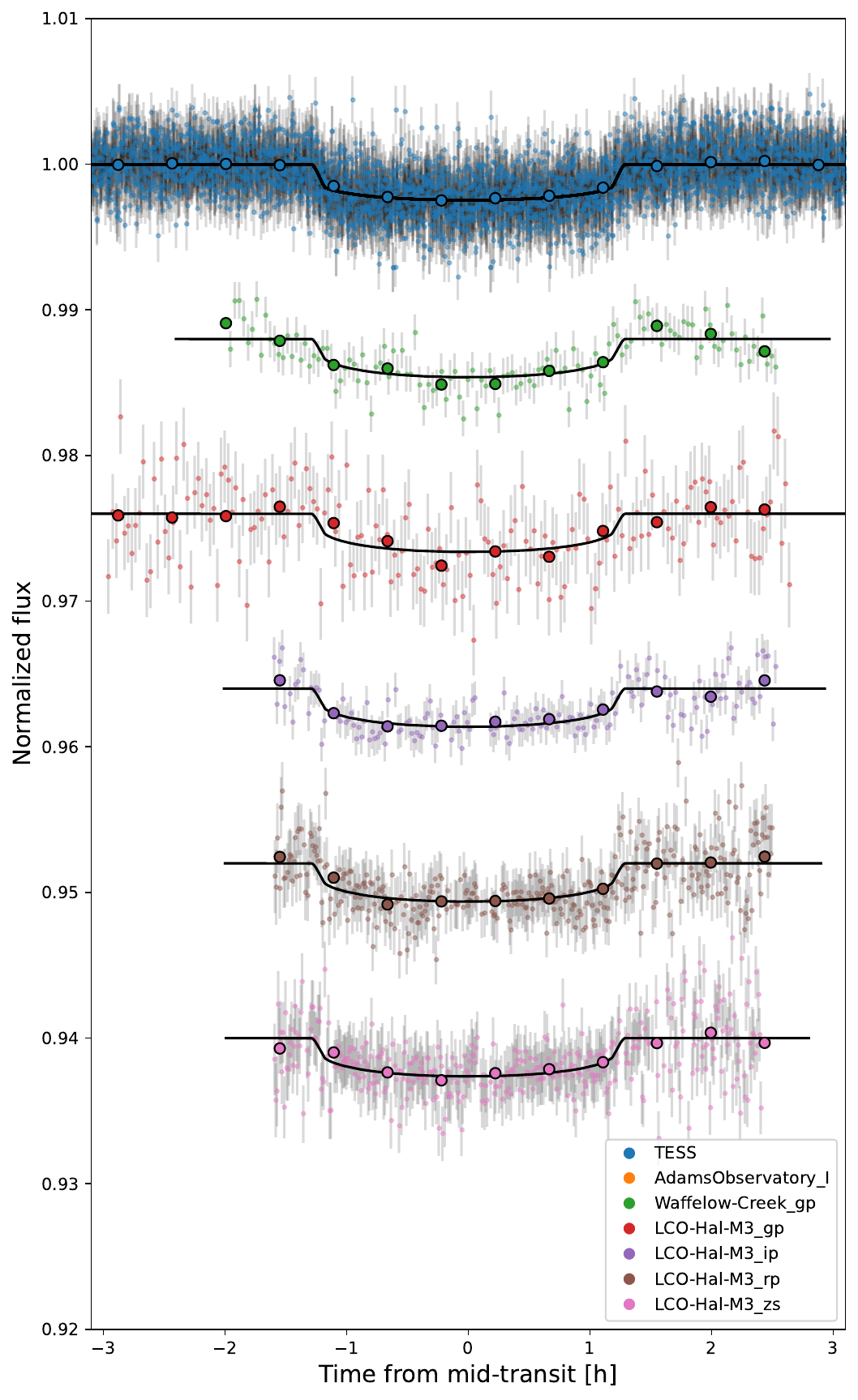}
 \end{subfigure}
 \hfill
 \begin{subfigure}[h]{0.50\textwidth}
     \centering
     \includegraphics[width=\textwidth]{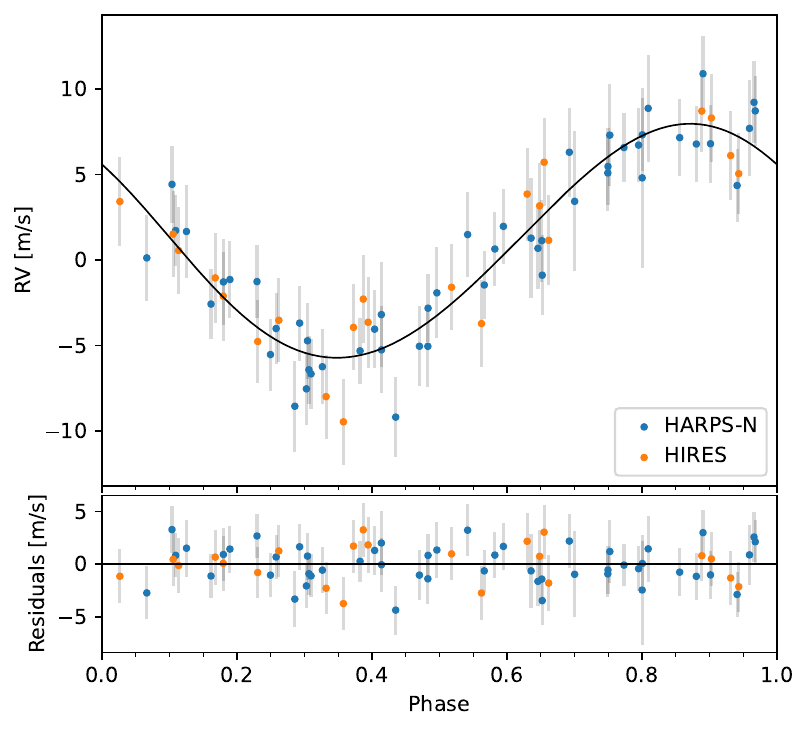}
 \end{subfigure}
    \caption{TESS light curves and binned data (upper), and HARPS-N and HIRES RV data (bottom) for TOI-1472\,b, with the 2p+GP model overplotted.}
    \label{fig:toi1472b}
\end{figure}
\begin{figure}
 \centering
 \hfill
 \begin{subfigure}[h]{0.49\textwidth}
     \centering
     \includegraphics[width=\textwidth]{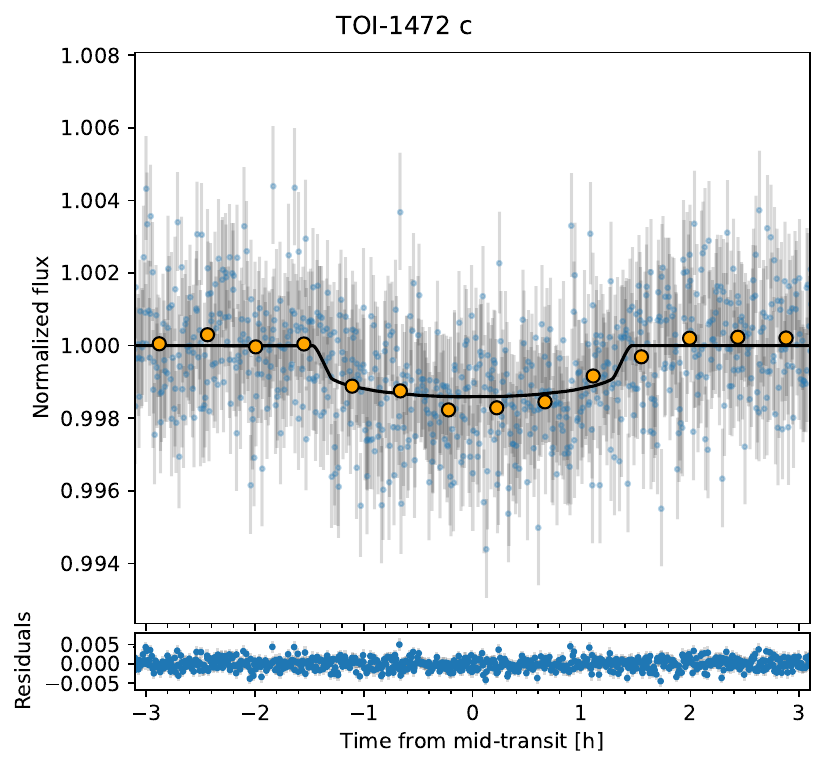}
 \end{subfigure}
 \hfill
 \begin{subfigure}[h]{0.50\textwidth}
     \centering
     \includegraphics[width=\textwidth]{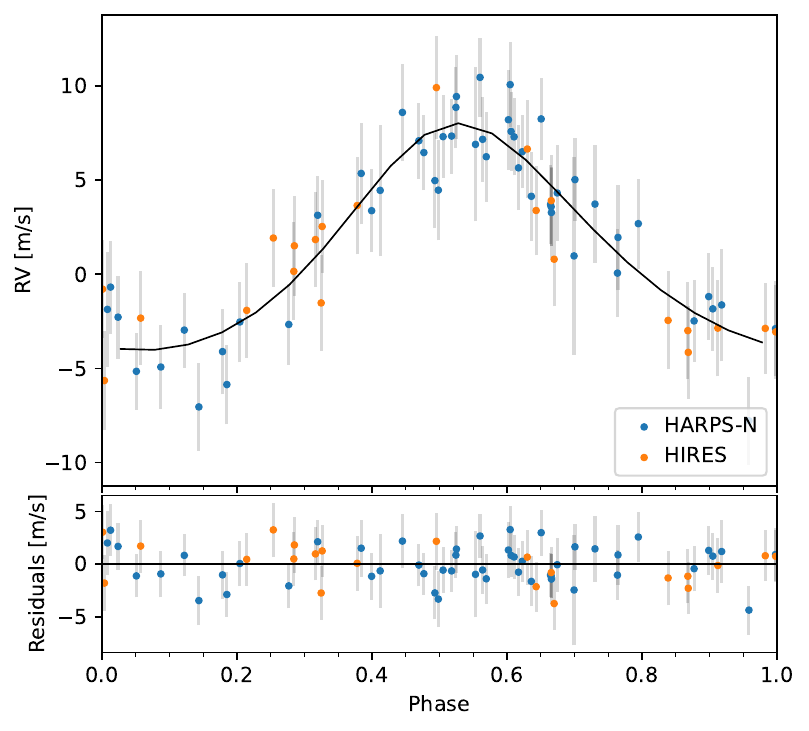}
 \end{subfigure}
    \caption{TESS light curves and binned data (upper), and HARPS-N and HIRES RV data (bottom) for TOI-1472\,c, with the 2p+GP model overplotted.}
    \label{fig:toi1472c}
\end{figure}

For TOI-1472, the planetary system parameters were derived through a 2-planet fit, since we identified a second planet from the RV data, combined with a GP. The results are detailed in Table \ref{tab:fit_params_toi1472}, and the radial velocity (RV) and transit fits, along with overlaid models, are depicted in Figures \ref{fig:toi1472b}  and \ref{fig:toi1472c}. TOI-1472\,b has a mass of 18.05$_{-0.85}^{+0.84}$ M$_{\oplus}$, a radius of 4.058 $\pm$ 0.098 R$_{\oplus}$, an orbital period of $\sim$6.36 days, and an eccentricity of 0.041\,$\pm$\,0.002. We find TOI-1472\,c to have a mass of 21.13$_{-0.99}^{+0.96}$~M$_{\oplus}$, a radius of $3.334 \pm 0.080$ R$_{\oplus}$, an orbital period of $\sim$15.54 days, and an eccentricity of 0.172\,$\pm$\,0.002.

\begin{figure}
 \centering
 \hfill
 \begin{subfigure}[h]{0.49\textwidth}
     \centering
     \includegraphics[width=\textwidth]{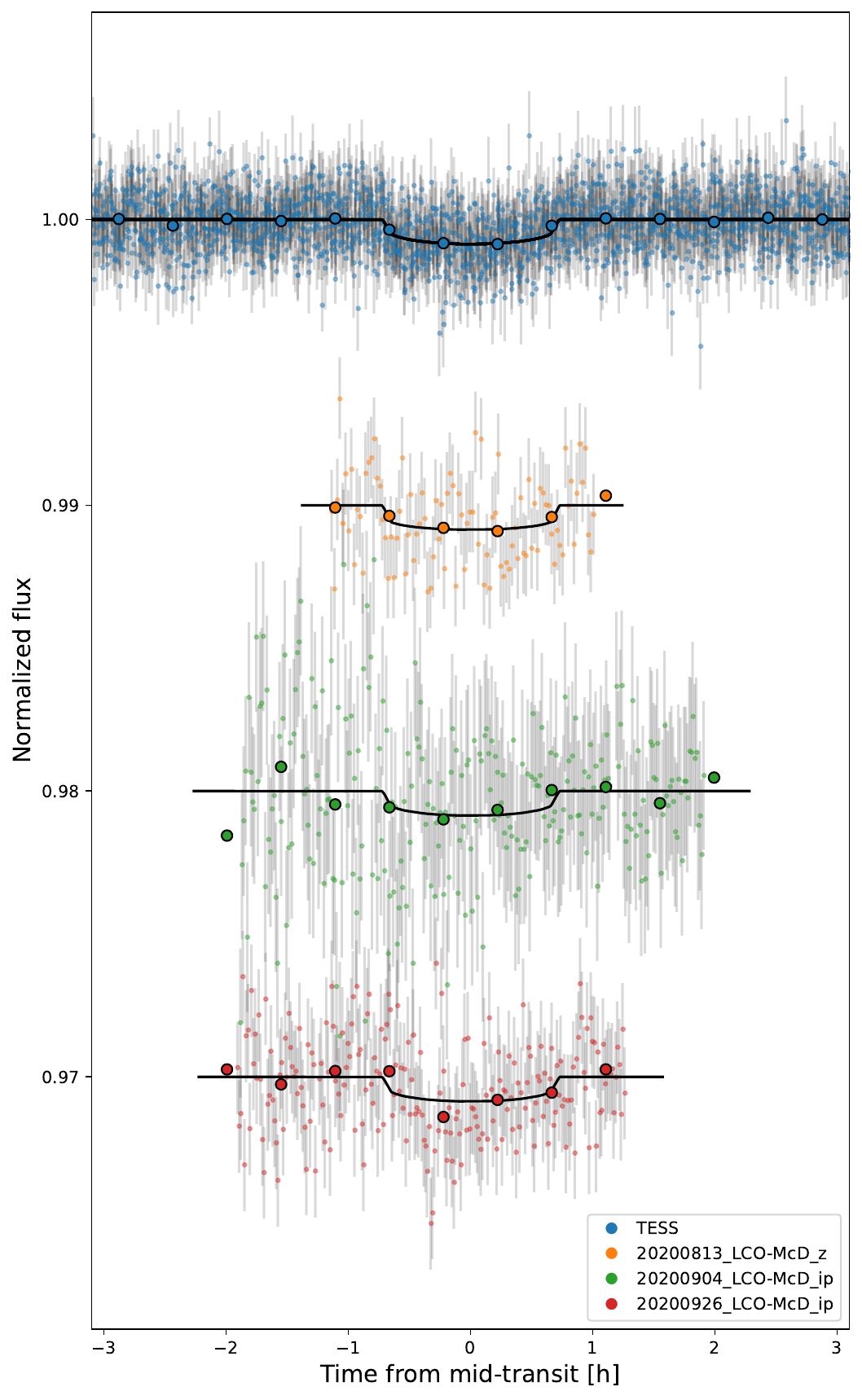}
 \end{subfigure}
 \hfill
 \begin{subfigure}[h]{0.50\textwidth}
     \centering
     \includegraphics[width=\textwidth]{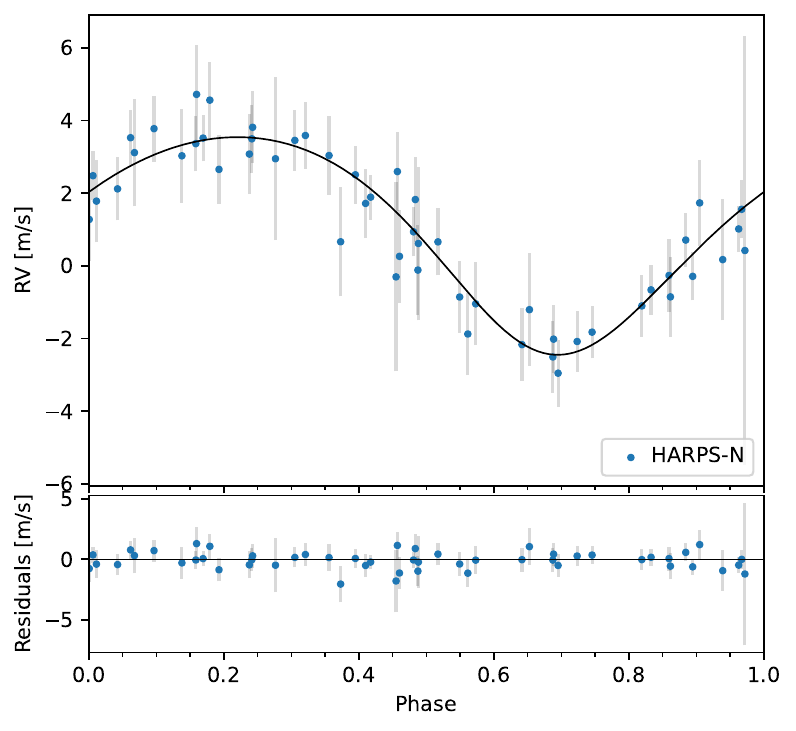}
 \end{subfigure}
    \caption{TESS light curves and binned data (upper), and HARPS-N RV data (bottom) for TOI-1648\,b, with the 1p+GP model overplotted.}
    \label{fig:toi1648}
\end{figure}

For TOI-1648, the system parameters obtained from a 1-planet plus GP fit are listed in Table \ref{tab:fit_params_toi1648}. The planet has a mass of 7.4$_{-1.3}^{+1.1}$ M$_{\oplus}$ and a radius of 2.54 $_{-0.12}^{+0.14}$ R$_{\oplus}$. It orbits its host star with a period of $\sim$7.33 days and has an eccentricity of 0.178$_{-0.053}^{+0.075}$. The transit and RV fits are shown in Figure \ref{fig:toi1648}.

\begin{table*}
  \footnotesize
  \caption{TOI-1472 parameters from the transit and RV joint fit, obtained with the model 2p+GP. \label{tab:fit_params_toi1472}}  
  \centering
  \begin{tabular}{lccc}
  \noalign{\smallskip}
  \hline
  \hline
  \noalign{\smallskip}
  Parameter & Prior$^{(\mathrm{a})}$ (Planet b / Planet c) & Value$^{(\mathrm{b})}$ Planet b & Value$^{(\mathrm{b})}$ Planet c \\
  \noalign{\smallskip}
  \hline
  \noalign{\smallskip}
  \multicolumn{3}{l}{\emph{ \bf Model Parameters }} \\
    Orbital period $P_{\mathrm{orb}}$ (days)  & $\mathcal{U}[5.9, 6.7]$ / \textbf{$\mathcal{U}[10.0, 19.0]$}  & $ 6.36338574\pm 1.1\times10^{-7}$ & $ 15.5381747\pm3.1\times10^{-6}$\\
      \noalign{\smallskip}
    Transit epoch $T_0$ (BJD - 2,450,000)  & $\mathcal{U}[9694.0, 9695.0]$ / $\mathcal{U}[9688.9, 9690.9]$  & $ 9694.66385\pm0.00002  $ & $9689.93567_{-0.00010}^{+0.00012}$\\
      \noalign{\smallskip}
    $\sqrt{e} \sin \omega_\star$ &  $\mathcal{U}(-1,1)$ & $ 0.182_{-0.004}^{+0.005}$  &   $-0.118_{-0.008}^{0.007}$\\
          \noalign{\smallskip}
    $\sqrt{e} \cos \omega_\star$  &  $\mathcal{U}(-1,1)$ & $0.088\pm0.004$ & $0.397\pm0.002$\\
          \noalign{\smallskip}
    Scaled planetary radius $R_\mathrm{p}/R_{\star}$ &  $\mathcal{U}[0,0.5]$ & $0.044\pm0.001$   & $0.036\pm0.001$\\
    Impact parameter, $b$ &  $\mathcal{U}[0,1]$  & $ 0.014_{-0.014}^{+0.029}$ & $0.657\pm0.003$\\
    Radial velocity semi-amplitude variation $K$ (m s$^{-1}$) &  $\mathcal{U}[0,300]$ & $ 6.841_{-0.019}^{+0.017} $  & $ 6.030\pm0.019$\\ 
          \noalign{\smallskip}
    \hline
          \noalign{\smallskip}
    \multicolumn{3}{l}{\textbf{Derived parameters}} \\
    Planet radius ($R_{\rm J}$)  & $\cdots$ & $ 0.362\pm0.009 $  & $0.298\pm0.007$\\
    Planet radius ($R_{\oplus}$)  & $\cdots$ & $ 4.058\pm0.098 $  & $ 3.334\pm0.080 $\\
    Planet mass ($M_{\rm J}$)  & $\cdots$ & $0.057\pm0.003$   & $0.067\pm0.003$ \\
    Planet mass ($M_{\oplus}$)  & $\cdots$ & $18.05_{-0.85}^{+0.84}$    & $21.13_{-0.99}^{+0.96}$\\
    Eccentricity $e$  & $\cdots$ & $0.041\pm0.002$   & $0.172\pm0.002$\\
    Scaled semi-major axis $a/R_\star$   & $\cdots$ & $ 18.565_{-0.034}^{+0.027} $  &$33.665_{-0.062}^{+0.050} $\\
    Semi-major axis $a$ (AU)  & $\cdots$ & $ 0.064\pm0.002 $  &$ 0.116\pm0.003$\\
    $\omega_{\rm P} $ (deg)  &  $\cdots$ &  $ 50.5\pm0.1$  & $164.7\pm0.2 $\\
    Orbital inclination $i$ (deg)  & $\cdots$ & $89.954_{-0.092}^{+0.044}$  & $88.906_{-0.005}^{+0.004} $\\
    Transit duration $T_{41}$ (days) & $\cdots$ & $ 0.114\pm0.001$ & $ 0.118\pm0.001$\\
    Transit duration $T_{32}$ (days) & $\cdots$ & $ 0.104\pm0.001$ & $ 0.104\pm0.001$\\
\noalign{\smallskip}
     \hline
\noalign{\smallskip}
    \multicolumn{3}{l}{\emph{\bf Calculated parameters}} \\
    Equilibrium Temperature$^{(\mathrm{c})}$ T$_{eq}$ (K) &  & $891\pm20$  & $662\pm13$\\
    Planetary density $\rho_P$ (g cm$^{-3}$) &  & $1.49\pm0.13$    & $3.14\pm0.27$\\
\noalign{\smallskip}
         \hline
\noalign{\smallskip}
    \multicolumn{3}{l}{\emph{\bf Other system parameters}} \\
    Jitter term $\sigma_{\rm HARPS-N}$ (\ms) & $\mathcal{U}[0,60]$ & \multicolumn{2}{c}{$1.678_{-0.016}^{+0.015}$  } \\
    Jitter term $\sigma_{\rm HIRES}$ (\ms) & $\mathcal{U}[0,60]$ & \multicolumn{2}{c}{$1.904_{-0.050}^{+0.045}$ } \\
    Limb darkening $q_1$  & $\mathcal{N}[0.451,0.1]$ & \multicolumn{2}{c}{$0.490\pm0.002$  } \\
    Limb darkening $q_2$ & $\mathcal{N}[0.121,0.1]$ & \multicolumn{2}{c}{$0.124\pm0.003$  } \\
\noalign{\smallskip}
         \hline
\noalign{\smallskip}
    \multicolumn{3}{l}{\emph{\bf Stellar activity GP model Parameters}} \\
    $h_{\rm HARPS-N}$  (\ms)  &  $\mathcal{U}[0, 100]$ & \multicolumn{2}{c}{$3.427_{-0.031}^{+0.028}$}  \\
    $h_{\rm HIRES}$  (\ms)  &  $\mathcal{U}[0, 100]$ &  \multicolumn{2}{c}{$3.798_{-0.040}^{+0.046}$ }  \\
    $\lambda$  (days)  &  $\mathcal{U}[5, 2000]$ & \multicolumn{2}{c}{$ 205.0_{-3.8}^{+3.3}$} \\
    $\omega$    &  $\mathcal{U}[0.01, 0.60]$ &\multicolumn{2}{c}{$0.217_{-0.002}^{+0.003}$ }\\
    $\theta$ (P$_{\rm rot}$)  (days)  &  $\mathcal{U}[2, 100]$ & \multicolumn{2}{c}{$69 _{-10}^{+13}$} \\
\noalign{\smallskip}
  \hline
   \noalign{\smallskip}
    \multicolumn{3}{l}{\emph{\bf Results from internal structure retrievals}} & $\cdots$ \\
    Envelope-to-planet mass fraction (\%) & $\cdots$ & $6.3_{-1.5}^{+1.8}$  & $\cdots$ \\
    Core and mantle mass fraction (\%) & $\cdots$ & $93.7_{-1.8}^{+1.5}$  & $\cdots$ \\
    \noalign{\smallskip}
  \hline
   \noalign{\smallskip}
    \multicolumn{3}{l}{\emph{\bf Results from atmospheric evolution}} \\
    Maximum initial mass ($M_{\oplus}$) & $\cdots$ & $27.7_{-1.2}^{+1.2}$ & $\cdots$ \\
    Initial envelope mass ($M_{\oplus}$) & $\cdots$ & $10.7_{-0.7}^{+0.6}$ & $\cdots$ \\
    Initial envelope-to-planet mass fraction (\%) & $\cdots$ & $38.6_{-4.0}^{+4.1}$ & $\cdots$ \\
    Initial radius ($R_{\oplus}$) & $\cdots$ & $10.2_{-0.2}^{+0.2}$ & $\cdots$ \\
    \noalign{\smallskip}
  \hline
   \noalign{\smallskip}
  \end{tabular}
~\\
  \emph{Note} -- $^{(\mathrm{a})}$ $\mathcal{U}[a,b]$ refers to uniform priors between $a$ and $b$, $\mathcal{N}[a,b]$ to Gaussian priors with median $a$ and standard deviation $b$.\\  
  $^{(\mathrm{b})}$ Parameter estimates and corresponding uncertainties are defined as the median and the 16th and 84th percentiles of the posterior distributions.\\
  $^{(\mathrm{c})}$ $T_{\rm eq}$ = $T_\star \, \biggl(\dfrac{R_\star}{2a}\biggr)^{1/2} \, [f(1-A_{\rm B})]^{1/4}$, assuming efficient recirculation f=1 and a null Bond albedo ($A_{\rm B}$ = 0).\\
\end{table*}

\begin{table*}
  \footnotesize
  \caption{TOI-1648 parameters from the transit and RV joint fit, obtained with the model 1p+GP. \label{tab:fit_params_toi1648}}  
  \centering
  \begin{tabular}{lcc}
  \noalign{\smallskip}
  \hline
  \hline
  \noalign{\smallskip}
  Parameter & Prior$^{(\mathrm{a})}$  & Value$^{(\mathrm{b})}$  \\
  \noalign{\smallskip}
  \hline
  \noalign{\smallskip}
  \multicolumn{3}{l}{\emph{ \bf Model Parameters }} \\
    Orbital period $P_{\mathrm{orb}}$ (days)  & $\mathcal{U}[6.8, 7.8]$   & $ 7.331602_{-0.000020}^{+0.000015}$ \\
      \noalign{\smallskip}
    Transit epoch $T_0$ (BJD - 2,450,000)  & $\mathcal{U}[9925.1,  9925.5]$   & $ 9925.3431_{-0.0010}^{+0.0008}$  \\
      \noalign{\smallskip}
    $\sqrt{e} \sin \omega_\star$ &  $\mathcal{U}(-1,1)$ & $ 0.11\pm0.18$ \\
          \noalign{\smallskip}
    $\sqrt{e} \cos \omega_\star$  &  $\mathcal{U}(-1,1)$ & $-0.380_{-0.070}^{+0.092}$ \\
          \noalign{\smallskip}
    Scaled planetary radius $R_\mathrm{p}/R_{\star}$ &  $\mathcal{U}[0,0.5]$ & $0.0287_{-0.0011}^{+0.0015}$   \\
    Impact parameter, $b$ &  $\mathcal{U}[0,1]$  & $ 0.736_{-0.060}^{+0.067}$ \\
    Radial velocity semi-amplitude variation $K$ (m s$^{-1}$) &  $\mathcal{U}[0,50]$ & $ 2.83_{-0.47}^{+0.38} $  \\ 
          \noalign{\smallskip}
    \hline
          \noalign{\smallskip}
    \multicolumn{3}{l}{\textbf{Derived parameters}} \\
    Planet radius ($R_{\rm J}$)  & $\cdots$ & $ 0.227_{-0.011}^{+0.013} $  \\
    Planet radius ($R_{\oplus}$)  & $\cdots$ & $ 2.54_{-0.12}^{+0.14} $  \\
    Planet mass ($M_{\rm J}$)  & $\cdots$ & $0.0233_{-0.0040}^{+0.0035}$    \\
    Planet mass ($M_{\oplus}$)  & $\cdots$ & $7.4_{-1.3}^{+1.1}$    \\
    Eccentricity $e$  & $\cdots$ & $0.178_{-0.053}^{+0.075}$   \\
    Scaled semi-major axis $a/R_\star$   & $\cdots$ & $ 26.4_{-1.9}^{+0.6} $  \\
    Semi-major axis $a$ (AU)  & $\cdots$ & $ 0.0694_{-0.0017}^{+0.0016} $ \\
    $\omega_{\rm P} $ (deg)  &  $\cdots$ &  $ 286_{-6}^{+7}$  \\
    Orbital inclination $i$ (deg)  & $\cdots$ & $88.29_{-0.26}^{+0.11}$ \\
    Transit duration $T_{41}$ (days) & $\cdots$ & $ 0.0642_{-0.0048}^{+0.0055}$ \\
    Transit duration $T_{32}$ (days) & $\cdots$ & $ 0.0566_{-0.0063}^{+0.0058}$ \\
\noalign{\smallskip}
     \hline
\noalign{\smallskip}
    \multicolumn{3}{l}{\emph{\bf Calculated parameters}} \\
    Equilibrium Temperature$^{(\mathrm{c})}$ T$_{eq}$ (K) &  & $799\pm16$ \\
    Planetary density $\rho_P$ (g cm$^{-3}$) &  & $2.49\pm0.60$   \\
\noalign{\smallskip}
         \hline
\noalign{\smallskip}
    \multicolumn{3}{l}{\emph{\bf Other system parameters}} \\
    Jitter term $\sigma_{\rm HARPS-N}$ (\ms) & $\mathcal{U}[0,60]$ & $0.65_{-0.40}^{+0.57}$   \\
    Limb darkening $q_1$  & $\mathcal{N}[0.5239,0.1]$ & $0.53\pm0.1$   \\
    Limb darkening $q_2$ & $\mathcal{N}[0.0985,0.1]$ & $0.107_{-0.096}^{+0.099}$   \\
\noalign{\smallskip}
         \hline
\noalign{\smallskip}
    \multicolumn{3}{l}{\emph{\bf Stellar activity GP model Parameters}} \\
    $h_{\rm HARPS-N}$  (\ms)  &  $\mathcal{U}[0, 100]$ & $2.27_{-0.34}^{+0.47}$  \\
    $\lambda$  (days)  &  $\mathcal{U}[5, 2000]$ $ 706_{-210}^{+392}$ \\
    $\omega$    &  $\mathcal{U}[0.01, 0.60]$ $0.085_{-0.019}^{+0.022}$ \\
    $\theta$ (P$_{\rm rot}$)  (days)  &  $\mathcal{U}[2, 100]$ & $31.8 _{-7}^{+32}$$^{(\mathrm{(d)})}$ \\
\noalign{\smallskip}
  \hline
   \noalign{\smallskip}
  \end{tabular}
~\\
  \emph{Note} -- $^{(\mathrm{a})}$ $\mathcal{U}[a,b]$ refers to uniform priors between $a$ and $b$, $\mathcal{N}[a,b]$ to Gaussian priors with median $a$ and standard deviation $b$.\\  
  $^{(\mathrm{b})}$ Parameter estimates and corresponding uncertainties are defined as the median and the 16th and 84th percentiles of the posterior distributions.\\
  $^{(\mathrm{c})}$ $T_{\rm eq}$ = $T_\star \, \biggl(\dfrac{R_\star}{2a}\biggr)^{1/2} \, [f(1-A_{\rm B})]^{1/4}$, assuming efficient recirculation f=1 and a null Bond albedo ($A_{\rm B}$ = 0).\\
  $^{(\mathrm{d})}$ The posterior distribution exhibits a double peak: one around 32 days and another around 64 days. When the prior range is extended to 200 days, an additional peak appears near 160 days (five times 32 days). \\
\end{table*}

\section{Detailed characterization of the systems} \label{sec:characterization}

In the previous sections, we presented the discovery of two new TESS planets, TOI-1472\,c and TOI-1648\,b, and an improved mass determination for the previously known planet TOI-1472\,b, all orbiting G-type stars. In the following subsections, we detail and discuss the bulk properties and system characterization of each planet.

\subsection{TOI-1472\,b and c}
\begin{figure}         
    \includegraphics[width=0.50\textwidth]{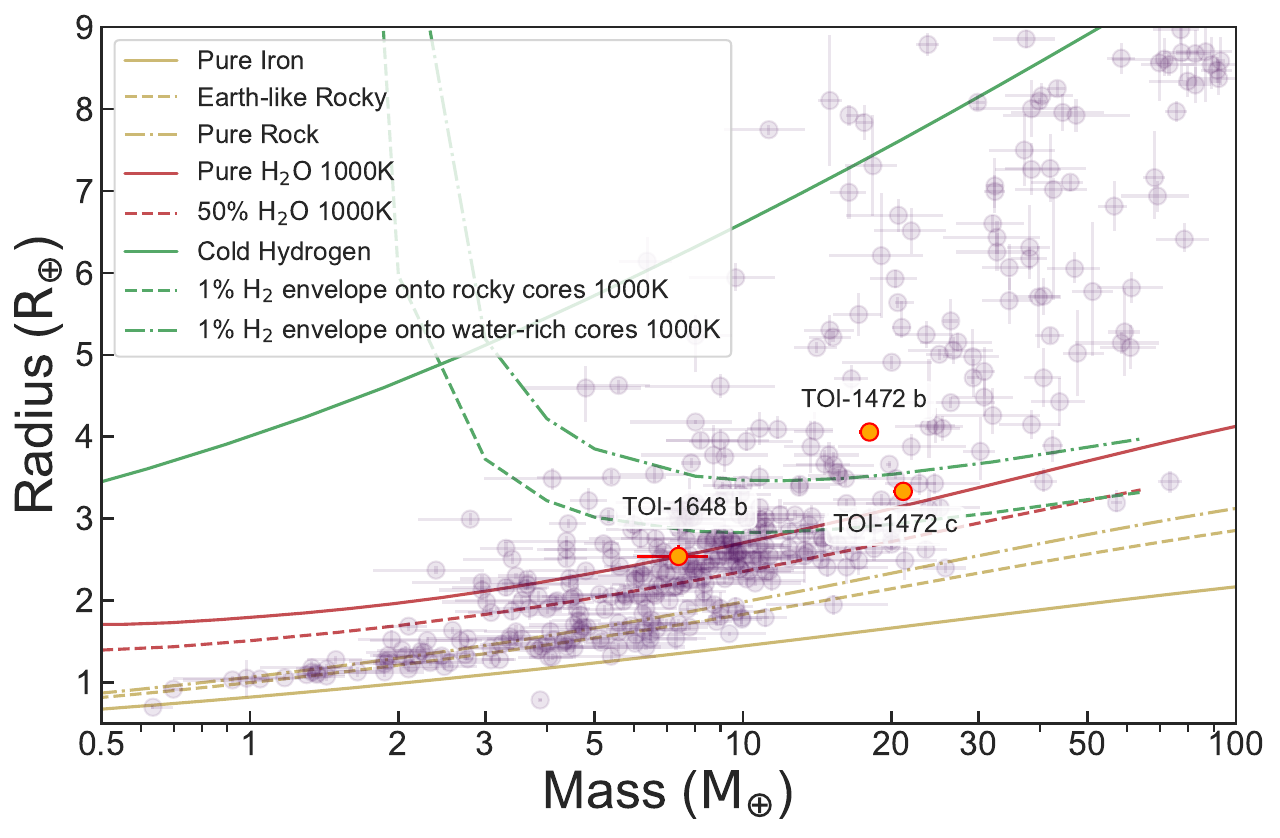}
    \caption{Mass–radius diagram for known exoplanets with measured masses and radii with precision better than 30\%. The theoretical composition models from \citet{Zeng2016} are showed (solid and dashed curves). Our planets, TOI-1472\,b and c, and TOI-1648\,b, are overplotted in color.}
    \label{fig:massradius}
\end{figure}

We determined the mass of TOI-1472\,b to be 18.05$_{-0.85}^{+0.84}$ M$_{\rm \oplus}$ and the radius to be 4.058 $\pm$ 0.098 R$_{\rm \oplus}$ (see mass-radius diagram in Figure \ref{fig:massradius}), which are consistent with the values reported in the discovery paper \citep{Polanskietal2024}, although we refined the planetary parameters with significantly enhanced precision. 
Compared to the previous estimates 
($16.5 \pm 5.1\,M_\oplus$ and $4.16 \pm 0.16\,R_\oplus$), our mass and radius measurements improve the precision by factors of $\sim$6 and $\sim$1.6, respectively; the detection significance increases from $3.2\sigma$ to $21.2\sigma$ for the mass and from $26.0\sigma$ to $41.4\sigma$ for the radius; and the relative uncertainties are reduced from 31\% to 4.7\% for the mass and from 3.85\% to 2.41\% for the radius. 
The improvement in the radius measurement is largely due to the availability of additional TESS sectors, which provided more transits and better phase coverage, while the improvement in the mass measurement is due to more precise radial velocities and a larger data set.

\begin{figure*}         
    \includegraphics[width=0.45\textwidth]{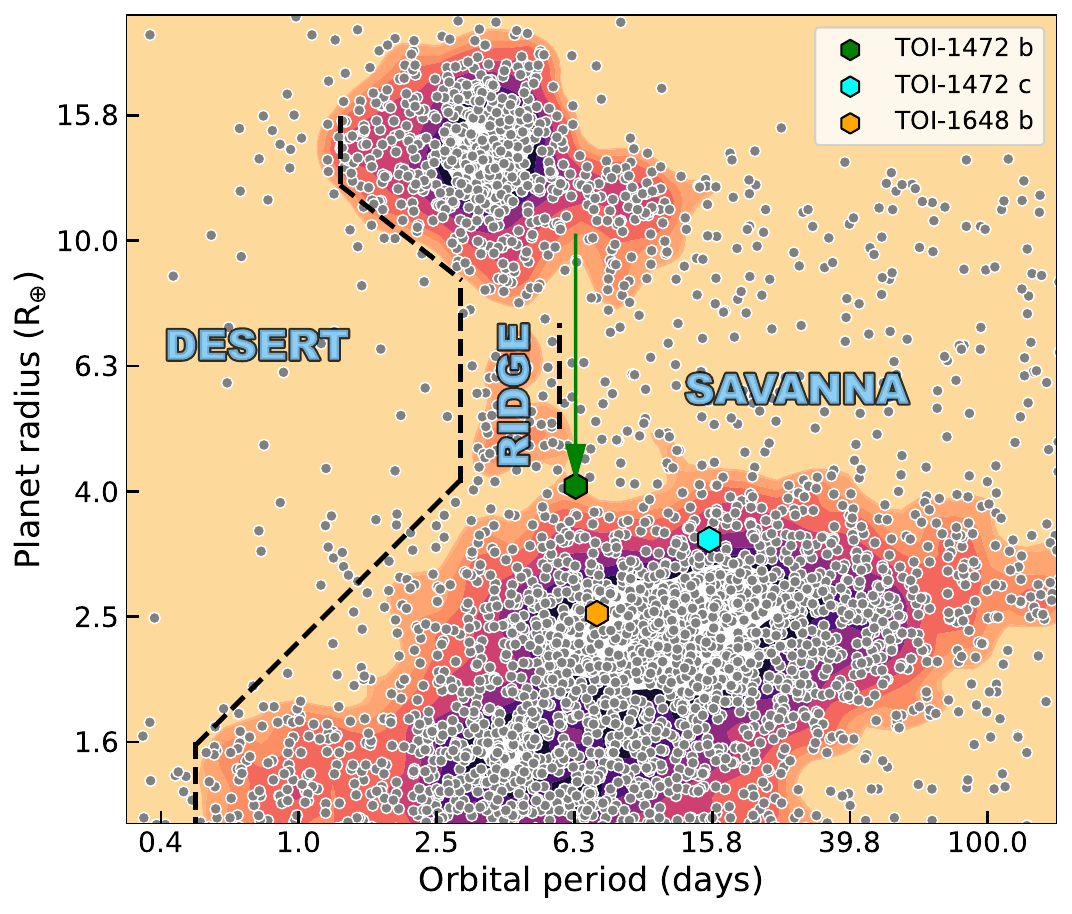}
    \includegraphics[width=0.45\textwidth]{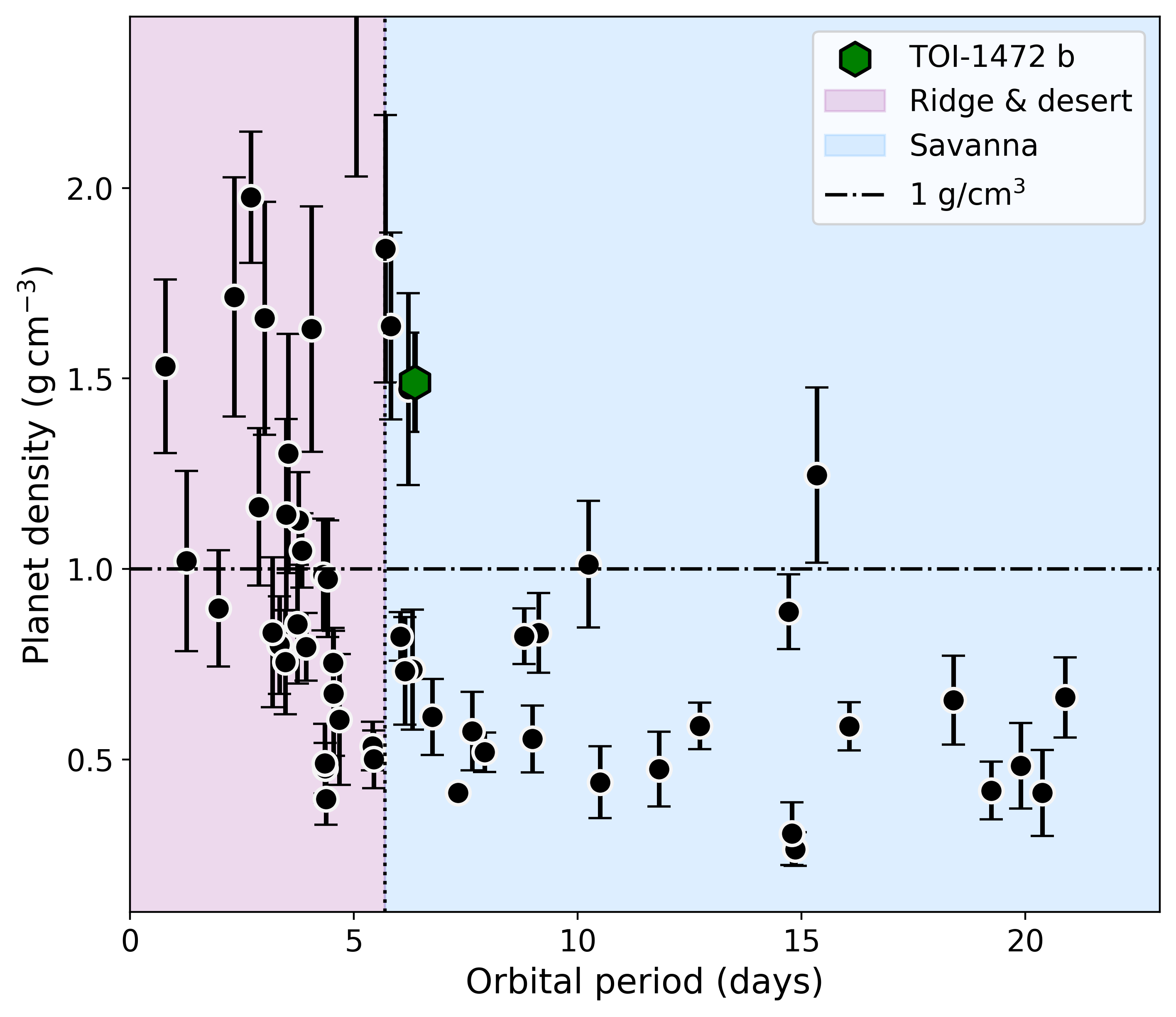}
    \caption{\textit{Left:} Planet radius vs. orbital period for all known exoplanets, highlighting the Neptunian desert, ridge, and savanna derived by \citealt{CastroGonzalez2024}. The arrows represent the evolution of TOI-1472\,b' radius from the initial value to the current value. \textit{Right:} Density-period diagram of all Neptunian planets (4.5 and 8.5 R$_{\rm \oplus}$) with density precision better than 33\%, with TOI-1472\,b overplotted. These plots were generated with \texttt{nep-des} (\url{https://github.com/castro-gzlz/nep-des}).}
    \label{fig:density_period_diagram_amedeo}
\end{figure*}

Planet b lies at the external edge of the Neptunian ridge, in the estimated transition between the ridge and savanna (Figure  \ref{fig:density_period_diagram_amedeo}, left). With a density of 1.49 $\pm$ 0.13 g\,cm$^{-3}$, this planet is compatible with the recently identified population of high-density ridge planets \citep[][see Figure \ref{fig:density_period_diagram_amedeo}, right]{CastroGonzalez2024b}, which motivated us to perform internal structure and atmospheric evolution simulations (see Section \ref{sec-toi-1471b-atm}.) 

We also discovered an additional transiting planet, TOI-1472\,c, with an orbital period of 15.54 days, a mass of $21.13_{-0.99}^{+0.96}$ M$_{\rm \oplus}$, and a radius of 3.334 $\pm$ 0.080 R$_{\rm \oplus}$. 

\begin{figure}  \includegraphics[width=0.47\textwidth]{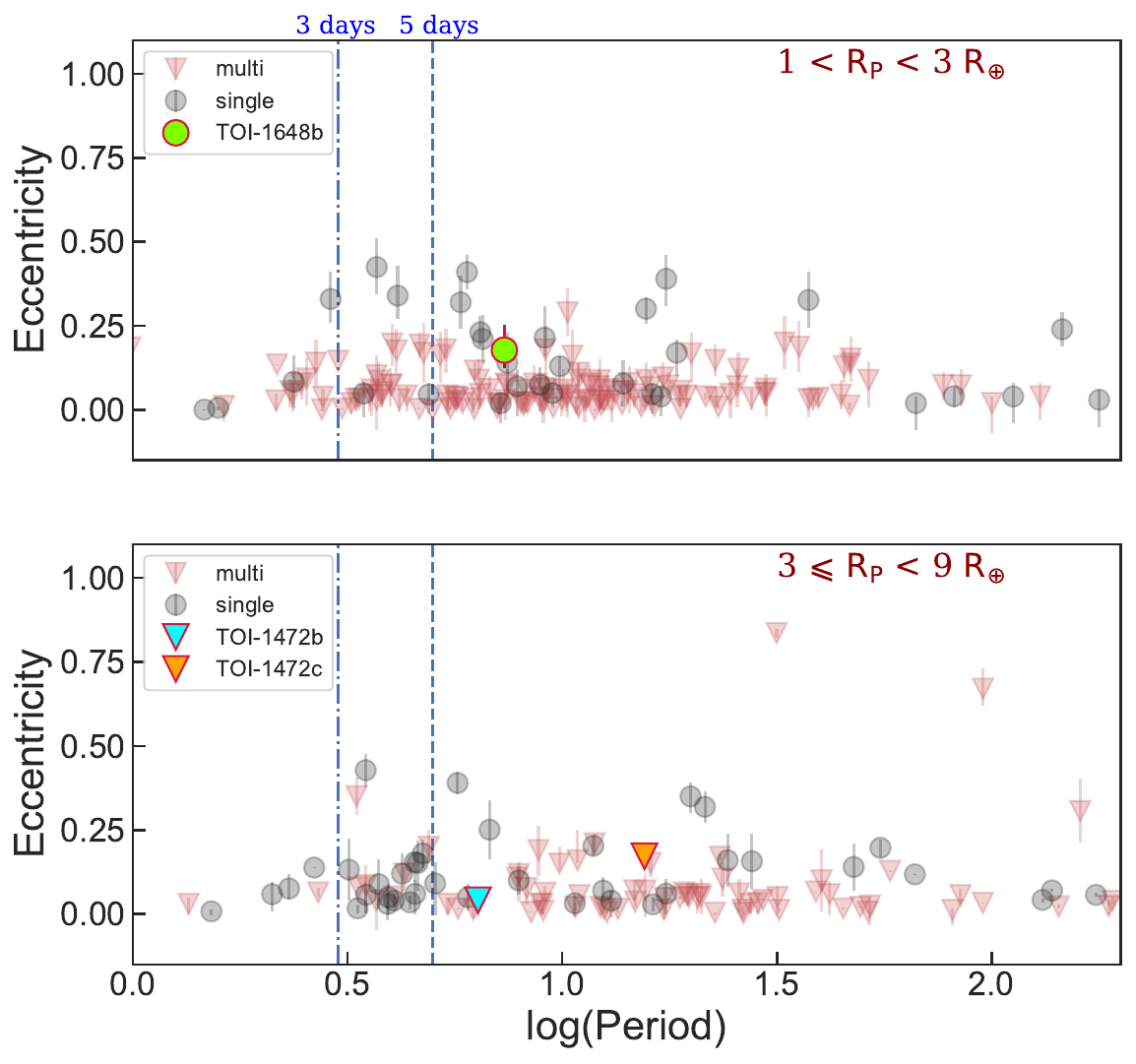}
    \caption{Distribution of eccentricities as a function of the orbital period for Earth-sized planets (upper) and Neptune-sized planets (bottom). The dashed blue lines represents the 3-day and 5-day boundaries. The red triangles represent the planets in multi-planetary systems. Data taken as of  June 2025.}
    \label{fig:eccentricity}
\end{figure}

The TOI-1472's planets have measured eccentricities of 0.041$\pm$0.002 and 0.172$\pm$0.002, respectively. To place the eccentricities of our planets in context, we produced a plot analogous to Figure~1 in \citet{2020A&A...635A..37C}, showing eccentricity as a function of orbital period for the population of super-Earths and sub-Neptunes (Figure \ref{fig:eccentricity}). On this diagram, we overplotted our three planets - TOI-1472\,b, TOI-1472\,c (and TOI-1648\,b, see Section \ref{sec-toi-1648b}) - highlighting how their eccentricities compare to the broader sample. 

TOI-1472\,b's low eccentricity lies below the typical eccentricity ($e$\,$\sim$\,0.15) observed for warm Neptunes \citep{2020A&A...635A..37C}, suggesting that tidal circularization has played a significant role at its relatively short orbital period. This is consistent with theoretical expectations where the tidal circularization timescale scales steeply with orbital period as $\tau_{\mathrm{circ}} \propto P_{\mathrm{orb}}^{13/3}$, leading to more efficient damping at shorter periods. In contrast, TOI-1472\,c exhibits a higher eccentricity, close to the median value found in \citet{2020A&A...635A..37C}'s sample. Its longer orbital period implies weaker tidal damping, allowing eccentricity to persist over long timescales. This elevated eccentricity may be maintained by mechanisms such as secular excitation from other companions or atmospheric processes that delay circularization, as discussed by \citet{2020A&A...635A..37C}.
Together, the differing eccentricities of the two planets highlight a system where tidal effects effectively damp eccentricity for the inner planet but are less efficient for the outer one, consistent with the broader trends illustrated in Figure~1 of \citet{2020A&A...635A..37C}.

\subsubsection{Internal structure and atmospheric evolution for TOI-1472\,b}
\label{sec-toi-1471b-atm}
We modeled the internal structure and atmospheric evolution of TOI-1472\,b using the \textsc{jade} code\footnote{Version 1.0.0, available at \url{https://gitlab.unige.ch/spice_dune/jade}}. This framework allows us to place constraints on the planet’s current interior composition and estimate its mass at the time of protoplanetary disk dispersal.

\textsc{jade} uses a Markov Chain Monte Carlo (MCMC) approach to infer key structural parameters of the planet—namely, the fraction of the total mass attributed to the silicate mantle, to the gaseous envelope, and the envelope’s metallicity. The retrieval is informed by the measured planetary mass and radius (Table~\ref{tab:fit_params_toi1472}) and the stellar age derived in Section~\ref{subsec:thomas}, while also accounting for stellar irradiation (including X-ray and extreme UV) and internal heat from the planet’s core.

The planet is modeled with a layered interior structure, consisting of an iron core, a silicate mantle, and an overlying hydrogen-helium envelope. The helium mass fraction is set to $Y = 0.2$, consistent with the composition of Neptune \citep{Hubbard1995, Helled2020}. The envelope includes a radiative upper layer and a deeper convective zone. Envelope opacities are adjusted by including trace metallic elements with solar abundances, parameterized by a metallicity term ($Z_{\rm met}$). The internal structure is calculated by integrating from the top of the atmosphere inward using a one-dimensional thermal model and polytropic equations of state for the rocky layers \citep{Seager2007}, iterating until the integrated mass reaches zero at the core \citep[e.g.,][]{Lopez2013, Jin2014}.

To account for the uncertainty in stellar age, we ran three separate retrievals using the nominal age (4.1 Gyr) and its $\pm3\sigma$ bounds. Each run consisted of 30,000 MCMC steps with 30 walkers and a burn-in phase of 6,000 steps. The retrieved internal parameters showed consistency across age ranges within their 1$\sigma$ uncertainties, so we adopt the results from the nominal-age run, reported in Table~\ref{tab:fit_params_toi1472}.

Following the internal structure modeling, we simulated the long-term atmospheric evolution. We fixed the mantle and core mass fractions, and adopted a small non-zero envelope metallicity ($Z_{\rm met} \sim 0.04$) inferred from the previous step. This small value ensures a physically realistic envelope structure without significantly affecting the model outcome \citep[see][]{Attia2025}.

Unlike scenarios involving late high-eccentricity migration — which allow planets to temporarily avoid high stellar irradiation and delay atmospheric loss \citep{Attia2021, Attia2025} — we assume TOI-1472\,b underwent early disk-driven migration. In this case, the planet settled near its current close-in orbit early in its evolution, experiencing prolonged and intense stellar irradiation. This assumption places the planet in a configuration that maximises its exposure to atmospheric erosion, enabling us to estimate its maximum initial mass. Stellar bolometric and high-energy (XUV) luminosity histories were computed with the GENEC stellar evolution models \citep{Eggenberger2008}, to capture how changing irradiation shaped the planet’s atmospheric evolution.

Under this early migration assumption, we held the planet’s orbital configuration constant and switched off dynamical evolution in the \textsc{jade} runs. The simulations began at the expected time of disk dispersal and extended to 13 Gyr, encompassing the full plausible age range of the system within $3\sigma$. These simulations trace the mass and radius evolution of the planetary envelope over time, enabling us to infer the maximum initial mass the planet could have had while still matching present-day observations.

\begin{figure}
\includegraphics[width=1.00\linewidth,trim=10 63 0 80,clip]{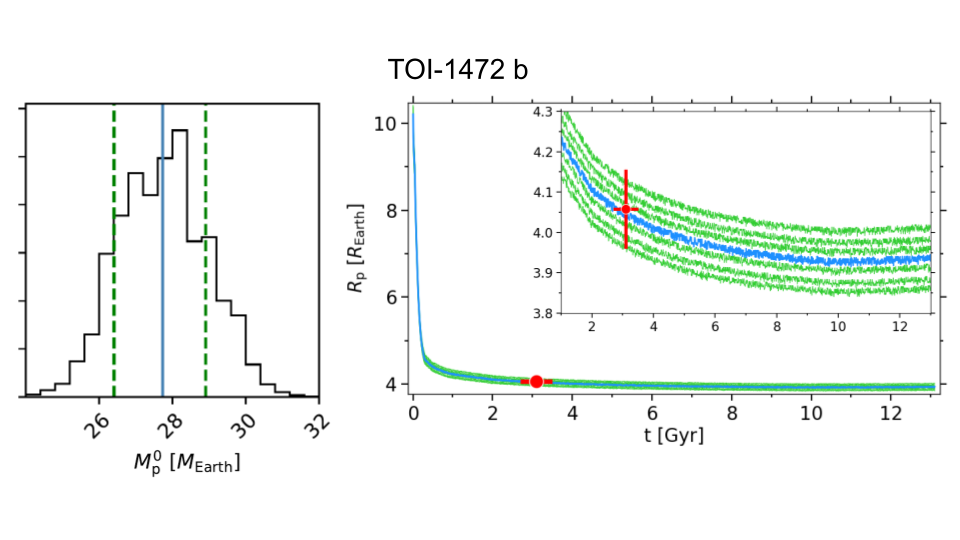}
\caption{\textsc{jade} simulations of atmospheric evolution for TOI-1472\,b. \textit{Left}: Posterior probability distribution for the planet's initial mass. The median value is indicated by a solid blue line, while the dashed green lines mark the 1$\sigma$ highest density interval (HDI). \textit{Right}: Time evolution of the planet’s radius from the best-fit simulation (blue) alongside a selection of representative models within the 1$\sigma$ HDI (green). The observed planetary radius and system age, including their uncertainties, are shown as a red data point. The inset provides a closer view of the radius evolution over a more limited time range.}  
\label{fig:atm-mass}
\end{figure}

We initialized a series of simulations with increasing masses, starting from the observed value of 0.06 $M_{\rm J}$ up to 1 $M_{\rm J}$. For each case, the planet’s radius at the system's current age was compared to the observed radius. Using importance sampling, we constructed a posterior distribution of initial masses, adopting the median and highest-density interval as the best-fit value and uncertainty (see Figure~\ref{fig:atm-mass}, left panels). This estimate represents the maximum feasible initial mass—any higher mass would retain too much atmosphere under the assumed high-irradiation conditions and yield a planet larger than observed, making it incompatible with current measurements.

Figure~\ref{fig:atm-mass} illustrates the atmospheric evolution history of TOI-1472\,b under the assumption that it migrated early to its current close-in orbit. During the first few hundred million years—when the host star's XUV emission was at its peak—the planet experienced substantial atmospheric erosion. As stellar high-energy output declined over time, the rate of atmospheric escape decreased, leading to a gradual contraction of the planetary radius over several gigayears.

We carried out atmospheric evolution simulations for a range of initial planetary masses, from 0.06 to 1.0 $M_{\rm J}$, keeping all other planetary parameters constant. The simulation that best matches the observed radius (depicted as the blue curve in the right panel of Fig~\ref{fig:atm-mass}) indicates that, at an estimated system age of 3 Gyr, TOI-1472\,b is likely still undergoing atmospheric loss, with its radius continuing to decrease below its current observed size of approximately 4 $R_{\rm \oplus}$. 
Under the assumption of early, rapid inward migration—which maximises mass loss—the planet is inferred to have initially formed with a mass of $27.7\pm1.2~M_{\rm \oplus}$ (equivalent to 0.087 $M_{\rm J}$) and a radius of about 10.2 $R_{\oplus}$. In this scenario, the envelope mass was around $10.7~M_{\oplus}$, or roughly 39\% of the total mass. We note that the uncertainty on this initial mass primarily reflects the system age under this migration assumption; different migration histories would likely lead to larger variations than indicated by this error.
The left panel of Figure~\ref{fig:density_period_diagram_amedeo} shows the planet's current location and its inferred evolutionary track from its initial, larger radius to its present-day size.

Following \cite{Kempton_et_al-2018PASP..130k4401K} we calculated the transmission spectroscopy metric (TSM) and the emission spectroscopy metric (ESM) for the planets of TOI-1472. For these calculations, we used stellar parameters ($R_\star$ and $T_{\mathrm{eff}}$) from Table \ref{Table: stellar spectroscopic parameters} (BACCHUS+PARAM), photometric and kinematic data ($m_J$, $m_K$, and RV) from Table \ref{t:star_param}, and planetary parameters ($R_p$, $M_p$, and $T_{\mathrm{eq}}$) from Table \ref{tab:fit_params_toi1472}.

\begin{figure*}
\centering
\includegraphics[width=\linewidth]{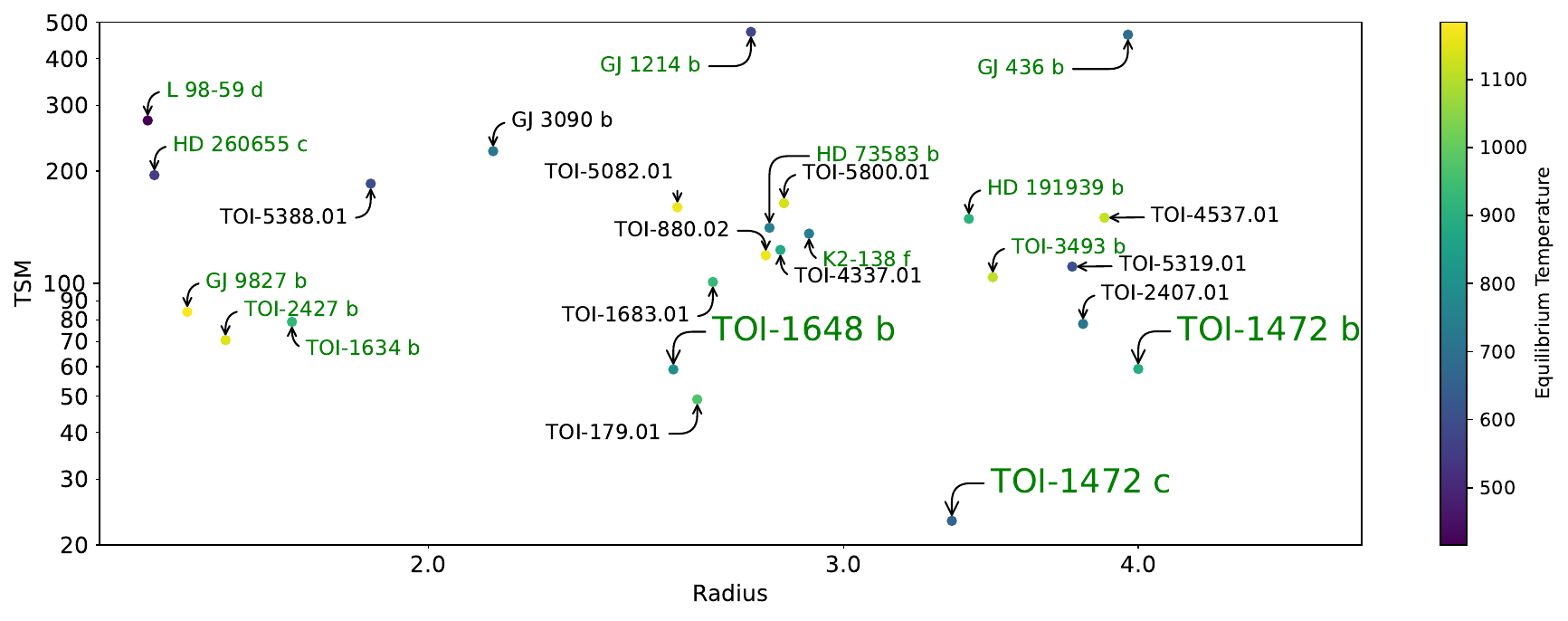}
\caption{TSM versus planet radius for hot (350\,K < T$_{eq}$ < 1250\,K) small sub-Neptunes (1.50 < R$_{\rm p}$ < 2.75\,\rearth) and large sub-Neptunes (2.75 < R$_{\rm p}$ < 4.00\,\rearth).
    Data come from \protect\cite{Hordetal2024} and include planets with high TSM and mass determination above $5\sigma$ recently confirmed by \protect\citet[][TOI-3439\,b]{2025arXiv250412884C} and \protect\citet[][TOI-2427\,b]{2025arXiv250412884S}. Names of the confirmed planets with mass measurements $> 5\sigma$ are in green. Planets presented in this paper are annotated with larger fonts.}
\label{figure-TOI-1472-TOI-1648-pr-tsm-coloured_annotations}
\end{figure*}

Figure \ref{figure-TOI-1472-TOI-1648-pr-tsm-coloured_annotations} displays all the sub-Neptunes (with radius $\leq$ 4 $R_{\oplus}$) and their TSM values with overplotted the planets from this work. The TSM values are 59.1 $\pm$ 6.0 and 23.2 $\pm$ 2.3 for planets b and c, respectively. The ESM is 8.8 $\pm$ 0.6 for TOI-1472\,b, and 2.8 $\pm$ 0.2 for TOI-1472\,c. The transmission and emission spectroscopy metrics of TOI-1472\,b are favorable for JWST follow-up observations. Unfortunately, those of TOI-1472\,c are too low. 

For TOI-1472\,b, three RV measurements were obtained during transit. Although these in-transit data are not analyzed in this work, they present a valuable opportunity for future studies. In-transit RVs can be used to probe atmospheric escape, particularly of species like hydrogen and helium, through high-resolution transmission spectroscopy. Additionally, such measurements are influenced by the Rossiter–McLaughlin (RM) effect, which for TOI-1472\,b is expected to produce a semi-amplitude of approximately 3.9\,m\,s$^{-1}$. Targeted observations of the RM effect could help constrain the projected spin–orbit alignment, offering further insight into the system's formation and migration history.

\subsubsection{Dynamical analysis of TOI-1472 system}
\label{sec-dynamical_analysis}

To ascertain the dynamical stability of the solution presented in Table~\ref{tab:fit_params_toi1472} we employed the Reversibility Error Method \citep[REM;][]{Panichi_Gozdziewski_Turchetti-2017MNRAS.468..469P}, which has been demonstrated to be a close analogue of the Maximum Lyapunov Exponent (MLE). In the analysis of multi-body systems, it relies on numerical integration schemes that are time-reversible, in particular symplectic algorithms. This method is based on calculating the difference between the initial state vector and the final state vector, which is obtained by integrating the system of equations at a specific time and returning to the initial epoch. The difference thus defined will depend on the dynamic nature of the system. $\mathrm{\widehat{REM}}=1$ or $\log\mathrm{\widehat{REM}}=0$ means that the difference reaches the size of the orbit.

\begin{figure}
\centering
\includegraphics[width=1.000\linewidth]{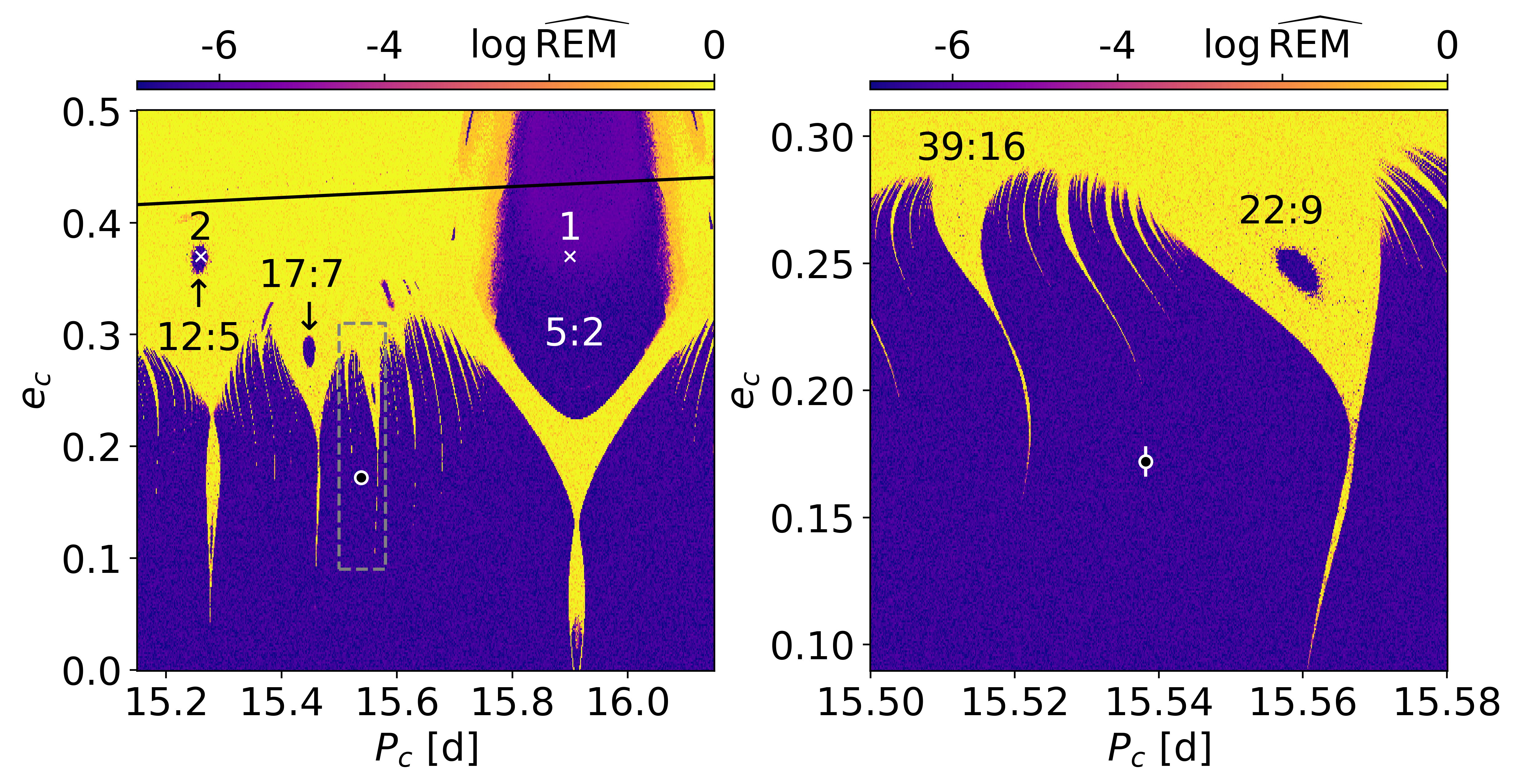}
    \caption{Dynamical maps for a wide range of orbital periods and eccentricities of the TOI-1472\,c (right panel is a zoomed version of left plot). Small values of the fast indicator $\log \mathrm{\widehat{REM}}$ characterises regular (long-term stable) solutions, which are marked with black/dark blue colour. Chaotic solutions are marked with brighter colours, up to yellow. The black line represents the so-called collision curve of orbits, defined by the condition: $a_b(1 + e_b) = a_c(1 - e_c)$. The black filled circle with a white rim indicates the position of the solution for the outer planet c presented in Table~\ref{tab:fit_params_toi1472}. Numbers show identified high-order resonances. Points 1 and 2 refer to Figure~\ref{figure-TOI-1472-critical_angles}. The resolution for the plots is 601 $\times$ 301 points.
    \label{figure-TOI-1472-dynamical_map}}
\end{figure}

The dynamical stability of the solution was tested using the \verb|whfast| integrator with the 17th order corrector (with a fixed time step of 0.15 d) as implemented within the \verb|REBOUND| package \citep{Rein_Liu-2012A&A...537A.128R,Rein2015MNRAS.452..376R, Rein2019MNRAS.489.4632R} for $\simeq 2\times 10^5$ orbital periods of the outermost planet ($\simeq 8500$ years). As illustrated in Figure~\ref{figure-TOI-1472-dynamical_map}, we obtained $\log\mathrm{\widehat{REM}}$ $\leq$ -5, indicating that the solution and its surrounding area are stable. The $\pm 3\sigma$ uncertainty of the orbital period and eccentricity of the planet c, placed on the dynamical maps, shows a safe distance from the chaotic structures. 

\begin{figure}
\centering
\includegraphics[width=0.49\linewidth]{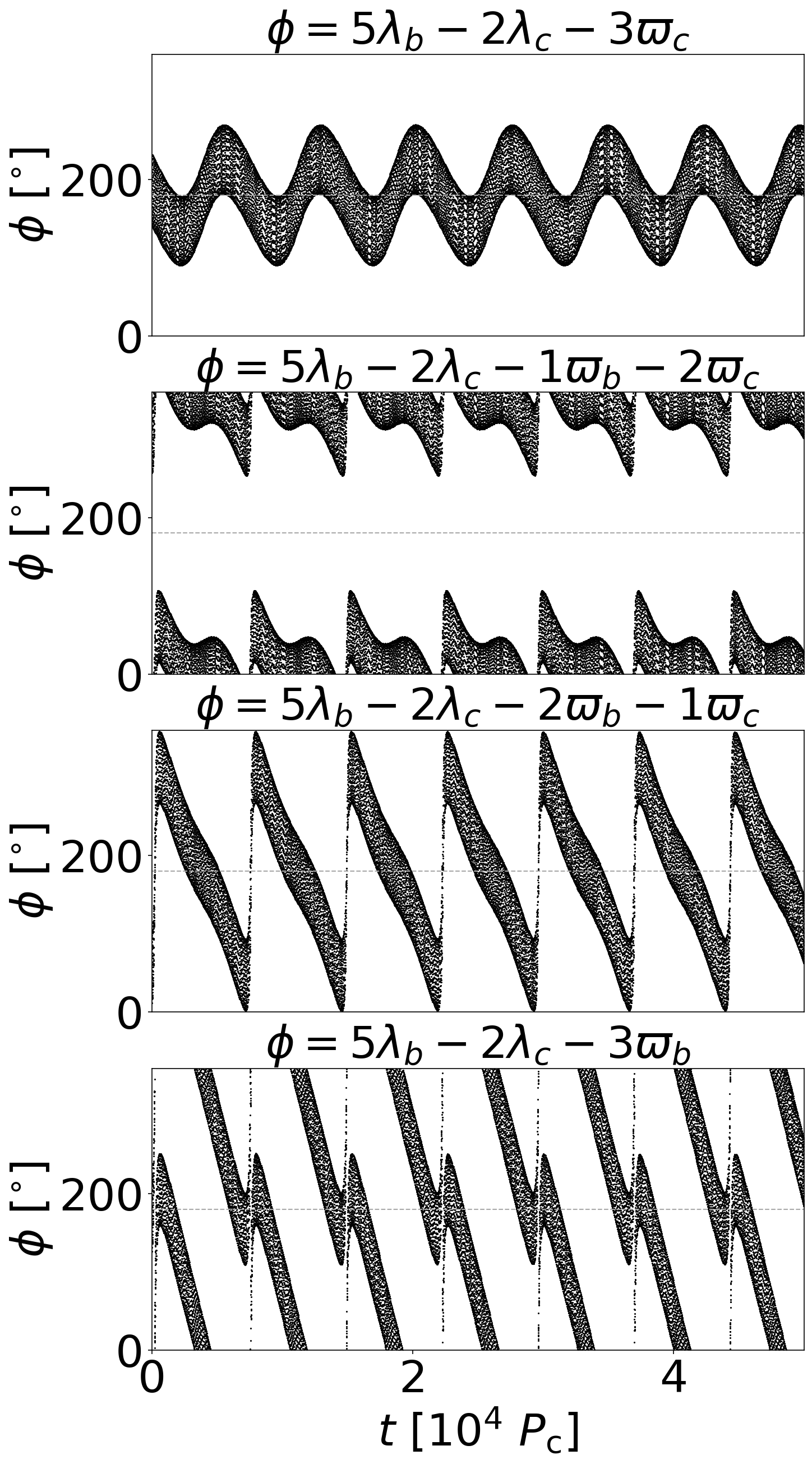}
\includegraphics[width=0.49\linewidth]{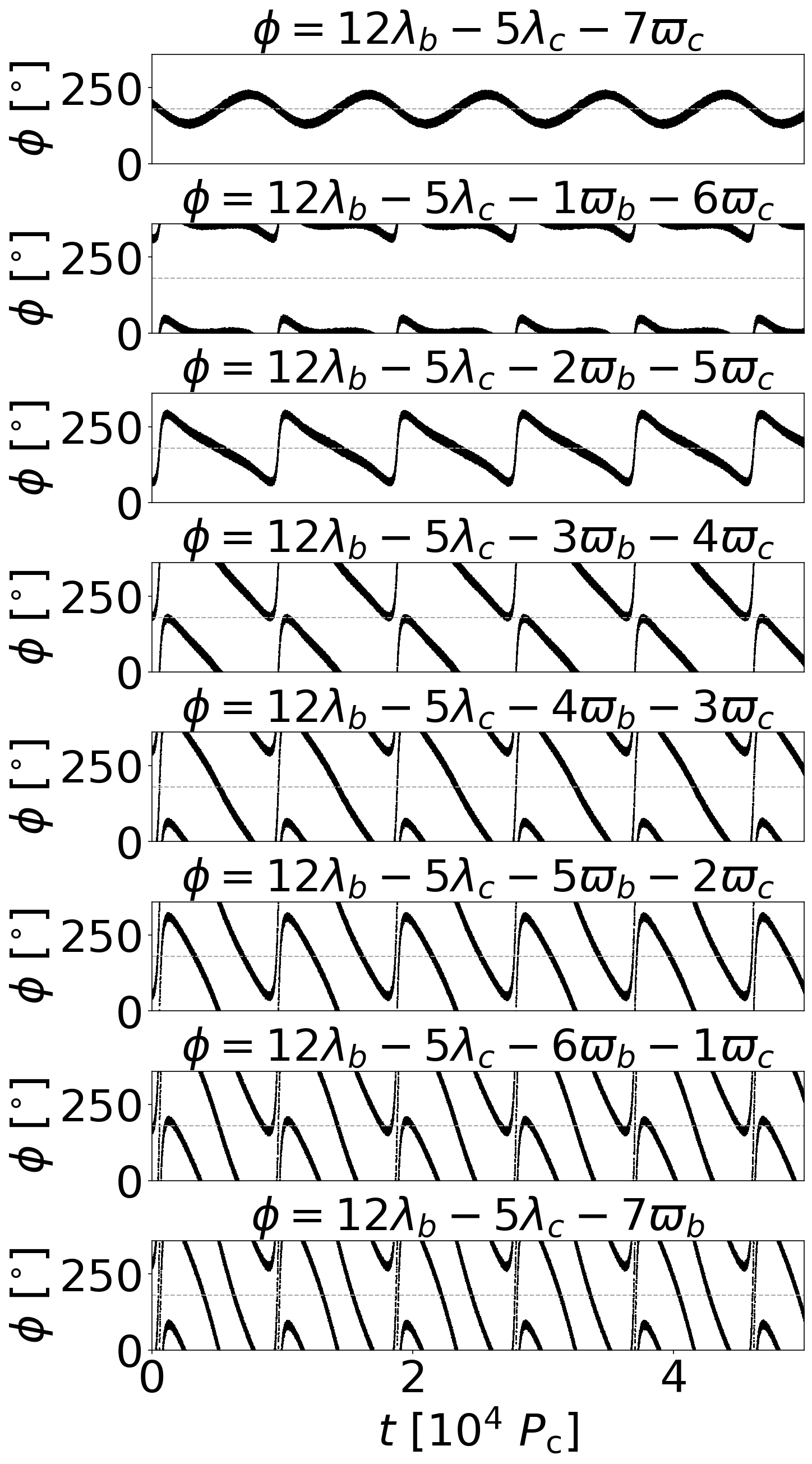}
    \caption{Evolution of critical angles for selected orbital period and eccentricity values of TOI-1472\,c, corresponding to the points marked in Figure~\ref{figure-TOI-1472-dynamical_map}. The remaining orbital elements are adopted from Table~\ref{tab:fit_params_toi1472}. The left column shows the 5:2 MMR (point 1: $P_c$ = 15.9 days, $e_c$ = 0.37), while the right column shows the 12:5 MMR (point 2: $P_c$ = 15.26 days, $e_c$ = 0.37).
    \label{figure-TOI-1472-critical_angles}}
\end{figure}

We identified several mean motion resonances (MMRs) within the examined orbital period and eccentricity parameter space of TOI-1472\,c, as illustrated in Figure~\ref{figure-TOI-1472-dynamical_map}. Using the \verb|ias15| integrator from the \verb|REBOUND| package \citep{Rein_Spiegel-2015MNRAS.446.1424R}, we computed the temporal evolution of critical angles for each resonance. Figure~\ref{figure-TOI-1472-critical_angles} presents representative examples of the 5:2 MMR and 12:5 MMR, where the upper panels demonstrate libration of the critical angles, confirming the resonant character of the investigated solutions. The ranges of $P_c$ and $e_c$ parameter values explored in these tests extend well beyond the $3 \sigma$ range. The nominal set of initial conditions for the TOI-1472 system listed in Table~\ref{tab:fit_params_toi1472} corresponds to a non-resonant solution.

The dominance of the number of stable over unstable solutions in the study area (Figure~\ref{figure-TOI-1472-dynamical_map}) results in faster calculations using the REM factor compared to the popular Mean Exponential Growth factor of Nearby Orbits \citep{Panichi_Gozdziewski_Turchetti-2017MNRAS.468..469P}. CPU time can be at least doubled by using the MEGNO method. The application of the REM method for the TOI-1472 is therefore justified.

\begin{figure}
\centering
\includegraphics[width=1.000\linewidth]{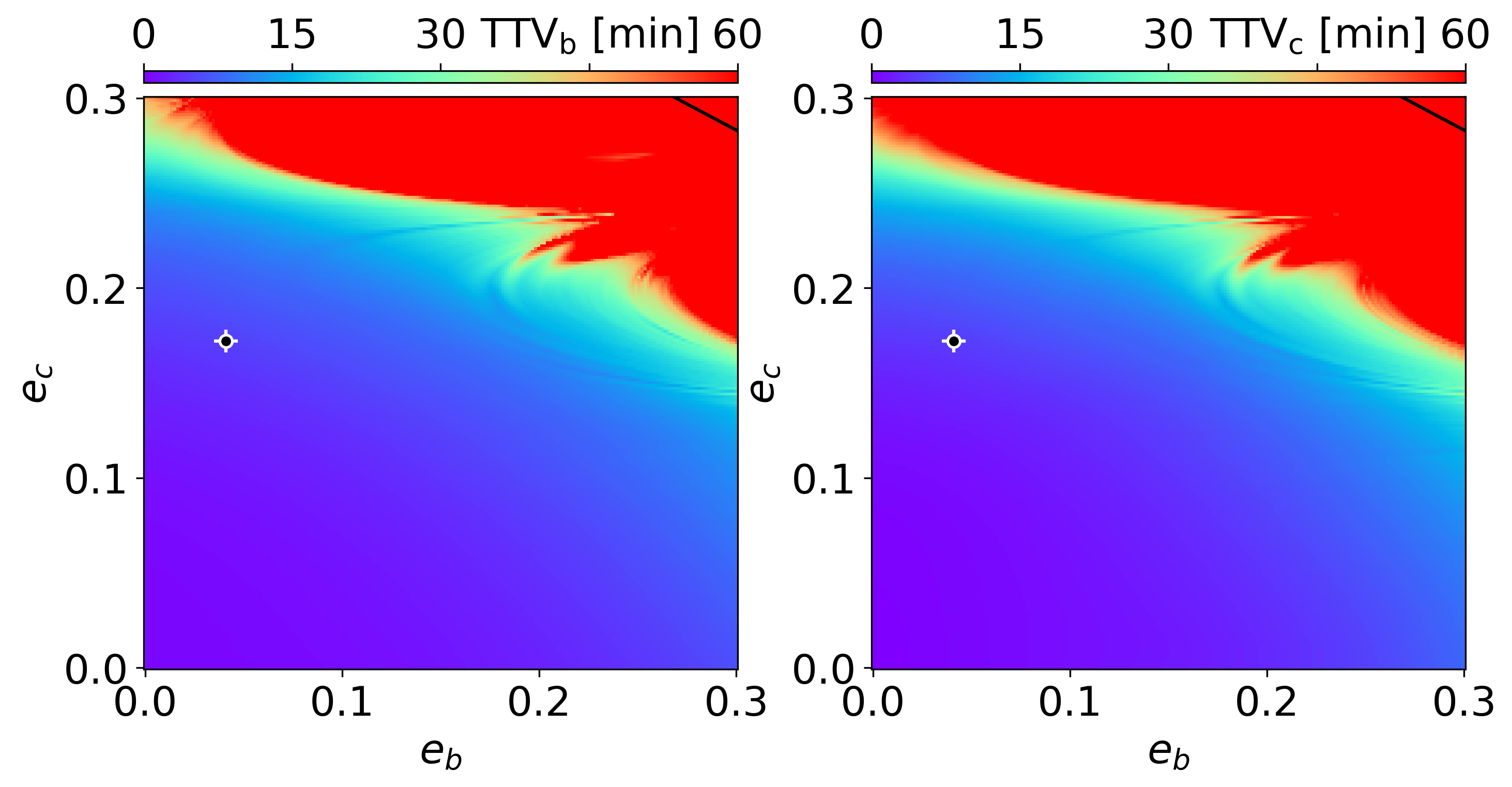}
    \caption{TTV amplitude in the ($e_b$, $e_c$)-plane calculated over the temporal span of the TOI-1472 photometric data. The black filled circle with a white rim shows the location of the solution from our modeling (Table~\ref{tab:fit_params_toi1472}). The error bars represent $3 \sigma$ uncertainties.
    \label{figure-TOI-1472-ttvs}}
\end{figure}

Due to gravitational interactions between the planets in the TOI-1472 system, arising from their close proximity and non-zero eccentricities, we evaluated the transit timing variation (TTV) amplitude. For these calculations, we again employed the \verb|ias15| integrator from the \verb|REBOUND| package. The results presented in Figure~\ref{figure-TOI-1472-ttvs}, displayed in the eccentricity space of both planets, were computed over $\sim$ 300 transits for planet b and $\sim$ 120 transits for planet c, spanning the temporal coverage of TESS photometric observations from sector 17 to sector 85. The TTV amplitudes obtained for the nominal solution given in Table~\ref{tab:fit_params_toi1472} reach $\sim$5 minutes, making them nearly undetectable with the available photometric precision.

\subsection{TOI-1648\,b}
\label{sec-toi-1648b}

We find TOI-1648\,b to have a slightly eccentric orbit with a period of $\sim$7.33 days, and we measure a mass of 7.4$_{-1.3}^{+1.1}$ M$_{\rm \oplus}$ and a radius of 2.54$_{-0.12}^{+0.14}$ R$_{\rm \oplus}$ (Table \ref{tab:fit_params_toi1648}, Figure \ref{fig:massradius}).  
TOI-1648\,b belongs to the group of small sub-Neptunes defined by \cite{Kempton_et_al-2018PASP..130k4401K} as planets with radii between 1.5 and 2.75\,$R_{\oplus}$. Recently, \cite{Hordetal2024} presented a sample of {\it TESS} planets and planet candidates optimal for transmission and emission spectroscopy with {\it JWST} in a given range of planetary sizes and equilibrium temperatures. With T$_{eq}$ = $799 \pm 16$\,K, TOI-1648\,b lies exactly at the boundary of two of the equilibrium temperature ranges defined by \cite{Hordetal2024} between 350\,K and 800\,K and between 800\,K and 1250\,K. Therefore, we have plotted all {\it TESS} small sub-Neptune planets and planetary candidates from the two above-mentioned equilibrium temperature groups in Figure~\ref{figure-TOI-1472-TOI-1648-pr-tsm-coloured_annotations}. With $\mathrm{TSM} = 58.9 \pm 14.5$ TOI-1648\,b ranks sixth in the group of small sub-Neptunes with equilibrium temperatures between 350\,K and 1250\,K and mass determination $> 5\sigma$. With the value of $5.0 \pm 0.6$ of the ESM, TOI-1648\,b is also suitable for atmospheric characterization with the emission spectroscopy.

TOI-1648\,b presents a relatively high eccentricity of 0.178$_{-0.053}^{+0.075}$ (Figure \ref{fig:eccentricity}, upper panel). This suggests that tidal circularization alone cannot fully explain its current orbital configuration (assuming that the planet moved inward during the early stages of the system's formation). Possible explanations include interactions with distant companions, which could maintain or excite its eccentricity via secular perturbations, as proposed in the companion-driven excitation scenario. Alternatively, atmospheric tides or sustained atmospheric evaporation could delay orbital circularization \citep{2018MNRAS.479.5012O}. Thus, TOI-1648\,b may exemplify a system where tidal damping has not yet circularized the orbit, aligning with \citet{2020A&A...635A..37C}’s observation that some warm Neptunes retain moderate eccentricities even after billions of years.

\subsection{Mass-Radius Diagram}
\label{mass-radius-section}
The three planets in our sample occupy distinct positions on the mass–radius diagram (Figure \ref{fig:massradius}), highlighting a diversity of internal structures and atmospheric outcomes. TOI-1648\,b (2.5 R$_\oplus$, 7.4 M$_\oplus$) lies directly on the pure-H$2$O composition curve from \citet{Zeng2016}, consistent with a volatile-rich interior and little to no H/He envelope. Its bulk density of $\sim$2.6 g cm$^{-3}$ disfavors a rocky super-Earth interpretation and instead points to an ice-dominated planet that has either lost its primordial atmosphere or failed to accrete one of significant size. TOI-1472\,c (3.33 R$_\oplus$, 21.1 M$_\oplus$) is located slightly above the pure-H$2$O line, indicating the presence of a modest H/He layer atop a water-rich core. Its density of $\sim$3.2 g cm$^{-3}$ suggests that this envelope constitutes only a small fraction of the total planetary mass. By contrast, TOI-1472\,b (4.06 R$_\oplus$, 18.1 M$_\oplus$) lies well above the 1\% H/He-on-water-rich-core track, implying a more extended gaseous envelope and a substantially lower density of $\sim$1.5 g cm$^{-3}$.

These structural inferences are consistent with our atmospheric evolution simulations. For TOI-1472\,b, JADE modeling under the assumption of early disk-driven migration shows that the planet likely formed with an initial mass of $27.7 \pm 1.2$ M$_\oplus$ and an envelope comprising $\sim$39\% of its mass. The model further predicts that the planet remains in a state of ongoing atmospheric escape, with its radius still contracting from an initially inflated $\sim$10.2 R$_\oplus$. This explains its present-day position above the 1\% H/He envelope curve. TOI-1472\,c, by contrast, is consistent with a history of more efficient atmospheric loss, leaving behind only a thin gaseous layer. Meanwhile, TOI-1648\,b’s alignment with the pure-H$_2$O track implies that it is either an almost atmosphere-free water world or has retained only a negligible volatile component after prolonged erosion on its short-period orbit.

Taken together, the three planets trace an evolutionary sequence across the sub-Neptune regime: from a nearly atmosphere-free water world (TOI-1648\,b), to a water-rich planet with a tenuous H/He veneer (TOI-1472\,c), to a volatile-rich sub-Neptune still undergoing mass loss (TOI-1472\,b). Their relative positions in mass–radius space emphasize how small differences in early envelope accretion and irradiation history can drive divergent present-day outcomes in bulk density and atmospheric observability.

\section{Conclusions} \label{sec:concl}
As part of the KESPRINT collaboration, we have presented a comprehensive characterization of three transiting planets identified by TESS, combining precise radial velocity measurements and transit modeling to refine their fundamental parameters and investigate their internal structure, atmospheric evolution, and dynamical stability.

We provide an improved mass measurement and detailed analysis of the previously known \citep{Polanskietal2024} warm sub-Neptune TOI-1472\,b (4.06~R$_\oplus$, 18.1~M$_\oplus$), which orbits a G-type star with a period of 9.2~days. Our analysis places it at the outer edge of the Neptunian ridge and near the transition into the Neptunian savanna—an underpopulated region in radius-period-density space. The combination of low density and its location near this transitional zone makes it a compelling target for atmospheric evolution modeling. Using internal structure and evolutionary simulations, we infer that TOI-1472\,b likely formed with a significantly larger radius and envelope mass fraction ($\sim$39\%), and has since undergone substantial atmospheric loss due to early and prolonged exposure to high stellar irradiation. Our forward modeling suggests the planet is still losing atmosphere and evolving toward a smaller radius.

In addition to refining the parameters of TOI-1472\,b, we discovered a second planet in the system, TOI-1472\,c, with a mass of 21.1 M$_\oplus$, a radius of 3.33 R$_\oplus$ and a slightly eccentric orbit ($e$=0.172). We performed a dynamical stability analysis and confirmed that the two-planet configuration is long-term stable and dynamically non-resonant.
If both planets formed simultaneously in the protoplanetary disk and migrated inward, one would expect near-circular, approximately resonant orbits and a lower density for the outer planet because of reduced irradiation. Instead, TOI-1472\,c appears unusually dense and on an eccentric, non-resonant orbit. This combination suggests a different evolutionary pathway — for example, one or more giant impacts could increase the planet's density and excite its eccentricity, producing the observed departure from resonance. High-eccentricity migration, however, is unlikely for planet c given its small radius and relatively large semi-major axis.

Among the two planets, TOI-1472\,b emerges as an excellent target for atmospheric characterization, with both TSM and ESM values that favor future JWST transmission and emission spectroscopy.

We also present the discovery and characterization of TOI-1648\,b, a 2.5 R$_\oplus$, 7.4 M$_\oplus$ sub-Neptune on a 7.3-day orbit, which lies at the temperature boundary between two key JWST spectroscopic windows. Its favorable TSM (58.9) and ESM (5.0) place it among the most promising small warm sub-Neptunes accessible for transmission and emission spectroscopy, especially among those with high-significance mass measurements. This makes TOI-1648\,b a valuable addition to the sample of well-characterized planets suitable for comparative atmospheric studies.

Together, these planets enrich the diversity of known sub-Neptunes and offer key test cases for models of planetary structure, migration, and atmospheric evolution.

\section*{Acknowledgements}

I.C. acknowledge financial contribution from the INAF Large Grant 2023 ``EXODEMO''.
C.M.P., gratefully acknowledge the support of the  Swedish National Space Agency (DNR 65/19). 
We acknowledge financial support from the Agencia Estatal de Investigaci\'on of the Ministerio de Ciencia e Innovaci\'on MCIN/AEI/10.13039/501100011033 and the ERDF “A way of making Europe” through project PID2021-125627OB-C32, and from the Centre of Excellence “Severo Ochoa” award to the Instituto de Astrofisica de Canarias.
T.M. acknowledges financial support from the Spanish Ministry of Science and Innovation (MICINN) through the Spanish State Research Agency, under the Severo Ochoa Program 2020-2023 (CEX2019-000920-S) and from the Spanish Ministry of Science and Innovation with the grant no. PID2023-146453NB-100 (\textit{PLAtoSOnG}). H.J.D acknowledges funding from the same source under grants PID2019-107061GB-C66 and PID2023-149439NB-C41. 
DRC acknowledges partial support from NASA Grant $18-2XRP18_2-0007$. This research has made use of the Exoplanet Follow-up Observation Program (ExoFOP; DOI: $10.26134/ExoFOP5$) website, which is operated by the California Institute of Technology, under contract with the National Aeronautics and Space Administration under the Exoplanet Exploration Program. Based on observations obtained at the Hale Telescope, Palomar Observatory, as part of a collaborative agreement between the Caltech Optical Observatories and the Jet Propulsion Laboratory operated by Caltech for NASA. The Observatory was made possible by the generous financial support of the W. M. Keck Foundation. The authors wish to recognize and acknowledge the very significant cultural role and reverence that the summit of Maunakea has always had within the Native Hawaiian community. We are most fortunate to have the opportunity to conduct observations from this mountain.
G.N. thanks for the research funding from the Ministry of Science and Higher Education programme the "Excellence Initiative - Research University" conducted at the Centre of Excellence in Astrophysics and Astrochemistry of the Nicolaus Copernicus University in Toru\'n, Poland.
D.J., K.G. and G.N. gratefully acknowledges the Centre of Informatics Tricity Academic Supercomputer and networK (CI TASK, Gda\'nsk, Poland) for computing resources (grant no. PT01016).
Some of the observations in this paper made use of the High-Resolution Imaging instrument ‘Alopeke and were obtained under Gemini LLP Proposal Number: GN/S-2021A-LP-105. ‘Alopeke was funded by the NASA Exoplanet Exploration Program and built at the NASA Ames Research Center by Steve B. Howell, Nic Scott, Elliott P. Horch, and Emmett Quigley. Alopeke was mounted on the Gemini North telescope of the international Gemini Observatory, a program of NSF’s OIR Lab, which is managed by the Association of Universities for Research in Astronomy (AURA) under a cooperative agreement with the National Science Foundation. on behalf of the Gemini partnership: the National Science Foundation (United States), National Research Council (Canada), Agencia Nacional de Investigaci\'on y Desarrollo (Chile), Ministerio de Ciencia, Tecnología e Innovación (Argentina), Ministério da Ciência, Tecnologia, Inovações e Comunicações (Brazil), and Korea Astronomy and Space Science Institute (Republic of Korea). 
J.L.-B. is partially funded by the NextGenerationEU/PRTR grant CNS2023-144309 and national projects PID2019-107061GB-C61 and PID2023-150468NB-I00 by the Spanish Ministry of Science and Innovation/State Agency of Research MCIN/AEI/10.13039/501100011033.
Support for this work was provided by NASA through the NASA Hubble Fellowship grant HST-HF2-51559.001-A awarded by the Space Telescope Science Institute, which is operated by the Association of Universities for Research in Astronomy, Inc., for NASA, under contract NAS5-26555.
This project has received funding from the European Research Council (ERC) under the European Union's Horizon 2020 research and innovation programme (project {\sc Spice Dune}, grant agreement No 947634).
This work has been carried out in the frame of the National Centre for Competence in Research PlanetS supported by the Swiss National Science Foundation (SNSF) under grants 51NF40\_182901 and 51NF40\_205606. P.E. acknowledges support from the
SNF grant No 219745. 
The authors acknowledge support from the Swiss NCCR PlanetS and the Swiss National Science Foundation. This work has been carried out within the framework of the NCCR PlanetS supported by the Swiss National Science Foundation under grants 51NF40182901 and 51NF40205606. J.K. gratefully acknowledges the support of the Swedish National Space Agency (SNSA; DNR 2020-00104) and of the Swiss National Science Foundation under grant number TMSGI2\_211697.
R.L. acknowledges financial support from the Severo Ochoa grant CEX2021-001131-S funded by MCIN/AEI/10.13039/501100011033 and the European Union (ERC, THIRSTEE, 101164189). Views and opinions expressed are however those of the author(s) only and do not necessarily reflect those of the European Union or the European Research Council. Neither the European Union nor the granting authority can be held responsible for them.
G.M. acknowledges financial support from the Severo Ochoa grant CEX2021-001131-S and from the Ramón y Cajal grant RYC2022-037854-I funded by MCIN/AEI/1144 10.13039/501100011033 and FSE+.
J.V. acknowledges financial support from ANID / Fondo ALMA 2024 / No. 31240039 “On the origin of Warm-Giant Planets”. 
F.M. acknowledges the financial support from the Agencia Estatal de Investigaci\'on del Ministerio de Ciencia, Innovaci\'on y Universidades (MCIU/AEI) through grant PID2023-152906NA-I00.

This work makes use of observations from the LCOGT network. Part of the LCOGT telescope time was granted by NOIRLab through the Mid-Scale Innovations Program (MSIP). MSIP is funded by NSF.
This paper is based on observations made with the Las Cumbres Observatory’s education network telescopes that were upgraded through generous support from the Gordon and Betty Moore Foundation.
Funding for the TESS mission is provided by NASA's Science Mission Directorate. KAC acknowledges support from the TESS mission via subaward s3449 from MIT.
This paper is based on observations made with the MuSCAT3 instrument, developed by the Astrobiology Center and under financial supports by JSPS KAKENHI (JP18H05439) and JST PRESTO (JPMJPR1775), at Faulkes Telescope North on Maui, HI, operated by the Las Cumbres Observatory.

\section*{Data Availability}
The data underlying this article are available in the article and in its online supplementary material.



\bibliographystyle{mnras}
\bibliography{references} 




\appendix

\section{RV and activity indicators datasets}
\label{app:rv_data}

\onecolumn

\begin{longtable}[!ht]{l|cccccrc}
\caption{\label{tab:rvdata_toi1472} Time series of TOI-1472 from HARPS-N and HIRES data: Julian dates, RVs, S-index and their related uncertainties. For HARPS-N data the \logrhk values are listed as well. }\\
\hline
\noalign{\smallskip}
& \multirow{2}{*}{JD }  &      RV  & $\sigma_{\rm RV}$  &        S-index  & $\sigma_{\rm S-index}$  & $\rm log\,R^{\prime}_\mathrm{HK}$ & $\sigma_{\rm log\,R^{\prime}_\mathrm{HK}}$  \\ 
     &        & (\ms) &    (\ms)   & &  & & \\
\hline             
\noalign{\smallskip}
HARPS-N &   2458862.347322   &       -14857.16   &       1.09   &      0.253   &      0.004   &      -4.83   &       0.01  \\
     & 2458895.370240   &       -14848.93   &       2.07   &      0.301   &      0.008   &      -4.74   &       0.01  \\
     & 2458896.359494   &       -14846.33   &       1.72   &      0.284   &      0.006   &      -4.77   &       0.01  \\
     & 2458897.362492   &       -14839.18   &       2.64   &      0.275   &      0.007   &      -4.78   &       0.01  \\
     & 2458898.360083   &       -14839.95   &       1.71   &      0.279   &      0.005   &      -4.78   &       0.01  \\
     & 2459132.716760   &       -14854.00   &       1.38   &      0.279   &      0.010   &      -4.78   &       0.02  \\
     & 2459148.685964   &       -14868.38   &       1.48   &      0.275   &      0.006   &      -4.78   &       0.01  \\
     & 2459149.517373   &       -14874.18   &       1.62   &      0.281   &      0.004   &      -4.77   &       0.01  \\
     & 2459150.532833   &       -14861.89   &       1.43   &      0.286   &      0.005   &      -4.76   &       0.01  \\
     & 2459151.517034   &       -14855.56   &       1.46   &      0.293   &      0.012   &      -4.75   &       0.02  \\
     & 2459160.502878   &       -14855.70   &       1.20   &      0.258   &      0.018   &      -4.82   &       0.04  \\
     & 2459204.441437   &       -14850.15   &       1.88   &      0.262   &      0.005   &      -4.81   &       0.01  \\
     & 2459205.480395   &       -14848.11   &       1.29   &      0.281   &      0.012   &      -4.77   &       0.02  \\
     & 2459206.451903   &       -14852.12   &       0.99   &      0.289   &      0.009   &      -4.76   &       0.02  \\
     & 2459248.358036   &       -14839.09   &       1.21   &      0.302   &      0.016   &      -4.74   &       0.03  \\
     & 2459249.359368   &       -14844.27   &       2.12   &      0.287   &      0.009   &      -4.76   &       0.02  \\
     & 2459265.334407   &       -14851.75   &       3.06   &      0.276   &      0.006   &      -4.78   &       0.01  \\
     & 2459268.347483   &       -14847.50   &       1.23   &      0.279   &      0.007   &      -4.78   &       0.01  \\
     & 2459405.689938   &       -14842.43   &       1.98   &      0.318   &      0.011   &      -4.71   &       0.02  \\
     & 2459410.640158   &       -14859.12   &       1.64   &      0.323   &      0.008   &      -4.70   &       0.01  \\
     & 2459432.724961   &       -14853.55   &       1.29   &      0.312   &      0.006   &      -4.72   &       0.01  \\
     & 2459469.688165   &       -14842.09   &       1.49   &      0.312   &      0.007   &      -4.72   &       0.01  \\
     & 2459549.505157   &       -14854.55   &       1.43   &      0.303   &      0.008   &      -4.73   &       0.01  \\
     & 2459572.468757   &       -14854.81   &       1.54   &      0.301   &      0.009   &      -4.74   &       0.02  \\
     & 2459593.391779   &       -14847.19   &       1.48   &      0.323   &      0.008   &      -4.70   &       0.01  \\
     & 2459609.368066   &       -14846.28   &       3.73   &      0.333   &      0.030   &      -4.69   &       0.05  \\
      &2459610.358438   &       -14844.42   &       1.50   &      0.314   &      0.009   &      -4.72   &       0.01  \\
      &2459765.638588   &       -14852.24   &       1.19   &      0.308   &      0.006   &      -4.72   &       0.01  \\
      &2459766.632046   &       -14855.66   &       1.92   &      0.306   &      0.012   &      -4.73   &       0.02  \\
      &2459785.657277   &       -14862.89   &       1.57   &      0.311   &      0.009   &      -4.72   &       0.01  \\
     & 2459810.618441   &       -14850.64   &       1.41   &      0.322   &      0.007   &      -4.70   &       0.01  \\
     & 2459811.612679   &       -14851.81   &       1.72   &      0.323   &      0.009   &      -4.70   &       0.01  \\
     & 2459812.650632   &       -14849.92   &       1.67   &      0.322   &      0.009   &      -4.70   &       0.01  \\
     & 2459813.633064   &       -14848.98   &       4.99   &      0.338   &      0.038   &      -4.68   &       0.06  \\
     & 2459814.644726   &       -14848.58   &       2.26   &      0.317   &      0.014   &      -4.71   &       0.02  \\
     & 2459882.688144   &       -14860.05   &       1.61   &      0.312   &      0.010   &      -4.72   &       0.02  \\
     & 2459883.631847   &       -14850.99   &       1.33   &      0.296   &      0.007   &      -4.75   &       0.01  \\
     & 2459888.505341   &       -14844.75   &       1.08   &      0.314   &      0.005   &      -4.72   &       0.01  \\
     & 2459888.605016   &       -14842.50   &       1.38   &      0.319   &      0.007   &      -4.71   &       0.01  \\
     & 2459890.567454   &       -14834.69   &       1.40   &      0.325   &      0.007   &      -4.70   &       0.01  \\
     & 2459911.499567   &       -14860.87   &       1.88   &      0.293   &      0.011   &      -4.75   &       0.02  \\
     & 2459912.325684   &       -14866.14   &       1.17   &      0.291   &      0.006   &      -4.75   &       0.01  \\
     & 2459913.428690   &       -14860.74   &       1.08   &      0.315   &      0.005   &      -4.71   &       0.01  \\
     & 2459931.372061   &       -14863.51   &       1.32   &      0.333   &      0.007   &      -4.69   &       0.01  \\
     & 2459934.365966   &       -14841.16   &       1.13   &      0.300   &      0.005   &      -4.74   &       0.01  \\
     & 2459936.468371   &       -14843.89   &       1.52   &      0.296   &      0.009   &      -4.75   &       0.01  \\
     & 2459937.396886   &       -14856.30   &       1.28   &      0.287   &      0.006   &      -4.76   &       0.01  \\
     & 2459950.353060   &       -14860.33   &       2.08   &      0.311   &      0.013   &      -4.72   &       0.02  \\
     & 2459958.345769   &       -14857.41   &       1.81   &      0.347   &      0.011   &      -4.66   &       0.02  \\
     & 2459972.412831   &       -14849.02   &       2.48   &      0.311   &      0.017   &      -4.72   &       0.03  \\
     & 2459989.349576   &       -14861.51   &       2.55   &      0.340   &      0.018   &      -4.67   &       0.03  \\
     &  2460259.580937    &   -14846.58      &      1.43  &  0.320          &  0.008 & -4.71          & 0.01 \\
\noalign{\smallskip}
\hline
\noalign{\smallskip}
HIRES &  2459071.110466   &           -1.98   &       1.71   &      0.245   &      0.001   &   &  \\
     &  2459093.055865   &           -6.64   &       1.66   &      0.236   &      0.001   &   &  \\
     &  2459117.044165   &          -15.61   &       1.54   &      0.258   &      0.001   &   &  \\
     &  2459119.057556   &           -5.17   &       1.58   &      0.253   &      0.001   &   &  \\
     &  2459119.099490   &           -2.61   &       1.74   &      0.244   &      0.001   &   &  \\
     &  2459119.142826   &           -7.16   &       1.80   &      0.235   &      0.001   &   &  \\
     &  2459188.937172   &           14.56   &       1.93   &      0.256   &      0.001   &   &  \\
     &  2459232.768015   &            1.07   &       1.65   &      0.238   &      0.001   &   &  \\
     &  2459268.720687   &            7.62   &       1.80   &      0.230   &      0.001   &   &  \\
     &  2459378.106313   &           -8.99   &       1.62   &      0.270   &      0.001   &   &  \\
     &  2459409.115634   &           -3.24   &       1.50   &      0.306   &      0.001   &   &  \\
     &  2459445.118985   &            4.82   &       1.48   &      0.285   &      0.001   &   &  \\
     &  2459470.920928   &            8.93   &       1.42   &      0.279   &      0.001   &   &  \\
     &  2459497.877869   &            0.10   &       1.76   &      0.260   &      0.001   &   &  \\
     &  2459511.970049   &           -6.42   &       1.85   &      0.288   &      0.001   &   &  \\
     &  2459541.949301   &           -2.15   &       1.65   &      0.266   &      0.001   &   &  \\
     &  2459574.762635   &            1.26   &       1.55   &      0.270   &      0.001   &   &  \\
     &  2459598.718018   &            3.48   &       1.73   &      0.280   &      0.001   &   &  \\
     &  2459632.736802   &           -7.05   &       1.64   &      0.272   &      0.001   &   &  \\
     &  2459738.107008   &            3.89   &       1.76   &      0.291   &      0.001   &   &  \\
     &  2459760.096365   &            2.70   &       1.75   &      0.297   &      0.001   &   &  \\
     &  2459776.105728   &            8.93   &       1.76   &      0.295   &      0.001   &   &  \\
\noalign{\smallskip}
\hline
\noalign{\smallskip}
\end{longtable}

\begin{table*}
\caption{\label{tab:drsdata_toi3862} Time series of additional TOI-1472 activity indicators from HARPS-N Data Reduction Software: bisector, CCF contrast, CCF FWHM, and their related uncertainties.}
\begin{tabular}{crccccc}
\noalign{\smallskip}
\hline
\noalign{\smallskip}
JD & BIS & $\sigma_{\rm BIS}$ & CONT & $\sigma_{\rm CONT}$ & FWHM & $\sigma_{\rm FWHM}$ \\
 & (\ms) & (\ms) & (\kms) & (\kms) & (\kms) & (\kms) \\
 \noalign{\smallskip}
\hline
\noalign{\smallskip}
2458862.347322 & 6.81 & 1.55 & 47.88 & 0.48 & 6.30 & 0.01 \\
2458895.370240 & 7.25 & 2.93 & 47.65 & 0.48 & 6.31 & 0.01 \\
2458896.359494 & 6.00 & 2.43 & 47.63 & 0.48 & 6.33 & 0.01 \\
2458897.362492 & 13.45 & 3.73 & 47.60 & 0.48 & 6.33 & 0.01 \\
2458898.360083 & 11.97 & 2.42 & 47.59 & 0.48 & 6.33 & 0.01 \\
2459132.716760 & 9.21 & 1.95 & 47.74 & 0.48 & 6.31 & 0.01 \\
2459148.685964 & 13.70 & 2.09 & 47.65 & 0.48 & 6.32 & 0.01 \\
2459149.517374 & 8.89 & 2.29 & 47.55 & 0.48 & 6.31 & 0.01 \\
2459150.532833 & 9.35 & 2.03 & 47.56 & 0.48 & 6.31 & 0.01 \\
2459151.517034 & 9.17 & 2.07 & 47.51 & 0.48 & 6.32 & 0.01 \\
2459160.502878 & 15.68 & 1.70 & 47.67 & 0.48 & 6.33 & 0.01 \\
2459204.441437 & -2.40 & 2.66 & 47.70 & 0.48 & 6.30 & 0.01 \\
2459205.480395 & 5.54 & 1.83 & 47.72 & 0.48 & 6.31 & 0.01 \\
2459206.451903 & 9.62 & 1.40 & 47.77 & 0.48 & 6.31 & 0.01 \\
2459248.358036 & 4.97 & 1.71 & 47.64 & 0.48 & 6.33 & 0.01 \\
2459249.359368 & 4.16 & 3.00 & 47.51 & 0.48 & 6.34 & 0.01 \\
2459265.334407 & -9.51 & 4.33 & 47.18 & 0.47 & 6.33 & 0.01 \\
2459268.347483 & 8.19 & 1.75 & 47.74 & 0.48 & 6.30 & 0.01 \\
2459405.689938 & 4.49 & 2.80 & 47.59 & 0.48 & 6.40 & 0.01 \\
2459410.640158 & 9.55 & 2.32 & 47.59 & 0.48 & 6.40 & 0.01 \\
2459432.724961 & 7.16 & 1.82 & 47.68 & 0.48 & 6.39 & 0.01 \\
2459469.688165 & 11.85 & 2.11 & 47.63 & 0.48 & 6.39 & 0.01 \\
2459549.505157 & 7.40 & 2.02 & 47.74 & 0.48 & 6.37 & 0.01 \\
2459572.468757 & 11.18 & 2.18 & 47.74 & 0.48 & 6.37 & 0.01 \\
2459593.391779 & 17.27 & 2.09 & 47.63 & 0.48 & 6.39 & 0.01 \\
2459609.368066 & 21.35 & 5.28 & 47.69 & 0.48 & 6.35 & 0.01 \\
2459610.358438 & 13.91 & 2.13 & 47.79 & 0.48 & 6.36 & 0.01 \\
2459765.638588 & 14.07 & 1.68 & 47.76 & 0.48 & 6.39 & 0.01 \\
2459766.632046 & 11.94 & 2.72 & 47.66 & 0.48 & 6.39 & 0.01 \\
2459785.657277 & 8.01 & 2.22 & 47.77 & 0.48 & 6.39 & 0.01 \\
2459810.618441 & 13.44 & 1.99 & 47.70 & 0.48 & 6.39 & 0.01 \\
2459811.612679 & 14.45 & 2.43 & 47.67 & 0.48 & 6.39 & 0.01 \\
2459812.650632 & 19.56 & 2.37 & 47.64 & 0.48 & 6.40 & 0.01 \\
2459813.633064 & 8.83 & 7.05 & 47.46 & 0.47 & 6.38 & 0.01 \\
2459814.644726 & 11.34 & 3.19 & 47.58 & 0.48 & 6.40 & 0.01 \\
2459882.688144 & 11.44 & 2.28 & 47.77 & 0.48 & 6.39 & 0.01 \\
2459883.631846 & 11.95 & 1.88 & 47.79 & 0.48 & 6.38 & 0.01 \\
2459888.505341 & 9.26 & 1.52 & 47.67 & 0.48 & 6.39 & 0.01 \\
2459888.605016 & 11.11 & 1.95 & 47.63 & 0.48 & 6.39 & 0.01 \\
2459890.567454 & 12.63 & 1.99 & 47.42 & 0.47 & 6.41 & 0.01 \\
2459911.499567 & 16.89 & 2.65 & 47.78 & 0.48 & 6.36 & 0.01 \\
2459912.325684 & 15.46 & 1.65 & 47.85 & 0.48 & 6.37 & 0.01 \\
2459913.42869 & 8.73 & 1.53 & 47.83 & 0.48 & 6.37 & 0.01 \\
2459931.372061 & 28.71 & 1.87 & 47.74 & 0.48 & 6.39 & 0.01 \\
2459934.365966 & 20.30 & 1.59 & 47.84 & 0.48 & 6.36 & 0.01 \\
2459936.468371 & 16.31 & 2.16 & 47.80 & 0.48 & 6.35 & 0.01 \\
2459937.396886 & 15.08 & 1.81 & 47.87 & 0.48 & 6.35 & 0.01 \\
2459950.35306 & 9.86 & 2.94 & 47.48 & 0.47 & 6.39 & 0.01 \\
2459958.345769 & 24.79 & 2.56 & 47.55 & 0.48 & 6.41 & 0.01 \\
2459972.412831 & 10.17 & 3.51 & 47.56 & 0.48 & 6.40 & 0.01 \\
2459989.349576 & 8.42 & 3.60 & 47.48 & 0.47 & 6.41 & 0.01 \\
2460259.580937 & 13.23 & 2.02 & 47.82 & 0.48 & 6.38 & 0.01 \\
\noalign{\smallskip}
\hline
\noalign{\smallskip}
\end{tabular}
\end{table*}

\begin{table*}
\caption{\label{tab:servaldata_toi3862} Time series of TOI-1472 activity indicators from \texttt{serval}: chromospheric index CRX, differential line width dLW, H-alpha, the sodium lines Na$_1$ and Na$_2$, and their related uncertainties.}
\begin{tabular}{crcrccccccc}
\noalign{\smallskip}
\hline
\noalign{\smallskip}
JD & CRX & $\sigma_{\rm CRX}$ & DLW & $\sigma_{\rm DLW}$ & H$\alpha$ & $\sigma_{\rm H\alpha}$ & Na1 & $\sigma_{\rm Na1}$ & Na2 & $\sigma_{\rm Na2}$ \\
\noalign{\smallskip}
\hline
\noalign{\smallskip}
2458862.347322 & 23.63 & 8.27 & -28.15 & 1.70 & 0.51 & 0.01 &0.20 & 0.01 & 0.25 & 0.01\\
2458895.370240 & -1.60 & 17.34 & -11.38 & 2.71 & 0.51 & 0.01 &0.22 & 0.01 & 0.28 & 0.01 \\
2458896.359494 & 23.98 & 13.42 & -7.68 & 2.55 & 0.52 & 0.01 &0.21 & 0.01 & 0.27 & 0.01 \\
2458897.362492 & -16.75 & 17.86 & -13.48 & 2.66 & 0.52 & 0.01 & 0.23 & 0.01 & 0.28 & 0.01  \\
2458898.360083 & 9.59 & 12.92 & -8.03 & 2.56 & 0.51 & 0.01 &  0.21 & 0.01 & 0.27 & 0.01\\
2459132.716760 & 3.95 & 9.00 & -16.84 & 2.10 & 0.52 & 0.01 &  0.19 & 0.01 & 0.26 & 0.01 \\
2459148.685964 & 16.01 & 9.54 & -7.27 & 2.07 & 0.52 & 0.01 & 0.20 & 0.01 & 0.27 & 0.01 \\
2459149.517374 & -6.08 & 12.21 & -5.38 & 2.42 & 0.52 & 0.01 & 0.21 & 0.01 & 0.28 & 0.01 \\
2459150.532833 & -16.76 & 9.09 & -3.36 & 2.27 & 0.51 & 0.01 &  0.20 & 0.01 & 0.26 & 0.01 \\
2459151.517034 & -12.76 & 8.13 & -4.12 & 2.49 & 0.51 & 0.01 & 0.20 & 0.01 & 0.27 & 0.01 \\
2459160.502878 & 1.04 & 8.53 & -9.93 & 1.52 & 0.52 & 0.01 & 0.20 & 0.01 & 0.26 & 0.01 \\
2459204.441437 & 14.15 & 12.44 & -12.50 & 2.42 & 0.51 & 0.01 & 0.20 & 0.01 & 0.27 & 0.01 \\
2459205.480395 & 3.37 & 9.78 & -14.18 & 1.95 & 0.51 & 0.01 & 0.20 & 0.01 & 0.27 & 0.01 \\
2459206.451903 & -10.02 & 6.28 & -19.29 & 1.53 & 0.51 & 0.01 &  0.20 & 0.01 & 0.26 & 0.01 \\
2459248.358036 & 0.50 & 9.79 & -7.49 & 1.68 & 0.51 & 0.01 & 0.19 & 0.01 & 0.26 & 0.01 \\
2459249.359368 & -9.59 & 13.18 & -1.72 & 2.80 & 0.51 & 0.01 &  0.20 & 0.01 & 0.26 & 0.01 \\
2459265.334407 & 44.26 & 15.77 & 61.77 & 4.86 & 0.52 & 0.01 & 0.24 & 0.01 & 0.29 & 0.01 \\
2459268.347483 & 7.05 & 7.65 & -17.69 & 1.75 & 0.51 & 0.01 & 0.20 & 0.01 & 0.26 & 0.01 \\
2459405.689938 & -3.13 & 13.44 & -0.04 & 2.92 & 0.51 & 0.01 & 0.20 & 0.01 & 0.25 & 0.01 \\
2459410.640158 & 7.30 & 7.95 & -3.05 & 2.28 & 0.52 & 0.01 & 0.20 & 0.01 & 0.25 & 0.01 \\
2459432.724961 & 2.88 & 8.06 & -7.96 & 1.85 & 0.51 & 0.01 & 0.20 & 0.01 & 0.25 & 0.01 \\
2459469.688165 & -1.93 & 10.47 & -5.16 & 2.07 & 0.51 & 0.01 & 0.19 & 0.01 & 0.26 & 0.01 \\
2459549.505157 & 12.38 & 10.02 & -15.38 & 1.55 & 0.51 & 0.01 & 0.20 & 0.01 & 0.27 & 0.01 \\
2459572.468757 & -7.91 & 11.05 & -16.24 & 1.79 & 0.51 & 0.01 & 0.20 & 0.01 & 0.27 & 0.01 \\
2459593.391779 & -3.37 & 8.96 & -7.79 & 1.60 & 0.52 & 0.01 &0.20 & 0.01 & 0.27 & 0.01 \\
2459609.368066 & -14.65 & 18.32 & -10.66 & 4.38 & 0.51 & 0.01 & 0.22 & 0.01 & 0.29 & 0.01 \\
2459610.358438 & 4.67 & 10.11 & -20.35 & 1.53 & 0.51 & 0.01 &  0.20 & 0.01 & 0.27 & 0.01 \\
2459765.638588 & 1.94 & 6.40 & -12.62 & 1.65 & 0.51 & 0.01 &  0.19 & 0.01 & 0.25 & 0.01 \\
2459766.632046 & 6.07 & 9.86 & -14.33 & 2.04 & 0.52 & 0.01 &  0.19 & 0.01 & 0.25 & 0.01 \\
2459785.657277 & -6.57 & 11.67 & -20.23 & 2.13 & 0.52 & 0.01 &  0.20 & 0.01 & 0.25 & 0.01 \\
2459810.618441 & -3.43 & 9.42 & -10.54 & 1.56 & 0.51 & 0.01 &  0.19 & 0.01 & 0.25 & 0.01 \\
2459811.612679 & -10.00 & 8.94 & -8.06 & 2.38 & 0.51 & 0.01 &  0.20 & 0.01 & 0.26 & 0.01\\
2459812.650632 & 7.24 & 11.38 & -7.55 & 1.96 & 0.52 & 0.01 &  0.20 & 0.01 & 0.26 & 0.01\\
2459813.633064 & -24.04 & 21.47 & -3.28 & 4.56 & 0.52 & 0.01 &  0.19 & 0.01 & 0.25 & 0.01 \\
2459814.644727 & 15.21 & 13.02 & -3.57 & 2.18 & 0.52 & 0.01 & 0.20 & 0.01 & 0.26 & 0.01\\
2459882.688144 & -2.98 & 11.45 & -16.69 & 2.11 & 0.52 & 0.01 &  0.20 & 0.01 & 0.27 & 0.01 \\
2459883.631846 & 5.28 & 9.23 & -18.41 & 1.83 & 0.52 & 0.01 &  0.20 & 0.01 & 0.27 & 0.01 \\
2459888.505341 & -0.91 & 7.99 & -9.26 & 1.21 & 0.52 & 0.01 & 0.19 & 0.01 & 0.27 & 0.01 \\
2459888.605016 & 5.99 & 8.84 & -3.54 & 1.97 & 0.52 & 0.01 &  0.20 & 0.01 & 0.26 & 0.01\\
2459890.567454 & -11.51 & 10.74 & 9.92 & 1.97 & 0.52 & 0.01 &  0.20 & 0.01 & 0.27 & 0.01\\
2459911.499567 & 16.49 & 11.24 & -19.50 & 2.15 & 0.51 & 0.01 & 0.20 & 0.01 & 0.26 & 0.01 \\
2459912.325684 & -8.29 & 9.56 & -23.53 & 1.24 & 0.51 & 0.01 &  0.19 & 0.01 & 0.26 & 0.01 \\
2459913.42869 & -3.73 & 7.59 & -23.40 & 1.36 & 0.52 & 0.01 &  0.19 & 0.01 & 0.26 & 0.01 \\
2459931.372061 & -3.25 & 8.73 & -12.81 & 1.46 & 0.51 & 0.01 & 0.20 & 0.01 & 0.26 & 0.01 \\
2459934.365966 & -3.90 & 7.93 & -21.81 & 1.21 & 0.51 & 0.01 & 0.19 & 0.01 & 0.26 & 0.01 \\
2459936.468371 & 5.25 & 10.16 & -25.49 & 1.96 & 0.51 & 0.01 &0.20 & 0.01 & 0.26 & 0.01\\
2459937.396886 & 2.11 & 8.61 & -24.46 & 1.72 & 0.51 & 0.01 &  0.19 & 0.01 & 0.26 & 0.01 \\
2459950.35306 & -11.10 & 13.07 & 1.22 & 2.56 & 0.51 & 0.01 & 0.20 & 0.01 & 0.27 & 0.01\\
2459958.345769 & -32.08 & 10.25 & 2.52 & 2.17 & 0.52 & 0.01 &  0.20 & 0.01 & 0.27 & 0.01 \\
2459972.412831 & 2.46 & 15.27 & -5.60 & 2.92 & 0.52 & 0.01 & 0.21 & 0.01 & 0.27 & 0.01 \\
2459989.349576 & -22.20 & 15.77 & 4.76 & 3.11 & 0.52 & 0.01 & 0.21 & 0.01 & 0.28 & 0.01\\
2460259.580937 & 12.21 & 11.37  & -19.00 &  1.92  & 0.50 &   0.01 &   0.20 &   0.01   &   0.27   &  0.01   \\
\noalign{\smallskip}
\hline
\noalign{\smallskip}

\end{tabular}
\end{table*}

\begin{table*}
\caption{\label{tab:rvdata_toi1648} Time series of TOI-1648 from HARPS-N data: Julian dates, RVs, S-index, \logrhk\ and their related uncertainties.}
\begin{tabular}{cccccrc}
\hline
\noalign{\smallskip}
JD   &      RV  & $\sigma_{\rm RV}$  &        S-index  & $\sigma_{\rm S-index}$  & $\rm log\,R^{\prime}_\mathrm{HK}$ & $\sigma_{\rm log\,R^{\prime}_\mathrm{HK}}$  \\ 
        & (\ms) &    (\ms)   & &  & & \\
\hline             
\noalign{\smallskip}
 2459132.740420   &       -32044.46   &       0.80   &      0.255   &      0.004   &      -4.94   &       0.01  \\
 2459148.710324   &       -32051.06   &       1.14   &      0.270   &      0.003   &      -4.92   &       0.00  \\
 2459149.605233   &       -32053.28   &       0.92   &      0.257   &      0.004   &      -4.94   &       0.01  \\
 2459150.615213   &       -32049.72   &       0.69   &      0.258   &      0.005   &      -4.94   &       0.01  \\
 2459151.599589   &       -32048.24   &       0.80   &      0.306   &      0.006   &      -4.86   &       0.01  \\
 2459160.585645   &       -32047.62   &       0.95   &      0.302   &      0.004   &      -4.86   &       0.01  \\
 2459204.475924   &       -32044.22   &       1.05   &      0.314   &      0.006   &      -4.85   &       0.01  \\
 2459205.513825   &       -32043.77   &       0.91   &      0.286   &      0.003   &      -4.89   &       0.00  \\
 2459206.513098   &       -32042.40   &       1.09   &      0.271   &      0.006   &      -4.92   &       0.01  \\
 2459248.392205   &       -32045.97   &       0.64   &      0.272   &      0.003   &      -4.91   &       0.00  \\
 2459265.360498   &       -32043.46   &       1.17   &      0.255   &      0.007   &      -4.95   &       0.01  \\
 2459268.373439   &       -32049.35   &       0.66   &      0.241   &      0.004   &      -4.97   &       0.01  \\
 2459451.739271   &       -32045.06   &       1.19   &      0.255   &      0.006   &      -4.94   &       0.01  \\
 2459452.748155   &       -32048.88   &       0.87   &      0.276   &      0.015   &      -4.91   &       0.03  \\
 2459469.708267   &       -32048.50   &       1.08   &      0.262   &      0.006   &      -4.93   &       0.01  \\
 2459470.678768   &       -32050.08   &       2.12   &      0.276   &      0.006   &      -4.91   &       0.01  \\
 2459549.526062   &       -32044.14   &       0.99   &      0.276   &      0.005   &      -4.91   &       0.01  \\
 2459571.512639   &       -32045.60   &       0.94   &      0.260   &      0.017   &      -4.94   &       0.03  \\
 2459573.533827   &       -32048.01   &       0.92   &      0.258   &      0.006   &      -4.94   &       0.01  \\
 2459574.527225   &       -32049.44   &       1.55   &      0.310   &      0.004   &      -4.85   &       0.01  \\
 2459610.427829   &       -32046.98   &       0.98   &      0.276   &      0.006   &      -4.91   &       0.01  \\
 2459765.667457   &       -32049.90   &       0.83   &      0.255   &      0.005   &      -4.94   &       0.01  \\
 2459766.664999   &       -32050.12   &       0.99   &      0.233   &      0.006   &      -4.99   &       0.01  \\
 2459785.695006   &       -32048.95   &       2.60   &      0.202   &      0.023   &      -5.06   &       0.06  \\
 2459810.670124   &       -32048.42   &       1.09   &      0.248   &      0.007   &      -4.96   &       0.01  \\
 2459811.689098   &       -32046.79   &       1.13   &      0.252   &      0.007   &      -4.95   &       0.01  \\
 2459812.695747   &       -32046.59   &       1.30   &      0.262   &      0.009   &      -4.93   &       0.02  \\
 2459813.712035   &       -32047.06   &       2.24   &      0.260   &      0.019   &      -4.94   &       0.03  \\
 2459814.690624   &       -32048.86   &       0.96   &      0.257   &      0.005   &      -4.94   &       0.01  \\
 2459851.716994   &       -32052.63   &       1.29   &      0.252   &      0.008   &      -4.95   &       0.02  \\
 2459882.709453   &       -32055.26   &       0.99   &      0.230   &      0.006   &      -4.99   &       0.01  \\
 2459883.675674   &       -32053.75   &       0.84   &      0.223   &      0.004   &      -5.01   &       0.01  \\
 2459888.527053   &       -32048.81   &       0.67   &      0.238   &      0.003   &      -4.98   &       0.01  \\
 2459892.415132   &       -32045.97   &       1.13   &      0.239   &      0.007   &      -4.98   &       0.01  \\
 2459894.570115   &       -32047.86   &       0.83   &      0.245   &      0.004   &      -4.96   &       0.01  \\
 2459913.479118   &       -32048.92   &       0.74   &      0.227   &      0.003   &      -5.00   &       0.01  \\
 2459932.564677   &       -32045.99   &       1.23   &      0.246   &      0.008   &      -4.96   &       0.02  \\
 2459934.456529   &       -32046.83   &       0.72   &      0.244   &      0.003   &      -4.97   &       0.01  \\
 2459937.483906   &       -32045.17   &       0.76   &      0.242   &      0.004   &      -4.97   &       0.01  \\
 2459943.448201   &       -32047.42   &       5.89   &      0.265   &      0.067   &      -4.93   &       0.12  \\
 2459950.538722   &       -32050.46   &       1.67   &      0.236   &      0.013   &      -4.98   &       0.03  \\
 2459959.487041   &       -32042.58   &       1.36   &      0.250   &      0.009   &      -4.95   &       0.02  \\
 2459962.433216   &       -32052.73   &       1.13   &      0.225   &      0.007   &      -5.01   &       0.02  \\
 2459973.475613   &       -32046.86   &       1.48   &      0.243   &      0.011   &      -4.97   &       0.02  \\
 2459989.386713   &       -32047.47   &       1.11   &      0.235   &      0.007   &      -4.98   &       0.01  \\
 2459990.378141   &       -32050.18   &       1.50   &      0.230   &      0.011   &      -4.99   &       0.02  \\
 2459995.428123   &       -32041.57   &       0.78   &      0.244   &      0.004   &      -4.97   &       0.01  \\
 2460002.351716   &       -32045.54   &       0.67   &      0.247   &      0.003   &      -4.96   &       0.01  \\
 2460005.367204   &       -32047.99   &       0.61   &      0.244   &      0.003   &      -4.97   &       0.01  \\
 2460007.354761   &       -32049.50   &       0.94   &      0.243   &      0.006   &      -4.97   &       0.01  \\
 2460009.365213   &       -32050.37   &       0.64   &      0.231   &      0.003   &      -4.99   &       0.01  \\
 2460014.342538   &       -32052.81   &       1.00   &      0.234   &      0.006   &      -4.99   &       0.01  \\
 2460259.620739    &    -32044.95      &     0.90  &   0.237      &       0.005 &  -4.98    &         0.01 \\
\noalign{\smallskip}
\hline
\noalign{\smallskip}
\end{tabular}
\end{table*}

\begin{table*}
\caption{\label{tab:drsdata_toi3862} Time series of additional TOI-1648 activity indicators from HARPS-N Data Reduction Software: bisector, CCF contrast, CCF FWHM, and their related uncertainties.}
\begin{tabular}{crccccc}
\noalign{\smallskip}
\hline
\noalign{\smallskip}
JD & BIS & $\sigma_{\rm BIS}$ & CONT & $\sigma_{\rm CONT}$ & FWHM & $\sigma_{\rm FWHM}$ \\
 & (\ms) & (\ms) & (\kms) & (\kms) & (\kms) & (\kms) \\
 \noalign{\smallskip}
\hline
\noalign{\smallskip}
2459132.740420 & 14.85 & 1.13 & 50.90 & 0.51 & 6.28 & 0.01 \\
2459148.710324 & 14.73 & 1.61 & 51.05 & 0.51 & 6.26 & 0.01 \\
2459149.605233 & 13.76 & 1.30 & 51.06 & 0.51 & 6.26 & 0.01 \\
2459150.615213 & 12.08 & 0.98 & 51.08 & 0.51 & 6.25 & 0.01 \\
2459151.599589 & 13.66 & 1.13 & 51.07 & 0.51 & 6.25 & 0.01 \\
2459160.585645 & 7.81 & 1.35 & 51.04 & 0.51 & 6.25 & 0.01 \\
2459204.475924 & 9.43 & 1.49 & 50.92 & 0.51 & 6.28 & 0.01 \\
2459205.513825 & 11.39 & 1.29 & 50.95 & 0.51 & 6.28 & 0.01 \\
2459206.513098 & 13.38 & 1.54 & 50.90 & 0.51 & 6.28 & 0.01 \\
2459248.392205 & 10.29 & 0.91 & 51.04 & 0.51 & 6.27 & 0.01 \\
2459265.360498 & 5.94 & 1.65 & 50.99 & 0.51 & 6.26 & 0.01 \\
2459268.373439 & 11.62 & 0.93 & 51.01 & 0.51 & 6.26 & 0.01 \\
2459451.739271 & 11.90 & 1.68 & 51.03 & 0.51 & 6.32 & 0.01 \\
2459452.748155 & 11.14 & 1.23 & 50.91 & 0.51 & 6.32 & 0.01 \\
2459469.708267 & 6.37 & 1.53 & 51.19 & 0.51 & 6.32 & 0.01 \\
2459470.678768 & 4.01 & 2.99 & 51.06 & 0.51 & 6.31 & 0.01 \\
2459549.526062 & 7.63 & 1.40 & 51.30 & 0.51 & 6.30 & 0.01 \\
2459571.512639 & 10.28 & 1.33 & 51.28 & 0.51 & 6.30 & 0.01 \\
2459573.533827 & 17.46 & 1.30 & 51.31 & 0.51 & 6.30 & 0.01 \\
2459574.527225 & 11.67 & 2.19 & 51.30 & 0.51 & 6.28 & 0.01 \\
2459610.427829 & 13.86 & 1.38 & 51.29 & 0.51 & 6.30 & 0.01 \\
2459765.667457 & 10.19 & 1.18 & 51.38 & 0.51 & 6.29 & 0.01 \\
2459766.664999 & 11.75 & 1.40 & 51.36 & 0.51 & 6.30 & 0.01 \\
2459785.695006 & 13.14 & 3.68 & 51.33 & 0.51 & 6.29 & 0.01 \\
2459810.670124 & 11.35 & 1.55 & 51.33 & 0.51 & 6.30 & 0.01 \\
2459811.689098 & 15.39 & 1.60 & 51.35 & 0.51 & 6.30 & 0.01 \\
2459812.695747 & 7.32 & 1.84 & 51.35 & 0.51 & 6.30 & 0.01 \\
2459813.712035 & 7.40 & 3.17 & 51.24 & 0.51 & 6.30 & 0.01 \\
2459814.690624 & 11.97 & 1.35 & 51.37 & 0.51 & 6.30 & 0.01 \\
2459851.716994 & 4.44 & 1.82 & 51.36 & 0.51 & 6.30 & 0.01 \\
2459882.709453 & 6.50 & 1.39 & 51.49 & 0.51 & 6.28 & 0.01 \\
2459883.675674 & -0.85 & 1.19 & 51.48 & 0.51 & 6.28 & 0.01 \\
2459888.527053 & 4.50 & 0.95 & 51.38 & 0.51 & 6.29 & 0.01 \\
2459892.415132 & 4.85 & 1.60 & 51.16 & 0.51 & 6.30 & 0.01 \\
2459894.570115 & 11.35 & 1.17 & 51.26 & 0.51 & 6.30 & 0.01 \\
2459913.479118 & 9.09 & 1.04 & 51.48 & 0.51 & 6.28 & 0.01 \\
2459932.564677 & 5.06 & 1.74 & 51.37 & 0.51 & 6.29 & 0.01 \\
2459934.456529 & 4.69 & 1.02 & 51.36 & 0.51 & 6.29 & 0.01 \\
2459937.483906 & 13.48 & 1.07 & 51.38 & 0.51 & 6.29 & 0.01 \\
2459943.448201 & 14.98 & 8.33 & 51.08 & 0.51 & 6.28 & 0.01 \\
2459950.538722 & 8.60 & 2.37 & 51.07 & 0.51 & 6.29 & 0.01 \\
2459959.487041 & 10.53 & 1.92 & 51.39 & 0.51 & 6.28 & 0.01 \\
2459962.433216 & 8.41 & 1.60 & 51.43 & 0.51 & 6.29 & 0.01 \\
2459973.475613 & 1.52 & 2.09 & 51.39 & 0.51 & 6.28 & 0.01 \\
2459989.386713 & 5.40 & 1.56 & 51.47 & 0.51 & 6.28 & 0.01 \\
2459990.378141 & 1.74 & 2.12 & 51.40 & 0.51 & 6.28 & 0.01 \\
2459995.428123 & 3.66 & 1.10 & 51.38 & 0.51 & 6.29 & 0.01 \\
2460002.351716 & 7.09 & 0.95 & 51.34 & 0.51 & 6.30 & 0.01 \\
2460005.367204 & 10.33 & 0.87 & 51.36 & 0.51 & 6.29 & 0.01 \\
2460007.354761 & 11.21 & 1.33 & 51.38 & 0.51 & 6.29 & 0.01 \\
2460009.365213 & 11.50 & 0.90 & 51.41 & 0.51 & 6.28 & 0.01 \\
2460014.342538 & 4.89 & 1.41 & 51.38 & 0.51 & 6.28 & 0.01 \\
2460259.620739 & 6.46 & 1.28 & 51.51 & 0.52 & 6.28 & 0.01 \\
\noalign{\smallskip}
\hline
\noalign{\smallskip}
\end{tabular}
\end{table*}

\begin{table*}
\caption{\label{tab:servaldata_toi3862} Time series of TOI-1648 activity indicators from \texttt{serval}: chromospheric index CRX, differential line width dLW, H-alpha, the sodium lines Na$_1$ and Na$_2$, and their related uncertainties.}
\begin{tabular}{crcrccccccc}
\noalign{\smallskip}
\hline
\noalign{\smallskip}
JD & CRX & $\sigma_{\rm CRX}$ & DLW & $\sigma_{\rm DLW}$ & H$\alpha$ & $\sigma_{\rm H\alpha}$ & Na1 & $\sigma_{\rm Na1}$ & Na2 & $\sigma_{\rm Na2}$ \\
\noalign{\smallskip}
\hline
\noalign{\smallskip}
2459132.740420 & -5.56 & 6.22 & 21.90 & 1.95 & 0.55 & 0.01 & 0.18 & 0.01 & 0.22 & 0.01\\
2459148.710324 & 1.90 & 8.76 & 5.41 & 1.65 & 0.55 & 0.01 & 0.18 & 0.01 & 0.23 & 0.01  \\
2459149.605233 & -12.80 & 6.20 & 5.27 & 1.53 & 0.55 & 0.01 & 0.17 & 0.01 & 0.22 & 0.01  \\
2459150.615213 & -3.54 & 6.13 & 3.68 & 1.27 & 0.55 & 0.01 & 0.18 & 0.01 & 0.22 & 0.01  \\
2459151.599589 & 1.85 & 6.42 & 3.71 & 1.30 & 0.55 & 0.01 & 0.18 & 0.01 & 0.22 & 0.01 \\
2459160.585645 & -4.68 & 8.81 & 5.62 & 1.56 & 0.55 & 0.01 &0.17 & 0.01 & 0.22 & 0.01\\
2459204.475924 & 18.16 & 8.66 & 16.15 & 1.67 & 0.55 & 0.01 & 0.17 & 0.01 & 0.22 & 0.01  \\
2459205.513825 & 9.42 & 7.20 & 16.43 & 1.60 & 0.55 & 0.01 &  0.17 & 0.01 & 0.22 & 0.01 \\
2459206.513098 & -14.48 & 7.43 & 18.86 & 1.86 & 0.55 & 0.01 & 0.17 & 0.01 & 0.23 & 0.01\\
2459248.392205 & -6.06 & 5.20 & 5.61 & 1.46 & 0.55 & 0.01 & 0.17 & 0.01 & 0.22 & 0.01 \\
2459265.360498 & 4.44 & 8.19 & 8.99 & 1.97 & 0.55 & 0.01 &  0.18 & 0.01 & 0.23 & 0.01 \\
2459268.373439 & -14.69 & 5.76 & 7.52 & 1.37 & 0.55 & 0.01 & 0.18 & 0.01 & 0.23 & 0.01\\
2459451.739271 & 0.83 & 8.89 & 7.91 & 2.51 & 0.55 & 0.01 & 0.17 & 0.01 & 0.22 & 0.01 \\
2459452.748155 & -0.98 & 6.94 & 24.15 & 1.91 & 0.55 & 0.01 &  0.17 & 0.01 & 0.22 & 0.01 \\
2459469.708267 & 0.51 & 7.88 & -0.06 & 1.57 & 0.55 & 0.01 & 0.17 & 0.01 & 0.22 & 0.01\\
2459470.678768 & -6.67 & 13.67 & 6.34 & 2.91 & 0.55 & 0.01 &  0.18 & 0.01 & 0.22 & 0.01 \\
2459549.526062 & 4.85 & 7.41 & -12.22 & 1.39 & 0.55 & 0.01 &  0.17 & 0.01 & 0.23 & 0.01 \\
2459571.512639 & 3.24 & 7.47 & -11.03 & 1.28 & 0.54 & 0.01 & 0.17 & 0.01 & 0.23 & 0.01 \\
2459573.533827 & 13.63 & 6.96 & -13.96 & 1.08 & 0.55 & 0.01 &0.17 & 0.01 & 0.23 & 0.01 \\
2459574.527225 & -18.74 & 10.97 & -16.49 & 1.62 & 0.55 & 0.01 & 0.17 & 0.01 & 0.23 & 0.01 \\
2459610.427829 & -5.00 & 7.90 & -10.43 & 1.21 & 0.54 & 0.01 & 0.17 & 0.01 & 0.23 & 0.01 \\
2459765.667457 & 2.74 & 7.80 & -18.40 & 0.88 & 0.55 & 0.01 & 0.17 & 0.01 & 0.22 & 0.01\\
2459766.664999 & 1.31 & 7.57 & -18.53 & 1.15 & 0.54 & 0.01 & 0.17 & 0.01 & 0.22 & 0.01  \\
2459785.695006 & 30.28 & 14.15 & -20.39 & 2.65 & 0.55 & 0.01 &  0.18 & 0.01 & 0.23 & 0.01\\
2459810.670124 & 16.24 & 8.04 & -17.62 & 1.46 & 0.54 & 0.01 & 0.17 & 0.01 & 0.22 & 0.01 \\
2459811.689098 & -8.98 & 7.00 & -16.47 & 1.56 & 0.55 & 0.01 & 0.17 & 0.01 & 0.22 & 0.01\\
2459812.695747 & -6.33 & 8.53 & -18.76 & 1.58 & 0.55 & 0.01 &  0.18 & 0.01 & 0.22 & 0.01\\
2459813.712035 & -6.98 & 13.68 & -7.89 & 2.59 & 0.54 & 0.01 & 0.18 & 0.01 & 0.23 & 0.01 \\
2459814.690624 & 4.44 & 6.70 & -20.07 & 1.40 & 0.55 & 0.01 &  0.17 & 0.01 & 0.22 & 0.01 \\
2459851.716994 & 5.35 & 8.31 & -20.11 & 1.87 & 0.55 & 0.01 &  0.18 & 0.01 & 0.22 & 0.01 \\
2459882.709453 & 11.01 & 6.54 & -31.02 & 1.45 & 0.55 & 0.01 &  0.18 & 0.01 & 0.22 & 0.01\\
2459883.675674 & 4.92 & 5.90 & -29.92 & 1.23 & 0.55 & 0.01 & 0.17 & 0.01 & 0.22 & 0.01  \\
2459888.527053 & 6.02 & 5.52 & -18.57 & 0.90 & 0.55 & 0.01 & 0.17 & 0.01 & 0.22 & 0.01 \\
2459892.415132 & 5.45 & 8.75 & -3.48 & 1.43 & 0.55 & 0.01 &  0.17 & 0.01 & 0.23 & 0.01 \\
2459894.570115 & -8.39 & 6.66 & -7.05 & 1.09 & 0.55 & 0.01 & 0.17 & 0.01 & 0.23 & 0.01 \\
2459913.479118 & -0.96 & 6.08 & -28.63 & 1.06 & 0.55 & 0.01 & 0.17 & 0.01 & 0.22 & 0.01  \\
2459932.564677 & 26.11 & 9.66 & -19.48 & 1.57 & 0.55 & 0.01 & 0.17 & 0.01 & 0.23 & 0.01 \\
2459934.456529 & 2.03 & 6.47 & -17.58 & 1.06 & 0.56 & 0.01 & 0.17 & 0.01 & 0.23 & 0.01 \\
2459937.483906 & 5.07 & 5.60 & -18.93 & 1.03 & 0.55 & 0.01 &  0.17 & 0.01 & 0.23 & 0.01 \\
2459943.448201 & 13.22 & 25.91 & -15.07 & 6.58 & 0.56 & 0.01 & 0.17 & 0.01 & 0.24 & 0.01 \\
2459950.538722 & -5.24 & 9.58 & 1.25 & 2.74 & 0.55 & 0.01 & 0.17 & 0.01 & 0.23 & 0.01 \\
2459959.487041 & -13.65 & 8.56 & -24.37 & 1.88 & 0.55 & 0.01 & 0.17 & 0.01 & 0.23 & 0.01\\
2459962.433216 & -13.88 & 7.81 & -25.37 & 1.32 & 0.56 & 0.01 & 0.17 & 0.01 & 0.23 & 0.01 \\
2459973.475613 & 7.82 & 8.86 & -22.48 & 2.22 & 0.55 & 0.01 & 0.18 & 0.01 & 0.23 & 0.01\\
2459989.386713 & -13.72 & 7.10 & -29.78 & 1.32 & 0.55 & 0.01 & 0.17 & 0.01 & 0.23 & 0.01 \\
2459990.378141 & 0.24 & 9.05 & -25.93 & 2.19 & 0.55 & 0.01 & 0.18 & 0.01 & 0.23 & 0.01 \\
2459995.428123 & 7.13 & 6.66 & -21.60 & 1.17 & 0.55 & 0.01 & 0.18 & 0.01 & 0.23 & 0.01\\
2460002.351716 & 4.15 & 6.00 & -15.06 & 0.72 & 0.55 & 0.01 & 0.17 & 0.01 & 0.23 & 0.01 \\
2460005.367204 & -5.41 & 5.54 & -17.27 & 0.82 & 0.55 & 0.01 & 0.17 & 0.01 & 0.23 & 0.01 \\
2460007.354761 & -8.37 & 7.65 & -19.78 & 1.38 & 0.55 & 0.01 & 0.17 & 0.01 & 0.23 & 0.01 \\
2460009.365213 & 2.92 & 6.54 & -23.82 & 0.87 & 0.56 & 0.01 &  0.17 & 0.01 & 0.23 & 0.01 \\
2460014.342538 & 15.00 & 6.85 & -6.57 & 2.83 & 0.55 & 0.01 & 0.18 & 0.01 & 0.23 & 0.01 \\
2460259.620739 & 5.95  & 8.53 & -26.77 & 1.57 & 0.52  &  0.01 & 0.17 & 0.01 & 0.22  & 0.01 \\
\noalign{\smallskip}
\hline
\noalign{\smallskip}

\end{tabular}
\end{table*}

\twocolumn


\bsp	
\label{lastpage}
\end{document}